\documentclass{aa}

\newcommand{\HIdef}{{\rm HI-def.}}
\newcommand{\Htwodef}{{\rm H$_2$-def.}}

\usepackage{placeins}
\usepackage{graphicx}
\usepackage{txfonts}
\usepackage{subfig}
\usepackage{natbib}
\usepackage{comment}
\bibpunct{(}{)}{;}{a}{}{,} 

\usepackage[normalem]{ulem}

\begin{document} 

   \title{Virgo Filaments I: Processing of gas in cosmological filaments around the Virgo cluster\thanks{Tables~\ref{tab:environmental_prop}-\ref{tab:summary_CO_HI_properties_all_galaxies} are available in electronic form at the CDS via anonymous ftp to cdsarc.u-strasbg.fr (130.79.128.5) or via http://cdsweb.u-strasbg.fr/cgi-bin/qcat?J/A+A/}}
   

   \author{G. Castignani
          \inst{1,2,3}\fnmsep\thanks{e-mail: gianluca.castignani@unibo.it}
          \and
          F. Combes\inst{4,5}
          \and
          P. Jablonka\inst{3,6}
          \and
          R. A. Finn\inst{7}
          \and
          G. Rudnick\inst{8}
          \and
          B. Vulcani\inst{9}
          \and
          V. Desai\inst{10}
          \and
          D. Zaritsky\inst{11}
          \and
          P. Salom\'e\inst{4}
          }

   \institute{Dipartimento di Fisica e Astronomia, Alma Mater Studiorum Università di Bologna, Via Gobetti 93/2, I-40129 Bologna, Italy   
   \and
   INAF - Osservatorio  di  Astrofisica  e  Scienza  dello  Spazio  di  Bologna,  via  Gobetti  93/3,  I-40129,  Bologna,  Italy
              \and
              Institute of Physics, Laboratory of Astrophysics, Ecole Polytechnique Fédérale de Lausanne (EPFL), Observatoire de Sauverny, CH-1290 Versoix, Switzerland 
              \and
              Observatoire de Paris, PSL university, Sorbonne University, CNRS, LERMA, F-75014, Paris, France            
               \and
             Coll\`{e}ge de France, 11 Place Marcelin Berthelot, 75231 Paris, France
             \and
             GEPI, Observatoire de Paris, Universit\'{e} PSL, CNRS, Place Jules Janssen, F-92190 Meudon, France
             \and
             Department of Physics \& Astronomy, Siena College, 515 Loudon Road, Loudonville, NY 12211, USA
             \and
             The University of Kansas, Department of Physics and Astronomy, Malott Room 1082, 1251 Wescoe Hall Drive, Lawrence, KS, 66045, USA
             \and
             INAF- Osservatorio astronomico di Padova, Vicolo Osservatorio 5, I-35122 Padova, Italy
             \and
             Spitzer Science Center, California Institute of Technology, MS 220-6, Pasadena, CA 91125, USA
             \and 
             Steward Observatory, University of Arizona, 933 North Cherry Avenue, Tucson, AZ 85721-0065, USA             }
                            \date{Received: December 16, 2020; Accepted: September 20, 2021}

  \abstract{It is now well established that galaxies have different morphologies, gas contents, and star formation rates { (SFR)} in dense environments like galaxy clusters. The impact of environmental density extends to several virial radii, and galaxies appear to be pre-processed in filaments and groups before falling into the cluster. Our goal is to quantify this pre-processing in terms of gas content and { SFR}, as a function of density in cosmic filaments. We have observed the two first CO transitions in 163 galaxies with the IRAM-30m telescope, and added 82 more measurements from the literature, thus forming a sample of 245 galaxies in
the filaments around the Virgo { cluster}. We gathered HI-21cm measurements from the literature and observed 69 galaxies with the Nançay telescope to complete our sample. We compare our filament galaxies with comparable samples from the Virgo cluster and with the
isolated galaxies of the AMIGA sample. We find a clear progression from field galaxies to filament and cluster galaxies for decreasing { SFR}, increasing fraction of galaxies in the quenching phase, an increasing proportion of early-type galaxies, and decreasing gas content.
Galaxies in the quenching phase, defined as having a { SFR} below one-third of that of the main sequence { (MS)}, are only between 0\% and 20\% in the isolated sample, according to local galaxy density, while they are 20\%-60\% in the filaments and 30\%-80\% in the Virgo cluster. Processes that lead to star formation quenching are already at play in filaments; they depend mostly on the local galaxy density, while the distance to the filament spine is a secondary parameter.
While the HI-to-stellar-mass ratio decreases with local density by an order of magnitude in the filaments, and two orders of magnitude in the Virgo cluster with respect to the field, the decrease is much less for the H$_2$-to-stellar-mass ratio. As the environmental density increases, the gas depletion time decreases, because the gas content decreases faster than the { SFR}. This suggests that gas depletion { precedes} star formation quenching.} 

   \maketitle
%

\section{Introduction}\label{sec:introduction}

There is strong observational evidence that dense local environments can have a large impact on the evolutionary path of galaxies. Following the pioneering work by \citet{Dressler1980}, numerous studies have shown that dense megaparsec-scale environments regulate star formation activity \citep{Butcher_Oemler1984,Peng2010} and gas content  \citep{Chung2009,Vollmer2012}. Galaxy clusters are also the sites where the dramatic morphological transformations of galaxies are observed, which are ultimately driven by galaxy--galaxy interactions within the complex cosmic web and give rise to the so-called morphology density relation in the cores of clusters \citep{Postman_Geller1984,Dressler1997,Goto2003}.

Environmental processes can remove gas through tidal heating and stripping that occurs in gravitational interactions and mergers between galaxies \citep{Merritt1983,Moore1998} or ram-pressure stripping due to a passage through the hot intra-cluster gas \citep{Gunn_Gott1972,Roediger_Henssler2005}; gas accretion from the cosmic web can also be suppressed, a process called starvation \citep{Larson1980,Balogh2000}. 
All these processes occur in clusters and in some cases dramatically, for example in the spectacular ram-pressure stripping reported by \citet{Jachym2014,Jachym2019}. Statistically, the HI deficiency in clusters has now been firmly established { \citep[][]{Giovanelli_Haynes1985,Haynes_Giovanelli1986a,Cayatte1990,Cayatte1994,Solanes2001,Gavazzi2005,Chung2009,Hess2015,Healy2021},} and molecular gas is also known to be depleted in dense environments \citep{Casoli1998,Lavezzi_Dickey1998,Vollmer2008,Scott2013}. \citet{Boselli2014,Boselli2014b} also reported tentative evidence of a correlation between HI and H$_2$ deficiencies for cluster galaxies.

There is also ample evidence that galaxy star formation rates are suppressed at distances
up to $\sim5$ virial radii from the cluster center  \citep[e.g.,][]{Lewis2002,Gomez2003,Vulcani2010,Finn2010,Haines2015}.
It is now clear that both field and group galaxies are being pre-processed before they fall into the cluster itself  { \citep[][]{Zabludoff_Mulchaey1998,Poggianti1999,Yutaka2004,Cortese2006,Kern2008,Kilborn2009,Catinella2013,Hess_Wilcots2013,Hou2014,Rudnick2017}.}

Large galaxy redshift surveys have revealed that galaxies are distributed in a complex network of matter ---with a large dynamic range of local density--- called the cosmic web or filamentary structures \citep{Haynes_Giovanelli1986b,Kitaura2009,Darvish2014,Darvish2017,Alpaslan2016,Chen2016,Chen2017,Malavasi2017,Kuutma2017,Kraljic2018,Laigle2018, Sarron2019,Luber2019,Salerno2019}. To determine the effect of environment on galaxy evolution, it is necessary to understand how galaxies are altered as they move through the cosmic web and enter the densest regions of clusters.

Hydrodynamic simulations of the cluster infall regions predict that the
density of gas in filaments is able to enhance the ram pressure by a factor of  up to about $100$ with respect to the pressure in the lower density regions \citep{Bahe2013}. According to the authors, this means that freshly infalling galaxies with stellar masses of $\log(M_\star/M_\odot)<9.5$ near a massive cluster can be stripped of their cold gas even well outside the virial radius. For more massive galaxies or those at larger distances from the cluster, the ram pressure in filaments is still sufficient to strip off the hot gas that will replenish the dense star-forming gas, although it will likely not directly
affect the densest cold gas. The latter will then be consumed on a timescale of $\sim2.3$~Gyr \citep{Bigiel2011}.

{In this work, we observationally quantify the amount of pre-processing of the cold gas of galaxies in cosmic filaments and investigate galaxy properties  as a function of the environment. 
To this aim, we report a multi-wavelength study of a stellar-mass-complete sample of 245 galaxies with $\log(M_\star/M_\odot)\sim9-11$  observed in cold gas, both atomic (HI) and molecular (CO). These sources live in cosmic filaments surrounding Virgo}, the benchmark cluster in the local Universe. Virgo has a distance of $\sim17$~Mpc \citep[][]{Mei2007} and a virial radius of $\sim6$~deg in projection. Due to the complex structure of Virgo, several estimates of around $\sim$2~Mpc have been reported in the literature for its virial radius, namely of 1.55~Mpc \citep{McLaughlin1999}, 1.72~Mpc \citep{Hoffman1980}, and 2.2~Mpc \citep{Fouque2001}. 

{The paper is structured as follows.  In Sect.~\ref{sec:cosmic_web} we characterize the filamentary structures around Virgo.  In Sect.~\ref{sec:our_sample} we describe our sample and present our cold gas observations.
In Sect.~\ref{sec:gas_masses} we derive gas properties such as H$_2$ and HI gas masses. In Sect.~\ref{sec:comparison_samples} we  introduce the comparison samples of field and Virgo cluster galaxies. 
In Sect.~\ref{sec:density_profiles} we provide a description of the environment of our sample of filament galaxies.}
In Sect.~\ref{sec:results} we describe our results. In Sect.~\ref{sec:summary_conclusions} we summarize the results and draw conclusions. In the Appendices~\ref{sec:tables}, \ref{sec:CO_HI_spectra}, and \ref{sec:SFR_gas_diagnostic} we report supplementary material including { tables, spectra, and diagnostic plots, respectively.}

Throughout this paper, we assume a Hubble constant of $H_0 = 100~h^{-1}$~km~s$^{-1}$~Mpc$^{-1}$, where $h=0.74$ \citep{Tully2008}. 
{Stellar masses and star formation rates adopted in this work rely on the \citet{Kroupa_Weidner2003} initial mass function.}

\section{The cosmic web around Virgo}\label{sec:cosmic_web}

Following early work by \citet{deVaucouleurs1953,deVaucouleurs1956} who noted an excess of nearby galaxies in the vicinity of the North Galactic pole, several studies attempted to provide a detailed characterization of the complex cosmic web around Virgo, which is indeed embedded in the Laniakea supercluster, including the local group \citep{Bahcall_Joss1976,Tully1982,Tully2014,Tully2016}. Simulations also contributed to these efforts, which were designed to reproduce the cosmic flow in the local Universe 
in detail \citep{Libeskind2018,Libeskind2020}.
{ As outlined in the following sections, we adopt an approach similar to that used by \citet[][]{Kim2016}, who identified the spines of several cosmological filaments around Virgo.}  \citet{Kim2016} provided a detailed characterization of the filamentary structures around Virgo,  but they released neither their catalog of galaxies nor the filament spines. This motivated us to build our sample of galaxies independently and characterize the filaments surrounding Virgo. We also consider a larger survey area than that covered by \citet{Kim2016}.

\subsection{The sample of galaxies around Virgo}
We have characterized the filamentary network surrounding Virgo  in detail. A release of the catalog of galaxies surrounding Virgo, of the associated filaments, and their  properties is presented in \citet{Castignani2021}. In the following we give an overview of our analysis.

We compiled a large sample of galaxies around Virgo, which have been selected within the J2000 coordinate ranges of 100~deg~<~R.A~<~280~deg and -35~deg<~Dec.~<75~deg, around Virgo cluster, whose center is at (R.A. ; Dec.) = (187.70; 12.34)~deg. We further limited ourselves to sources with heliocentric velocities $V_{\rm H}<3300$~km~s$^{-1}$. { Similarly to \citet{Kim2016}, these cuts ensure the inclusion of all main filamentary structures potentially associated with the Virgo cluster.}
The catalog was built primarily by cross-matching the NASA Sloan Atlas (NSA)\footnote{\url{http://www.nsatlas.org/}} and HyperLeda\footnote{\url{http://leda.univ-lyon1.fr/}}  \citep{Makarov2014} catalogs. { The cross-matching was done by galaxy name, when available, or adopting a search radius of 10~arcsec.}
Our selection yields 10 305~galaxies  in the cosmic web around Virgo, all with unique { NASA/IPAC Extragalactic Database (NED)}
counterparts. Among these sources, in this work we focus on a sample of 245 galaxies  for which observations are available in atomic and molecular gas, either from our own work or from the literature.  These galaxies belong to filamentary structures around Virgo, and are described in Sect.~\ref{sec:our_sample}.

\begin{figure*}[hbt]\centering
\captionsetup[subfigure]{labelformat=empty}
\subfloat[]{\hspace{0.cm}\includegraphics[trim={0cm 0cm 0cm 
0cm},clip,width=0.5\textwidth,clip=true]{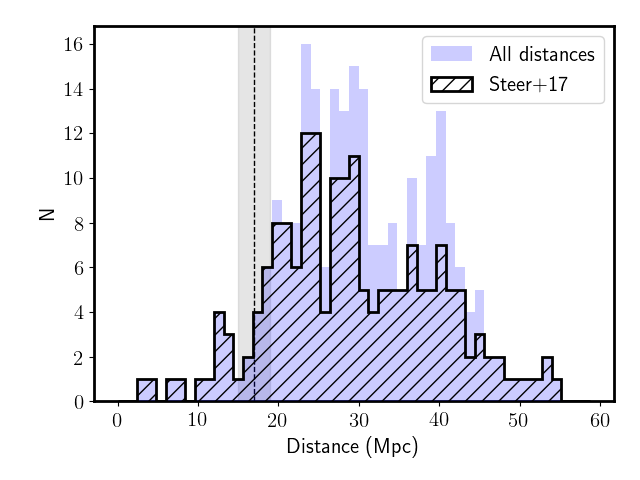}}
\subfloat[]{\hspace{0.7cm}\includegraphics[trim={0cm 0cm 0cm 
0cm},clip,width=0.5\textwidth,clip=true]{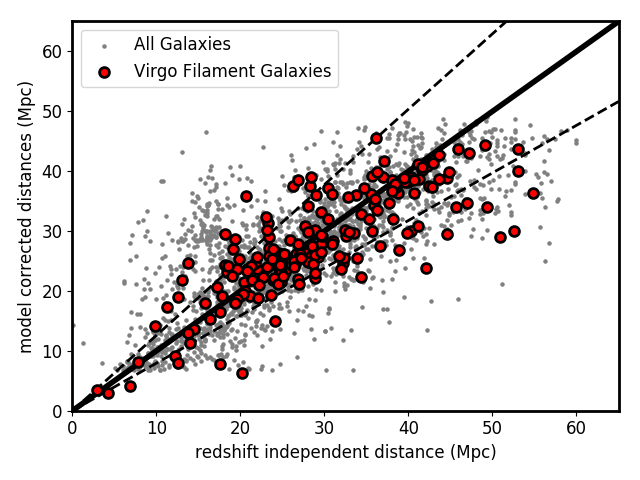}}
\caption{{ Left:} Distribution of distances for the 245 sources of our sample. Redshift-independent distances \citep{Steer2017} are reported in the foreground (dashed histogram), while in the background (blue filled histogram) we report all adopted distances, including those that are model-corrected. { The vertical dashed line shows the Virgo cluster distance, while the gray vertical rectangle corresponds to the $\pm r_{200}$ region.} 
{ Right:} Scatter plot with model-corrected distances (y-axis) vs. redshift-independent distances (x-axis). Red points refer to our sample of 245 filament galaxies, gray points refer to galaxies in the field of Virgo at distances in the range $\sim$(0-60)~Mpc. The solid black line shows the one-to-one relation, while the dashed lines correspond to an rms~=~0.1~dex.}\label{fig:distances}
\end{figure*}

\subsection{Distances}\label{sec:distances}

To characterize the cosmic web around Virgo  in 3D,{ we estimated the intrinsic distances of galaxies by applying the model by \citet{Mould2000} of the velocity field. According to this model,} we first derived the correction of the observed heliocentric velocities of the galaxies to the centroid of the Local Group. We then included an additional correction that takes into account the infall towards the Virgo attractor.
With this procedure, heliocentric radial velocities $V_{\rm H}$ were converted into cosmic radial velocities $V_{\rm cosmic}$.
{ Recession velocities $V_{\rm cosmic}$ were then translated into distances via the Hubble law.}

{ As our goal is to obtain the most accurate distances, we revised them by looking for redshift-independent estimates} within the \citet{Steer2017} catalog. { This is a large database containing distances of more than 28 000 galaxies based on indicators that are independent of cosmological redshift. }
Our search yielded redshift-independent distances for 2 763 sources out of the 10 305~galaxies (27\%) around Virgo, { and we homogenized them to $H_0 = 74$~km~s$^{-1}$~Mpc$^{-1}$ used in this work for better precision.}

{For these 2 763 sources, we replaced the model-dependent distances with redshift-independent ones.
As further outlined below, we verified that this replacement does not introduce any bias for our environmental analysis and, on the contrary, leads to a robust characterization of the cosmic web around Virgo. }

{ Limiting ourselves to our sample of 245 filament galaxies, we{ found redshift-independent distances for 181 of them (74\%). This higher fraction for our sample compared to that of the full catalog can be explained by the fact that our filament galaxies { are massive}, { with $\log(M_\star/M_\odot)\sim9-11$ (Sect.~\ref{sec:Mstar_SFR_our_sample})}, while the cosmic web and filaments in particular tend to be largely populated by less massive systems, including dwarfs \citep{Kim2016}. }

In Fig.~\ref{fig:distances} (left) we report the resulting distance distribution for the 245 filament galaxies in our sample (Sect.~\ref{sec:our_sample}), where redshift-independent distances are highlighted. Our filament sources have a median distance of $\sim29$~Mpc, and are thus behind Virgo cluster \citep[see also][]{Kim2016}. When considering the subsample of 181 sources with redshift-independent distances and the full sample of 245 sources, separately, we obtain median distances of $(28.0^{+12.3}_{-8.4})$~Mpc and $(29.3^{+11.3}_{-8.5})$~Mpc, respectively. The two are thus in agreement with each other.\footnote{Hereafter in this work we report the 1$\sigma$ confidence interval as uncertainty to the median value, unless otherwise stated.}

In Fig.~\ref{fig:distances} (right) we report a comparison between model-corrected distances ($D_{\rm model}$, y-axis) and redshift-independent distances ($D_{z-{\rm independent}}$, x-axis). For the full catalog of sources  around Virgo and the 181 filament galaxies in our sample with redshift-independent distances, the mean logarithmic difference is indeed found to be $\log(D_{\rm model}/D_{z-{\rm independent}})=0.004\pm0.15$ and $-0.015\pm0.10$, respectively. Here the reported uncertainty is the { root mean square (rms)} dispersion around the mean. { This implies a negligible bias and a limited rms scatter of $\sim0.1$~dex, which are values that are competitive with those of recent studies of the local Universe \citep{Leroy2019}. However, this leads to a statistical uncertainty of $\sim58\%$ in mass estimates.}

As seen in Fig.~\ref{fig:distances}, sources with redshift-independent distances $\gtrsim~40$~Mpc tend to deviate from the one-to-one line, showing { redshift-independent} distances that are greater than the corresponding model-corrected distances.  { The apparent deviation from the one-to-one line may be explained as a selection effect, considering that the sources have model-dependent distances not exceeding $\sim45$~Mpc, mainly due to the 3 300~km/s recessional velocity cut of the catalog. }
However, we find that for this subsample of distant sources the mean logarithmic difference of $\log(D_{\rm model}/D_{z-{\rm independent}})=-0.08\pm0.08$ is still limited and within the $1\sigma$ dispersion, for both the full catalog of sources  around Virgo and the subsample of 181 filament sources. 

\begin{figure*}[]\centering
\captionsetup[subfigure]{labelformat=empty}
\subfloat[]{\hspace{0.cm}\includegraphics[page=1, trim={0cm 0cm 0cm 0cm},clip,width=1.1\textwidth,clip=true]{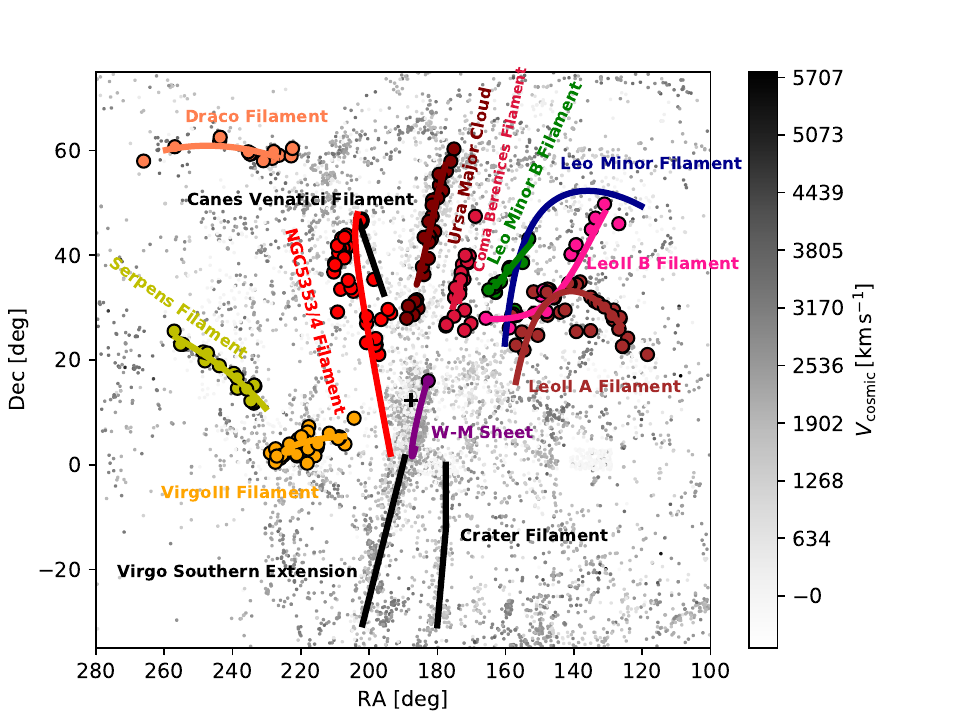}}
\caption{Cosmic web up to $\sim12$ virial radii in the projected sky from Virgo cluster, whose coordinates are (R.A.~;~Dec.)~=~(187.70; 12.34)~deg (J2000) and (SGX; SGY; SGZ)~=~(-2.26; 9.90; -0.42)~$h^{-1}$~Mpc, as denoted by the central cross. Gray points show galaxies color coded according to their{ cosmic velocity.} Large colored points show filament galaxies in our sample with CO and HI observations, while the curves are the locations of the filament spines. Different colors refer to different filaments.}\label{fig:Virgo_field}
\end{figure*}

\subsection{Filament spines}
{ In this section, we characterize the filamentary structures around Virgo by deriving their filament spines.} 
Figure~\ref{fig:Virgo_field} displays the sources in the cosmic web around Virgo,  color coded in gray scale according { to their $V_{\rm cosmic}$.} Our sample of 245 filament galaxies observed both in CO and HI are highlighted with different colors according to the associated filament. Similarly to \citet{Kim2016}, { we considered the following  well-known} filamentary structures in the Northern hemisphere: VirgoIII~filament, NGC~5353/4 filament, and the W-M sheet, as well as the Leo Cloud with the Leo Minor, LeoII~A, and LeoII~B filaments.
For our analysis, we also included the nearby Ursa Major Cloud  \citep{Tully1982,Tully_Fisher1988}.
Some of our 245 sources also belong to additional structures that were considered for our observations, after visual selection. These are the{ Serpens, Draco, Coma Berenices, and Leo Minor B filaments.} These structures were selected as filaments as they are elongated in the plane of the sky and exhibit a clustering of sources at similar recession velocities, which we indeed targeted with our observational campaign (Sect.~\ref{sec:observations}). All these filaments are reported in Fig.~\ref{fig:Virgo_field} { in projection,} in addition to Canes Venatici filament, Virgo Southern Extension, and Crater filament, {  which are shown in black in the figure} and have not been considered in this work.

To characterize the filaments, all galaxies were first mapped into the 3D Cartesian super Galactic (SGX, SGY, SGZ) coordinate system. { We refer to  \citet[][]{Tully2008,Tully2019}, for example, for similar analyses of the local Universe.}
For for each filamentary structure, to determine its spine, we first considered a cube, defined in the Cartesian (SGX,SGY,SGZ) coordinate frame, 
that is large enough to conservatively enclose all galaxies that define the structure.  The spines of the filaments were then determined by fitting the locations of galaxies with a third-order polynomial curve in super Galactic coordinates, as in \citet{Kim2016}. 


\subsection{Environmental parameters}\label{sec:environmental_properties}
We aim to fully characterize the environment of the 245 sources in our sample. Therefore, we estimated several environmental properties as outlined in the following. Quantities were computed in 3D within the (SGX,SGY,SGZ) Cartesian frame or in 2D, by projection onto the (SGX,SGZ) plane.
{ Given the location of Virgo in the sky, SGY coordinates can be considered as a distance proxy, at least at first order.}

{ The k-nearest neighbor densities ($n_k$, with $k=5$) have been used in this work,  both in 2D and in 3D. This is a widely used estimator for the local density. We refer to \citet{Muldrew2012} for a review.  These confer the advantage of probing local densities on approximately megaparsec scales, which are typical of the large-scale structures we are interested in, such as the filaments. We verified that for the 245 sources in our sample, the median distance of the fifth nearest neighbor is $(0.5^{+0.8}_{-0.3})~h^{-1}$~Mpc and $(1.0^{+1.0}_{-0.5})~h^{-1}$~Mpc in the 2D and 3D cases, respectively. The reported uncertainties correspond to a $1\sigma$ confidence interval. By adopting a lower value of $k$ we would probe densities at smaller scales, but at the cost of higher shot-noise uncertainties. By choosing higher values of $k$ we would instead probe densities at larger scales, with the risk of smoothing megaparsec-scale density fluctuations. }


For each galaxy we also estimated the minimum separation ---in the Super Galactic Coordinate frame--- from its corresponding filament spine ($d_{\rm fil}$) and from Virgo cluster center ($d_{\rm cluster}$), both in 3D and in projection (2D). 
When comparing 2D to 3D quantities (densities, $d_{\rm fil}$, and $d_{\rm cluster}$) we did not find any significant difference in our results.
The filamentary structures are elongated mostly along the plane of the sky,{ which can be approximated by} the (SGX,SGZ) plane (see Sect.~\ref{sec:filament_profiles} for further discussion).
{ Virgo is one of the few clusters for which we can obtain a fairly accurate characterization of the environment in 3D;} there are indeed many more redshift-independent distances available than for other clusters at larger distances. However, recession velocity uncertainties for galaxies around more distant clusters are higher.
{ Therefore, in this work we preferentially use 3D quantities rather than projected (2D) quantities.}
In Table~\ref{tab:environmental_prop} we report the environmental properties for our sample.

\section{Our sample of filament galaxies}\label{sec:our_sample}
\subsection{Sample selection}

To characterize the effects of the filament environment on the galaxy gas content, we selected 245 galaxies belonging to the longest filaments with the  highest contrast around Virgo.  
{ The filaments} extend up to several virial radii from the cluster center and span up to $\sim30$~Mpc in length \citep[e.g.,][]{Kim2016}. 



{ The initial selection of the filament galaxies was  done on the basis of their recession velocities and positions in the sky, by requiring proximity to the filamentary structures. { We sampled a wide range of local densities and distances, up to several megaparsec from the filament spine, to be able to determine the influence of the filament environment in the gas processing.} We also restricted to the stellar mass range between $\sim10^9~M_\odot$ and $\sim10^{11}~M_\odot$. Below this range, the low expected metallicity could prevent us from detecting CO { \citep[e.g.,][]{Bolatto2013}}, while at higher masses than this range we expect intrinsic quenching processes to play a more important role { \citep[e.g.,][]{Baldry2006,Peng2010}}. }

{ Our sample is supported by a wealth of existing data which enable robust estimates of the stellar mass, stellar population, and the  star formation rate (SFR) for our galaxies: { Sloan Digital Sky Survey (SDSS)} \textsf{ugriz} imaging, optical spectroscopy, and far-infrared fluxes (WISE, IRAS).}



\subsection{Galaxy properties}
Properties of our sample of filament galaxies are listed in Table~\ref{tab:general_properties_all_galaxies}, which {includes} galaxy coordinates and recession velocities, stellar masses, and star formation rates, both absolute and relative to the main sequence (MS). {We also list morphologies, sizes, inclinations, and position angles, all taken from HyperLeda.}

\subsubsection{Stellar masses and star formation rates}\label{sec:Mstar_SFR_our_sample}
{ In order to provide stellar mass and star formation estimates for our sample we looked for them within the $z=0$ Multiwavelength Galaxy Synthesis release provided by \citet{Leroy2019}.}
This catalog is based on a multi-wavelength dataset including WISE in infrared and GALEX in ultraviolet and provides accurate stellar masses and SFRs of local sources up to distances of $\sim50$~Mpc.

We found stellar masses for 235 sources out of 245 in \citet{Leroy2019}. For the remaining 10 sources, stellar masses were taken from the NASA Sloan Atlas, { revised} to $h=0.74$.
Out of 245 sources in our sample, 231 have $M_\star$ estimates from both \citet{Leroy2019} and the NASA Sloan Atlas. The median logarithmic difference between the two estimates is $-0.05^{+0.18}_{-0.17}$. The comparison thus yields a good agreement, with a negligible bias and a limited 1$\sigma$ scatter.

We found estimates of the SFR by \citet{Leroy2019} for a subsample of  234 sources. For the remaining 11 galaxies we gathered the SFRs from the literature, as follows. For six sources, namely NGC~4144, PGC~023706, PGC~031387, PGC~049002, PGC~1925809, and PGC~2151881,  we have taken the SFRs estimated by \citet{Chang2015}, which are based on both SDSS and WISE photometry,{ similarly to} \citet{Leroy2019}.  For NGC~2592, NGC~4214, and NGC~4244, the SFRs {have been taken} from the DustPedia archive\footnote{\url{http://dustpedia.astro.noa.gr/}}, which provides SFRs estimated by fitting multi-wavelength spectral energy distributions with {\sc Cigale} \citep{Burgarella2005,Noll2009,Boquien2019}. NGC~2793 is a ring galaxy and we have the SFR$\sim0.2~M_\odot$/yr estimated by \citet{Mayya_Romano2002} via H$_\alpha$ imaging. Last, for PGC~049386 (i.e., CGCG~219-021) we  find no estimates in the literature and have therefore converted its 22~$\mu$m W4 WISE emission into SFR using the \citet{Calzetti2007} relation.  We
also refer to Table~\ref{tab:general_properties_all_galaxies}, where the stellar masses and SFRs of all sources are reported. 

For 132 sources out of 245, we found SFR estimates from both \citet{Leroy2019} and \citet{Chang2015}. The median logarithmic difference between the two estimates is $0.48^{+0.93}_{-0.59}$~dex. Our comparison thus suggests that \citet{Leroy2019} SFRs are, on average, {  a factor of approximately three higher} than those estimated by \citet{Chang2015}, although the reported 1$\sigma$ scatter is not negligible.  { SFR estimates can be relatively uncertain, and can differ between analyses by a factor of three or higher \citep[e.g.,][]{daCunha2008,Calzetti2013}.}


\subsubsection{Morphology}
Figure~\ref{fig:histo_Hubble_type} shows the distribution in morphology taken from HyperLeda for our sample of filament galaxies. It can be noted that the barred galaxy fraction is about two-thirds of that normally found for spirals, except for Sb types, { for which the fraction is lower, possibly because of a statistical fluke.}
We note that the classification  likely comes from optical images,
while bars are better seen in the infrared \citep{Eskridge2000}. 
Hereafter, we denote the de Vaucouleurs morphological parameter as $T$, and early-type galaxies ($T<0$) are often distinguished from late-type galaxies ($T\geq0$).

A number of studies have suggested a link between the cessation (quenching) of star formation and the presence of bars \citep[e.g.,][]{James_Percival2016,Fraser-McKelvie2020,Newnham2020}. 
Indeed, bars favor gas inflow, while subsequent bar-induced shocks may inject substantial turbulent energy into a galactic disk which stabilizes the molecular gas against collapse \citep{Khoperskov2018}. 
As further outlined in the following sections, as part of the present study we investigated both the star formation and gas content of our sample of filament galaxies, but we have not found any statistical significant difference between 
barred { (SB, SAB)} and nonbarred galaxies. We suspect that the absence of correlations is due to the limited statistics of our sample, which comprises {only} 41 and 31 galaxies classified as{ barred (SB) and weakly barred (SAB),} respectively. Therefore, in the following sections, we do not distinguish between the two classes.

 \begin{figure}[]\centering
\captionsetup[subfigure]{labelformat=empty}
\subfloat[]{\hspace{0.cm}\includegraphics[trim={0cm 0cm 0cm 
0cm},clip,width=0.5\textwidth,clip=true]{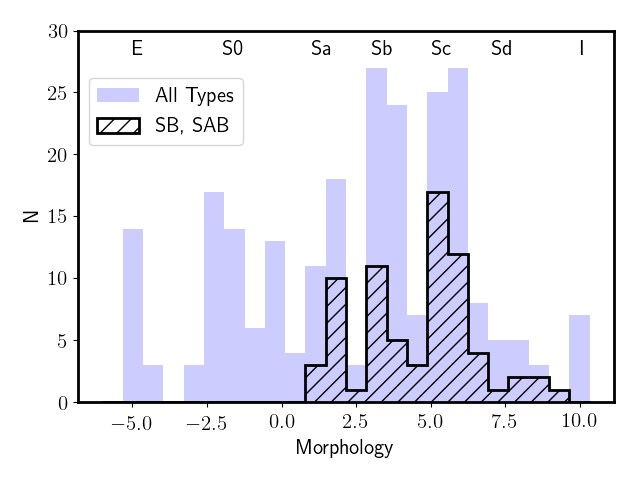}}
\caption{Hubble-type distribution for {our sample of filament galaxies (filled histogram) and for the subsample of barred galaxies (hatched histogram). The Hubble type (top) and 
the de Vaucouleurs classification (bottom) are reported on the x-axis.}}
\label{fig:histo_Hubble_type}
\end{figure}

\subsection{Observations}\label{sec:observations}
{ In this section we describe our cold gas observations of our sample of filament galaxies.} 
We observed the majority of our filament galaxies  in CO, and for the rest we gathered CO fluxes from the literature.
Archival HI masses are also available for many sources, while we observed the missing ones with the Nan\c{c}ay telescope, as  described below. 

\subsubsection{IRAM-30m CO observations}\label{sec:CO_IRAM30m_observations}
The large majority (163/245, i.e., 67$\%$) of the sources in our sample were observed by us at the IRAM-30m telescope at Pico Veleta, Granada, Spain, in October and December 2016, and then in March and July 2017. For each source, we observed both CO(1$\rightarrow$0) and CO(2$\rightarrow$1), simultaneously.
The full width at half power (FWHP) beam size is 21~arcsec and 10.5~arcsec at the CO(1$\rightarrow$0) and CO(2$\rightarrow$1) frequencies of 115 GHz and 230 GHz, respectively. Our targets have recession velocities between 1000
and 3000 km~s$^{-1}$. 

The SIS receivers (EMIR) were used for observations in wobbler-switching
mode, with reference positions offset by $\pm$120 arcsec in azimuth.
The {conversion} efficiency of IRAM-30m is $\rm \eta_{\rm mb}= B_{\rm eff}/F_{\rm eff} = T_{\rm A}^{*} / T_{\rm mb}$ = 0.83 and 0.64 at 115 GHz and 230 GHz, respectively. B$_{\rm eff}$ and F$_{\rm eff}$ are the main beam and forward efficiencies, respectively, while $T_{\rm A}^{*}$ and $T_{\rm mb}$ denote the antenna temperature and the {main beam} temperature, respectively.\footnote{ \url{https://publicwiki.iram.es/Iram30mEfficiencies}}
The system temperatures ranged
between 150~K and 400~K at 2.6~mm and between 200~K and 800~K at 1.3~mm.
The pointing was checked every 2~h on a nearby planet or a bright continuum source, and the focus was reviewed after each sunrise and at the beginning of each run. The on-source time typically ranged from 0.3~h to 2~h, according to the signal-to-noise ratio (S/N) already reached on a target and the weather.
Two backends were used simultaneously: the autocorrelator WILMA, and the Fourier transform spectrometer FTS. The $T_{\rm mb}$ { rms} noise level in 2~h integration was $\sigma _{\rm mb}\simeq1.4$~mK with a spectrometer resolution of 20~km~s$^{-1}$ for 115~GHz and $\sigma _{\rm mb}\simeq1.8$~mK for 230~GHz. { The upper limits reported in the following sections for the velocity-integrated CO fluxes  are computed at $3\sigma$ and are  equal to $3\,{\rm rms}_{300}~\times~$300~km~s$^{-1}$, where rms$_{300}$ is the rms (units of Jy) estimated in velocity bins equal to 300~km~s$^{-1}$, assuming a standard IRAM-30m  $T_{\rm mb}$-to-flux conversion of 5~Jy~K$^{-1}$.}

{ The data reduction and analysis were performed using the {\sc GILDAS-Class} software. The baseline was removed on each spectrum using {a linear fit}. When present, CO detections were then fit with a Gaussian curve. For all detections, we verified that {the CO velocity barycenter} is consistent with the galaxy recession velocity. Following this analysis, the majority (83\%) of our sources, i.e., { 136} out of 163, were detected at { S/N$>3$ } in CO(1$\rightarrow$0) or CO(2$\rightarrow$1), or both. 
Five sources, namely, UGC~05020 in CO(2$\rightarrow$1), PGC~035472 in CO(2$\rightarrow$1), IC~4263 in CO(2$\rightarrow$1), UGC~09556 in CO(1$\rightarrow$0), and UGC~10968 in both lines were tentatively detected at { S/N$\lesssim3$,} while for 22 galaxies we set 3$\sigma$ upper limits.

For the remaining 82 sources out of the 245 galaxies in our sample, CO observations were found in the literature. In case of multiple observations we gave preference to the those with higher { S/N.}
The Ursa Major Cloud is nearby and the corresponding cold gas observations  all come from the literature.
{Table~\ref{tab:CO_properties_literature}  summarizes the molecular gas properties for the subsample of 82 sources with CO observations from the literature, and we report the references.}

Seven additional sources were observed with the IRAM-30m because at the time of our observations they were considered as filament galaxies on the basis of their recession velocities and positions in the projected space. However, following { an environmental analysis}, we conservatively removed them {a posteriori} from our main sample, as described in the following. UGC~7039, UGC~7143, and PGC~38859 { fall in the field of the W-M sheet, but are located behind it. Indeed, while the W-M sheet spans the range SGY$\sim(16-25)~h^{-1}$~Mpc \citep{Kim2016} the sources have higher SGY$\sim$29, 28, and 27~$h^{-1}$~Mpc, respectively.
 IC~777 is in the field of, but is behind, the nearby Ursa Major Cloud, which spans the range SGY$\sim(2-16)~h^{-1}$~Mpc \citep{Tully1982,Tully_Fisher1988}, while the source has higher SGY$\sim41~h^{-1}$~Mpc. }
 Similarly, NGC~5240 and NGC~5089 are both located in the field of the NGC~5353/4 filament, but{ are at a much greater distance ---SGY$\sim(31-34)~h^{-1}$~Mpc--- than the filament,} because it spans SGY$\sim(22-27)~h^{-1}$~Mpc \citep{Kim2016}. These results suggest that all of the six sources are located behind the main filamentary structure of reference. The seventh source is PGC~214137, or equivalently  UGC~08656~ NOTES01. We erroneously targeted this object instead of its more massive companion UGC~08656 during our IRAM-30m observations. We conservatively decided not to consider
 { either} of the two sources for our main sample of 245 galaxies.

In Fig.~\ref{fig:30m_spectrum} we report the IRAM-30m spectra and the Gaussian fits for the CO lines for all{ 147  sources that we detected securely or tentatively with our campaign in at least one CO line}, including six out of the seven  sources considered separately.
We smoothed the spectra according to the { full width at half maximum (FWHM)} of the detected signal. In Table~\ref{tab:CO_properties_IRAM30m} we report the results of our IRAM-30m observations for the 163 sources of our main sample.  At the bottom of the table we also report the results for the 7 sources that were removed {a posteriori} from the main sample. Upper limits at $3\sigma$ are also reported, together with secure and tentative { (S/N$<$3) } detections, which are distinguished in the table. In Fig.~\ref{fig:histo_FWHM_CO_HI} we report the FWHM distribution for the sources detected with our IRAM-30m campaign. The figure { shows} that the large majority of FWHMs are below 300~km/s, and therefore this value used { to estimates the upper limits} is very conservative.

\subsubsection{Nan\c{c}ay HI observations}\label{sec:Nancay_HI_observations}
New  HI observations of  69 galaxies in the filamentary structures around Virgo were obtained using  the  Nan\c{c}ay  decimetric  radio telescope and  1024-channel autocorrelator spectrometer between January and December 2017. The Nan\c{c}ay telescope is a meridian transit-type instrument with an effective
collecting area of $\sim$7000~m$^2$. At 21~cm, the FWHP beam size is $3.6^\prime$~(East--West)~$\times~23^\prime$~(North--South) within the range of declinations spanned by our sources. The FWHP beam size changes only slightly with the source declination \citep{Fouque1990}.
Observations were obtained through position switching, with
an OFF at (14-20)~arcmin East, and alternating 2~min ON and 2~min OFF.
Tracking was limited  to  about  one hour  per  source  per  day.
For most of the sources, one track was sufficient; in some cases,
we repeated the track because of technical problems.  The system temperature
was typically 35~K. With an efficiency of 0.8~Jy/K, evaluated at the declination of our sources, we obtained
an rms of 2~mJy at a velocity resolution of 13~km~s$^{-1}$ for all spectra.
The total available bandwidth being nearly 10 000 km/s, we observed all sources with a common tuning, given their recession velocity range (1000-2500)~km~s$^{-1}$.

The spectra were first calibrated and reduced using the {\sc NAPS} reduction
package available at the Nan\c{c}ay site. The spectra were then exported
into fits files and analyzed using the {\sc Gildas-CLASS} software. { For each spectrum, the baseline was then removed with a linear fit. In case of detection, the velocity integrated flux, the velocity  width  at 50\%  of  the  maximum  flux (W$_{50}$), and the recession velocity at the HI emission peak were then estimated directly from the spectrum. }  For all detections,{ we verified that the HI velocity barycenter  is consistent with the galaxy recession velocity.}  The corresponding results are reported in Table~\ref{tab:HI_properties_all_galaxies}. 

Among the  69 sources observed in HI at Nan\c{c}ay, 58 are part of our sample of 245 filament galaxies. The HI results for the remaining 11 are reported separately at the bottom of the Table.
{ These galaxies were initially included as targets of our observational campaign.} However, while having been observed by us in HI at Nan\c{c}ay, they have not been observed in CO with our IRAM-30m campaign. We checked that they do not have CO observations from the literature either. Therefore, we preferred not to include them in our main sample of filament galaxies{, which comprises filament sources observed in both atomic and molecular gas. 
We also verified {a posteriori} that only 6 out of the 11 sources are part of the filamentary structures considered in this work, as follows. UGC~06326, PGC~034951, PGC~2139858, and PGC~035474 are within the{ Coma Berenices Filament}, while PGC~2578846 and NGC~6149 belong to the { Draco and Serpens filaments,} respectively. }
In Fig.~\ref{fig:Nancay_HI_spectra} we report the HI spectra for all  45 sources out of the 69 observed at Nan\c{c}ay with secure  (44) or tentative (PGC~2333993) HI detections. { We smoothed the spectra according to the width of the detected signal.} For the remaining  24 sources, we set 3$\sigma$ upper limits{ at a  resolution of 300~km~s$^{-1}$}, similarly to what we have{ done} for the upper limits in CO (Sect.~\ref{sec:CO_IRAM30m_observations}). 

\section{Molecular and atomic gas}\label{sec:gas_masses}

\subsection{H$_2$ gas masses}

We estimated molecular gas masses using a Galactic conversion factor $\alpha_{\rm CO}=4.3~M_\odot$~(K~km~s$^{-1}$~pc$^{2}$)$^{-1}$, equivalently,
$X_{\rm CO}=N_{\rm H_2}/I_{\rm CO}=2.0\times10^{20}$~cm$^{-2}$~$({\rm K~km~s}^{-1})^{-1}$ \citep[e.g.,][]{Dickman1986,Strong1988}, where $N_{\rm H_2}$ is the H$_2$ column density (in units of cm$^{-2}$)  and $I_{\rm CO}$ the  velocity  integrated  CO   line  intensity in units of K~km~s$^{-1}$. The molecular gas mass was then determined{ using the following expression \citep[see e.g.,][]{Bolatto2013}:}
\begin{equation}
 \label{eq:MH2_1}
\hspace{-0.2cm}\frac{M_{\rm H_2}}{M_\odot}=1.05\times10^4\frac{X_{\rm CO}}{2.0\times10^{20}~{\rm cm^{-2}~({\rm K~km~s}^{-1})^{-1}}}\frac{S_{\rm CO(1\rightarrow0)}\Delta\varv}{\rm Jy~km~s^{-1}}\Big(\frac{D}{\rm Mpc}\Big)^2\,,
\end{equation}
or, equivalently, as:
\begin{equation}
 \label{eq:MH2_2}
 \frac{M_{\rm H_2}}{M_\odot}=\frac{\alpha_{\rm CO}}{M_\odot~{\rm (K~km~s^{-1}~pc^{2})^{-1}}}\frac{L^\prime_{\rm CO(1\rightarrow0)}}{\rm K~km~s^{-1}~pc^2}\,.
\end{equation}

Here $S_{\rm CO(1\rightarrow0)}\Delta\varv$ and $L^\prime_{\rm CO(1\rightarrow0)}$ are the velocity-integrated CO$(1\rightarrow0)$ flux and luminosity, respectively, while $D$ is the distance of the source considered.
The $L^\prime_{\rm CO(J\rightarrow J-1)}$ luminosity for the generic ${\rm CO(J\rightarrow J-1)}$ transition can also be expressed as:

\begin{equation}
\label{eq:LpCO}
 \frac{L^{\prime}_{\rm CO(J\rightarrow J-1)}}{\rm K~km~s^{-1}~pc^2}=3.25\times10^7\,\frac{S_{\rm CO(J\rightarrow J-1)}\Delta\varv}{\rm Jy~km~s^{-1}}\,\bigg(\frac{\nu_{\rm CO(J\rightarrow J-1)}}{\rm GHz}\bigg)^{-2}\,\bigg(\frac{D}{\rm Mpc}\bigg)^2\,,
\end{equation}
where Eq.~3 by \citet{Solomon_VandenBout2005} has been used in the limit $z\ll1$, which is valid for nearby sources such as those considered in this work.
The frequency $\nu_{\rm CO(J\rightarrow J-1)}$ is that associated with the  ${\rm CO(J\rightarrow J-1)}$ transition. In this work, we use the $S_{\rm CO(1\rightarrow0)}$ flux to estimate H$_2$ molecular gas mass via Eqs.~(\ref{eq:MH2_1},~\ref{eq:MH2_2}). In the cases where the $S_{\rm CO(1\rightarrow0)}$ flux was at low { S/N$<$3} (tentative detections) or was unavailable, we used higher-J transitions via $M_{\rm H_2}=\alpha_{\rm CO}L^{\prime}_{\rm CO(J\rightarrow J-1)}/r_{J1}$. { The error of $M_{\rm H_2}$ is assumed to be proportional to the velocity-integrated CO flux error, that is, the error on the distance was not taken into account.}

Here $r_{J1}= L^{\prime}_{\rm CO(J\rightarrow J-1)}/L^{\prime}_{\rm CO(1\rightarrow0)}$ is the excitation ratio. We assume the following fiducial excitation ratios, typical of star forming galaxies, namely, $r_{21}=0.8$  \citep[][]{Bothwell2013,Daddi2015,Freundlich2019}  and $r_{31}=0.5$ \citep{Bothwell2013,Carilli_Walter2013}. We used the CO(3$\rightarrow$2) transition only for NGC3265 to set an upper limit on its H$_2$ gas mass using  JCMT observations by \citet{Wilson2012}. From the same authors, we also found CO(3$\rightarrow$2) observations  for NGC4559. However, for this source we used the CO($1\rightarrow 0$) flux by  \citet{Sage1993} to estimate $M_{\rm H_2}$. We refer to Table~\ref{tab:CO_properties_literature} for details. 
In the cases where only 3$\sigma$ upper limits to $M_{\rm H_2}$ were available, we used the most stringent one between those estimated via CO(2$\rightarrow$1) and CO(1$\rightarrow$0).

\subsection{Aperture corrections}\label{sec:aperture_correction}
Equations~(\ref{eq:MH2_1},~\ref{eq:MH2_2}) relate the $M_{\rm H_2}$ molecular gas mass to the total CO emission. However, as the sources in our sample are extended and nearby, their extension may be larger than the beam size $\Theta$, which is equal, for our IRAM-30m observations, to 21~arcsec and 10.5~arcsec in the case of CO(1$\rightarrow$0) and CO(2$\rightarrow$1), respectively. Therefore, following \citet{Lisenfeld2011}, when estimating the $M_{\rm H_2}$ mass, we multiplied the observed CO flux by 
the inverse of the filling factor, $f_{ap}$, which we estimate as follows:
\begin{small}
\begin{equation}
\label{eq:aperture_correction}
\hspace{-0.cm} f_{ap}=\frac{2}{\pi r_e^2}\hspace{-0.1cm}\int_0^\infty\hspace{-0.3cm} dx \int_0^\infty \hspace{-0.3cm} dy\,\exp \Bigg\{ -\ln(2) \bigg[\Big(\frac{2x}{\Theta}\Big)^2+\Big(\frac{2y\cos i}{\Theta}\Big)^2 \bigg] \Bigg\}\exp(\frac{-\sqrt{x^2+y^2}}{r_e})\,,
\end{equation}
\end{small}
where $i$ is the inclination angle between the line of sight and the polar axis of the galaxy.
The above equation also relies on the additional assumption that the CO line intensity has an exponential radial profile $I_{\rm CO}(r)\propto\exp(-r/r_e)$
\citep{Nishiyama2001,Regan2001,Leroy2008}, where $r_e$ is the CO scale length, which previous studies found to be well correlated with the optical exponential scale length \citep[e.g.,][]{Regan2001,Leroy2008,Saintonge2012}. Consistently with previous studies \citep{Leroy2008,Lisenfeld2011,Boselli2014} we assumed $r_e\simeq0.1~D_{25}$, where $D_{25}$ is the optical 25~mag~arcsec$^{-2}$ isophotal diameter.

\subsubsection{Datasets from the literature}
{ Table~\ref{tab:CO_properties_literature} summarizes the molecular  properties for the sources with CO observations from the literature. We also report the excitation ratio, the aperture correction adopted, and the corresponding reference and telescope.} We revised the CO fluxes from the literature according to the CO aperture correction described
 above in order to obtain the most homogeneous dataset possible, as outlined in the following.

We applied the same extrapolation as in Eq.~(\ref{eq:aperture_correction}) for the CO fluxes coming from single-dish and central pointing observations  \citep{Braine1993,WelchSage2003,Combes2007,Young2011,Vila-Vilaro2015,OSullivan2015,OSullivan2018}. According to the different telescopes used by these studies, we adopted the corresponding FWHP beam size $\Theta$. 

CO fluxes reported by \citet{Lisenfeld2011} were corrected for the aperture as in Eq.~\ref{eq:aperture_correction}. Therefore, we did not apply any additional correction. Similarly, \citet{Boselli2014} used an aperture correction very similar to that used by \citet{Lisenfeld2011}, with an additional non-negligible disk width in their formalism. However, a detailed comparison between their 3D extrapolated fluxes and those extrapolated via a 2D modeling, e.g., as in  Eq.~\ref{eq:aperture_correction}, leads to negligible statistical differences at  the level of a few percent, both in the mean values and in the rms dispersion \citep[see Table~7 of][]{Boselli2014}. Therefore, when considering their CO  fluxes we safely adopted their extrapolation. 

Observations by { \citet{Sage1993}, \citet{Young1995}, and \citet{Wilson2012}} took into account the extension of the sources when estimating CO fluxes by means of multiple pointings, when needed. Similarly, \citet{Alatalo2013} showed that their interferometric maps were able to recover more flux than the IRAM-30m,  on average, for the same targets (see their Table~3). Therefore, we preferred not to apply any aperture correction, limiting to the CO fluxes from these studies.








 \begin{figure}[h!]\centering
\captionsetup[subfigure]{labelformat=empty}
\subfloat[]{\hspace{0.cm}\includegraphics[trim={0cm 0cm 0cm 
0cm},clip,width=0.5\textwidth,clip=true]{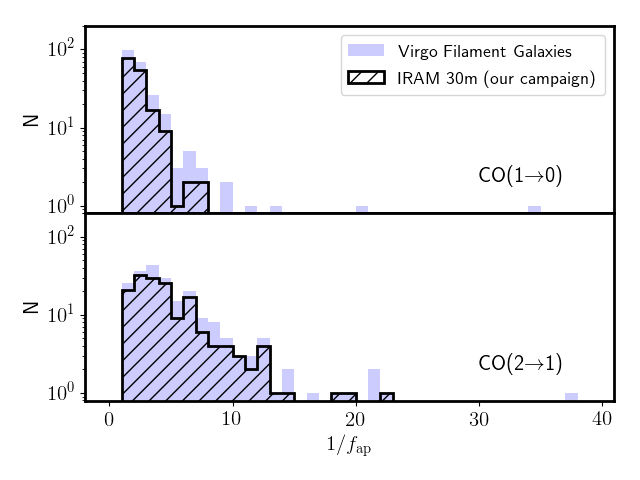}}
\caption{Distribution of the CO aperture correction for the filament sources in our sample (blue filled histogram). The subsample of sources from our IRAM-30m campaign is highlighted (dashed histogram).\\
}\label{fig:histo_CO_aperture_correction}
\end{figure}

\subsubsection{Aperture correction distribution}

In Fig.~\ref{fig:histo_CO_aperture_correction}  we report the aperture correction distribution for both CO(1$\rightarrow$0) and CO(2$\rightarrow$1) observations. The figure shows similar distributions for the extrapolation when considering the sources observed with our IRAM-30m campaign and the full sample of 245 filament galaxies, thus including also (extrapolated) CO fluxes from the literature. For the full sample of 245 sources, the aperture corrections span the ranges of $1/f_{\rm ap}\sim1-10$ and $\sim1-20$ for CO(1$\rightarrow$0) and CO(2$\rightarrow$1), with only a few exceptions with higher corrections. For our IRAM-30m observations, we have median aperture corrections $1/f_{\rm ap}=2.1^{+1.2}_{-0.6}$ and $1/f_{\rm ap}=4.0^{+3.5}_{-1.7}$, { respectively. The corrections are} of the order of unity, similarly to those previously found in other studies \citep[e.g.,][]{Lisenfeld2011,Boselli2014},{ which also showed that extrapolated CO fluxes provide good estimates for the total molecular mass when they are compared within objects that have multiple pointings of their disk.}
 
The highest aperture corrections $1/f_{\rm ap}\gtrsim10$ for CO(1$\rightarrow$0) and $\gtrsim25$ for  CO(2$\rightarrow$1) are found only in a few cases, when CO observations from the literature are considered. The higher aperture corrections estimated for CO(2$\rightarrow$1) are mainly due to the smaller FWHP than for CO(1$\rightarrow$0) because the FWHP scales with the inverse of the observed frequency. 
{ In addition to the fact that the excitation ratio for the first CO transition is equal to unity by definition, the smaller $1/f_{\rm ap}$ observed for CO(1$\rightarrow$0)  adds another piece of evidence in favor of its use  when estimating H$_2$ masses, as opposed to higher J transitions.

\subsubsection{Multiple IRAM-30m  pointings}
In this section, we consider four separate filament galaxies that were observed with our IRAM-30m campaign with multiple pointings, at their center and off-center. Figure~\ref{fig:COmultiple_pointings} displays the spiral morphologies of the four sources, namely NGC~5985, NGC~5350, NGC~5290, and UGC~09837, together with IRAM-30m spectra at the different pointings. The three NGC galaxies are almost perfectly face-on, while UGC~09837 is more edge-on. 

These four sources are among those of our sample with the largest extension in the projected sky and are therefore good test cases with which to  explicitly show the need for the flux extrapolations discussed above and evaluate the amount of CO coming from the center and the arms with ongoing star formation. 
However, because of their large size, the four sources have high aperture corrections, as further outlined below, and so they are not representative of the mean population, for which lower corrections are needed. The results of our observations are summarized in Table~\ref{tab:COmultiple_pointings}.

{ Our results show that the ratio of the sum of the fluxes of both central and off-center pointings to the flux at the central beam} ranges between $\sim$2.3 and 3.1 and between $\sim$1.6 and 4.4 for CO(1$\rightarrow$0) and CO(2$\rightarrow$1), respectively. We also note that these ratios are lower limits to the actual aperture correction, because our off-center pointings do not cover the whole galaxy disk in a uniform manner. This can be appreciated directly from the galaxy images in Fig.~\ref{fig:COmultiple_pointings}, where the { IRAM-30m beams} are also reported. Indeed, for these sources we estimated larger corrections than the ratios reported above. The aperture corrections for the four sources range between $1/f_{\rm ap}\simeq3.2$ and $6.3$ and { between} $\simeq7.7$ and $20.5$  for CO(1$\rightarrow$0) and CO(2$\rightarrow$1), respectively (see Table~\ref{tab:CO_properties_IRAM30m}).

{ The flux observed in the central beam is often higher than the fluxes from each of the off-center pointings (see Table~\ref{tab:COmultiple_pointings}). This result is} consistent with the declining disk profile adopted for the CO flux extrapolation. 
NGC~5985 may be an exception, because off-center pointings are associated with a large amount of CO, comparable to that observed from the center. This may be due to the fact that off-center pointings specifically targeted clumpy star forming regions in the spiral arms, as it is tentatively suggested by visual inspection of the galaxy (Fig.~\ref{fig:COmultiple_pointings}).
{ Our observations also show that the emission from the galaxy centers has a larger velocity dispersion than that from the outskirts. For the central beam, the velocity gradient is high, as the gravitational potential is deep, while in the outer parts the velocity gradient is lower, which results in a narrower spectrum than for the central regions.}




\begin{table}\centering
\begin{tabular}{cccc}
\hline\hline
ID & CO(J$\rightarrow$J-1) & $S_{\rm CO(J\rightarrow J-1)}\Delta\varv$ & FWHM\\
 &   &  (Jy~km~s$^{-1}$) & (km~s$^{-1})$ \\
 
(1) & (2)  & (3) & (4) \\
 \hline
&  & \\
NGC~5985  &  & \\
 center & 1$\rightarrow$0 & $15.10\pm2.77$ & $226\pm61$  \\
                  & 2$\rightarrow$1 & $6.70\pm2.53$  & $83\pm43$ \\
 North & 1$\rightarrow$0 & $19.05\pm1.64$ & $81\pm9$ \\
                 & 2$\rightarrow$1 & $16.53\pm3.60$ & $66\pm21$ \\
 South & 1$\rightarrow$0 & $12.86\pm1.54$ &  $89\pm12$\\
                 & 2$\rightarrow$1 & $6.23\pm2.43$   &  $58\pm23$ \\
& & \\
NGC~5350 & & \\
center & 1$\rightarrow$0 & $48.05\pm1.95$ &  $258\pm12$  \\
                  & 2$\rightarrow$1 & $71.26\pm3.11$  & $220\pm11$ \\
North  & 1$\rightarrow$0 & $35.29\pm2.10$  & $221\pm17$ \\
                  & 2$\rightarrow$1 & $18.18\pm2.73$  & $124\pm22$ \\
South & 1$\rightarrow$0 & $33.39\pm2.19$ & $178\pm15$ \\
                 & 2$\rightarrow$1 & $22.81\pm2.65$  & $148\pm18$ \\
& & \\
NGC~5290 & & \\
center  & 1$\rightarrow$0 & $99.00\pm2.25$ & $248\pm6$  \\
                 & 2$\rightarrow$1 & $126.23\pm3.09$   & $242\pm7$ \\
West  &  1$\rightarrow$0  & $70.32\pm2.74$ & $190\pm9$ \\
                & 2$\rightarrow$1 & $60.91\pm3.66$  & $160\pm12$  \\
East & 1$\rightarrow$0 & $59.12\pm2.49$ & $212\pm11$ \\
                & 2$\rightarrow$1 & $55.25\pm3.80$ & $193\pm17$ \\
& & & \\
UGC~09837  & & & \\
center & 1$\rightarrow$0 & $2.53\pm0.61$ & $73\pm21$ \\
       & 2$\rightarrow$1 & $<5.56$ & --- \\
West   & 1$\rightarrow$0 & $<4.30$ &  --- \\
       & 2$\rightarrow$1 & $5.41\pm1.76$ & $85\pm28$ \\
\hline
\end{tabular}
\caption{CO results for the four filament galaxies with multiple IRAM-30m pointings. Column description: (1) target ID and pointing; (2) CO(J$\rightarrow$J-1) transition; (3) observed velocity integrated CO(J$\rightarrow$J-1) flux, not aperture corrected; (4) FWHM of the CO(J$\rightarrow$J-1) line. { For UGC~09837 (center, west),}  the reported upper limits are at 3$\sigma$ and estimated at 300~km~s$^{-1}$ resolution.}
\label{tab:COmultiple_pointings}
\end{table}

\begin{figure*}[]\centering
\captionsetup[subfigure]{labelformat=empty}
\subfloat[]{\hspace{0.cm}\includegraphics[trim={0cm 0cm 0cm 
0cm},clip,width=0.5\textwidth,clip=true]{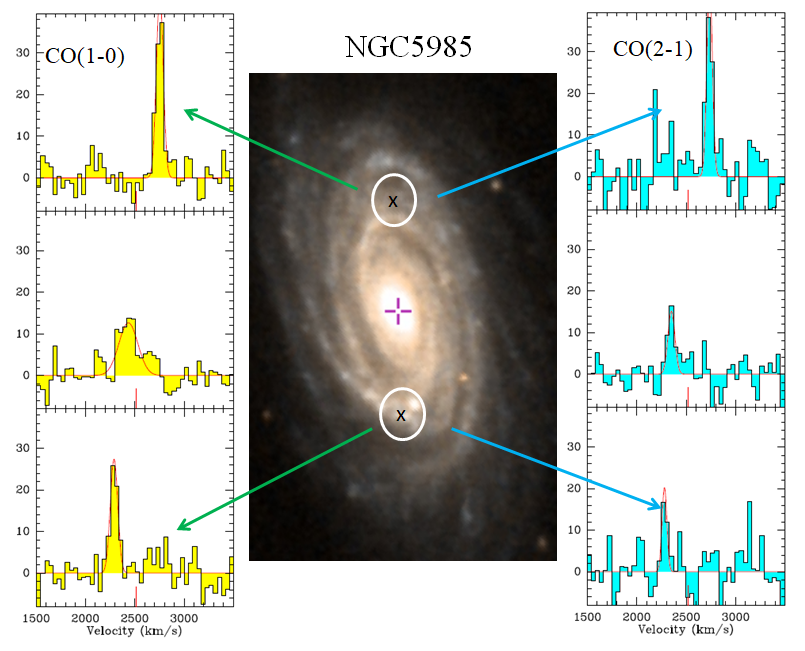}}
\subfloat[]{\hspace{0.7cm}\includegraphics[trim={0cm 0cm 0cm 
0cm},clip,width=0.51\textwidth,clip=true]{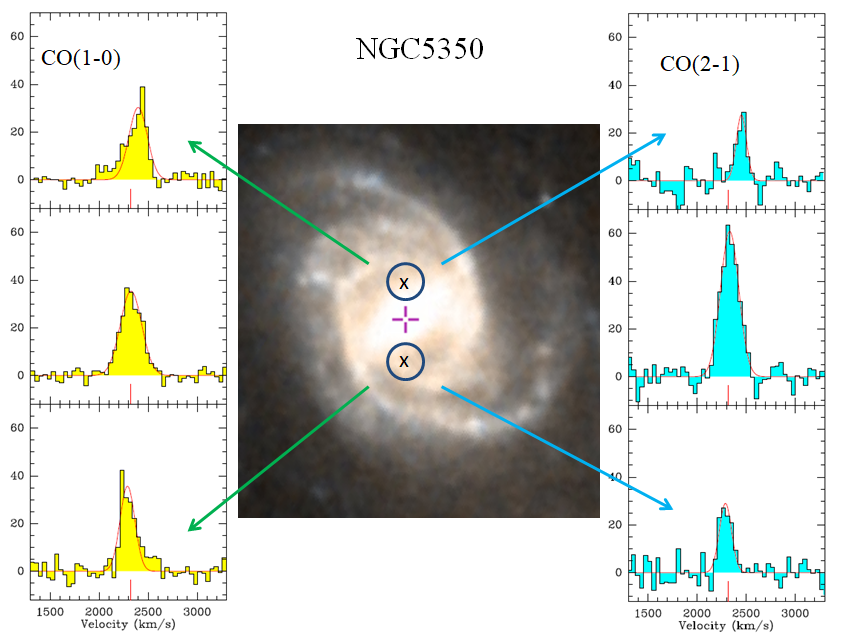}}\\
\subfloat[]{\hspace{0.cm}\includegraphics[trim={0cm 0cm 0cm 
0cm},clip,width=0.5\textwidth,clip=true]{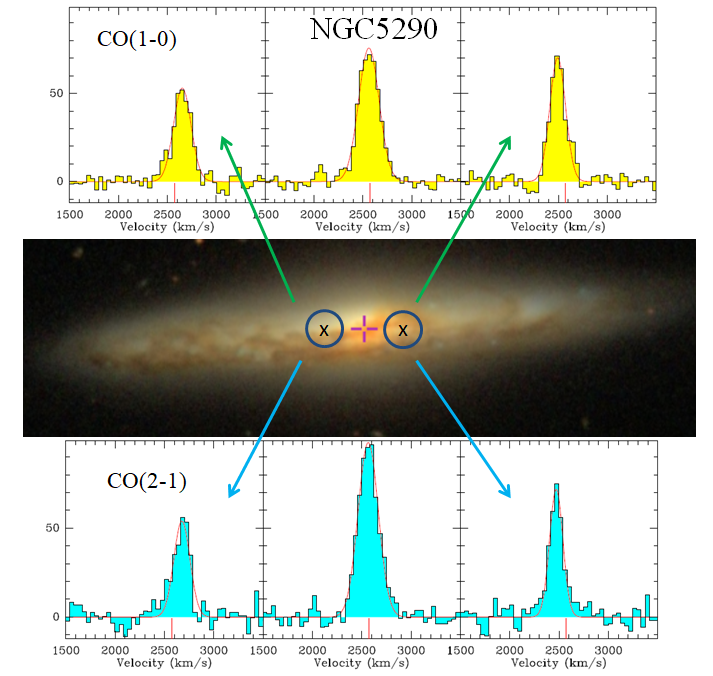}}
\subfloat[]{\hspace{3cm}\includegraphics[trim={0cm 0cm 0cm 
0cm},clip,width=0.3\textwidth,clip=true]{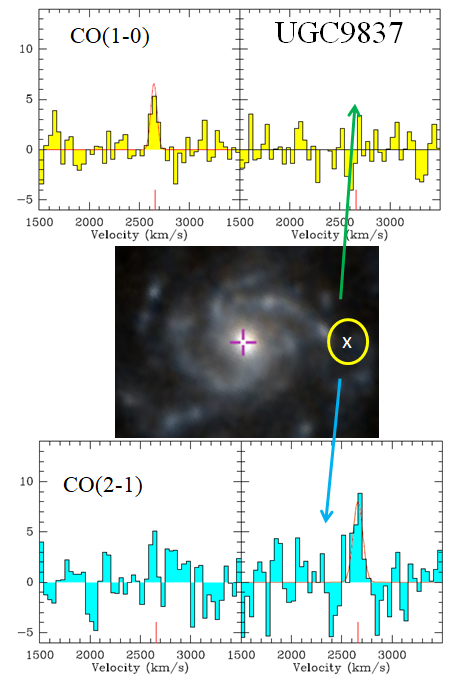}}\\
\vspace{-0.5cm}
\caption{Optical images and IRAM-30m spectra of the four sources in our sample with multiple pointings. In the images, north is up, east is left, while the crosses in magenta (+) and black/white (x) correspond to the central and offset  pointings, respectively. The circles correspond to the CO($2\rightarrow1$) beams of 10.5$^{\prime\prime}$ each, with the exception of NGC~5985, for which the CO($1\rightarrow0$) beams of 21$^{\prime\prime}$ each are instead reported. The reported spectra are baseline-subtracted, and the x- and y-axes show the relative velocity and $T_{\rm mb}$ (in mK), respectively. The solid lines in the spectra are the Gaussian fits to the CO($1\rightarrow0$) and CO($2\rightarrow1$) lines.}\label{fig:COmultiple_pointings}
\end{figure*}

\subsection{Excitation ratios}
We estimated excitation ratios $r_{21}$, or their upper limits, for 147 out of 245 sources (i.e., 60\%). They are all detected in CO(1$\rightarrow$0) and have detections{ (132)} or upper limits { (15)} in CO(2$\rightarrow$1). We set 3$\sigma$ upper limits to the latter { 15 sources, as reported in Tables~\ref{tab:CO_properties_literature} and \ref{tab:CO_properties_IRAM30m}.} For NGC~4151, NGC~4414, and NGC~4559, the CO fluxes reported in Table~\ref{tab:CO_properties_literature} imply low values of  $r_{21}\simeq0.01-0.02$. Such low excitation ratios{ for each of the three galaxies are possibly due to the fact that CO fluxes are associated with old observations \citep{Sage1993,Braine1993,Young1995}}
made with different telescopes and observational strategies { (i.e., single or multiple pointings)}. The corresponding excitation ratios are therefore very uncertain and we conservatively preferred {a posteriori} not to report any $r_{21}$ for the three galaxies.

In Fig.~\ref{fig:scatter_plot_r21} we report the excitation ratio $r_{21}$ as a function of galaxy morphology for the 147 sources. { The $r_{21}$ uncertainty is estimated by propagating in quadrature the uncertainties on the velocity integrated CO(2$\rightarrow$1) and CO(1$\rightarrow$0) fluxes.}
The associated median value is $r_{21}={0.53^{+0.42}_{-0.17}}$, which is formally lower that that of $r_{21}\sim0.8$ assumed for our sources. Excitation ratio estimates are also very uncertain, which is confirmed by the large reported uncertainties and the wide range of values within $r_{21}\sim0.1-3$ that are found for our sources. 

The scatter plot reported in Fig.~\ref{fig:scatter_plot_r21}, along with the binned median values (triangles), shows an increase of the excitation from late-type ($T\geq0$) to early-type ($T<0$) galaxies. { The excitation ratio $r_{21}$ and morphological parameter $T$ turn out to be anti-correlated. Spearman's test gives a probability of the null-hypothesis (no correlation) of ${p=~3.6\times10^{-3}}$, { where upper limits have  been discarded. However, only a tentative correlation is found, ${p=~3.6\times10^{-2}}$, when upper limits are included and considered as detections.} The observed trend may be a result of our selection, as in the figure we report only those galaxies that are at least detected in CO(1$\rightarrow$0). With this selection we miss the majority (68\%)  of early-type galaxies but only 30\% of late-type galaxies in our sample. 
No clear trend was instead} found when plotting $r_{21}$ against the SFR, { specific star formation rate (sSFR)}, stellar mass, distance to Virgo, distance to the filament spine, or local density.




 \begin{figure}[]\centering
\captionsetup[subfigure]{labelformat=empty}
\subfloat[]{\hspace{0.cm}\includegraphics[trim={0cm 0cm 0cm 
0cm},clip,width=0.5\textwidth,clip=true]{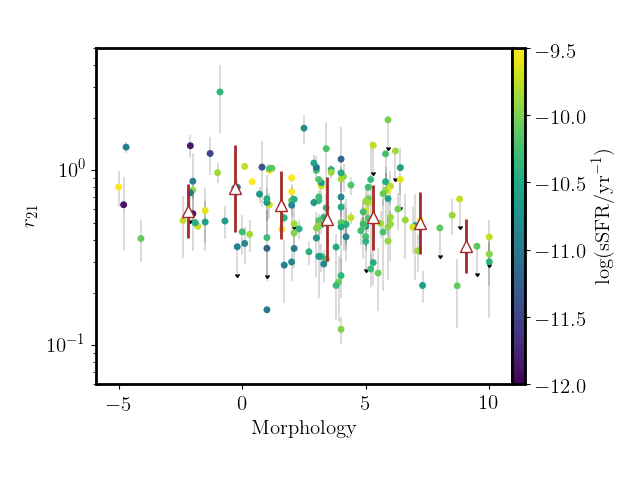}}
\caption{Excitation ratio ($r_{21}$) vs. morphological type for the filament galaxies in our sample. Points are color coded according to the sSFR, as illustrated in the color bar (right). 
Triangles show the binned median values, while their error bars correspond to  the rms dispersion around the median; { equally spaced bins with at least five points each} have been considered.}
\label{fig:scatter_plot_r21}
\end{figure}

\subsection{HI gas masses}

Our observations at Nan\c{c}ay targeted 58 out of the 245   filament galaxies in our sample, as described in  Sect.~\ref{sec:Nancay_HI_observations}. We complemented these by looking for HI observations  from the literature for the remaining 187. 
HI properties are summarized in Table~\ref{tab:HI_properties_all_galaxies}. 
At variance with the CO fluxes, no flux extrapolation was needed for the HI fluxes from our campaign at Nan\c{c}ay. This is due to the large FWHP beam size of $3.6^\prime$~(east--west)~$\times~23^\prime$~(north--south) of the Nan\c{c}ay telescope (see Sect.~\ref{sec:Nancay_HI_observations}). It is safely larger than the size $D_{25}\leq4.1$~arcmin of the sample galaxies that we observed at Nan\c{c}ay, while our filament sources have overall a median diameter of $D_{25} =(1.67^{+1.30}_{-0.76})$~arcmin, as reported in Table~\ref{tab:general_properties_all_galaxies}. 

Gas masses were then homogeneously estimated for all sources in our sample using the following formula:

\begin{equation}
 \label{eq:MHI}
 \hspace{-0.cm}M_{\rm HI}=2.36\times10^5\frac{S_{\rm HI} \Delta\varv}{\rm Jy~km~s^{-1}}\Big(\frac{D}{\rm Mpc}\Big)^2\,M_\odot,
\end{equation}
where $S_{\rm HI}\Delta\varv$ is the velocity-integrated HI flux and $D$ is the distance of the source considered. 
We refer { to \citet{Wild1952} and \citet{Roberts1962} for } further details. { The error on $M_{\rm HI}$ is assumed to be proportional to the velocity integrated HI flux error, that is, the error on the distance was not taken into account.}


We have also estimated the correction to $M_{\rm HI}$ due to self-absorption of HI, which affects the densest regions in the galaxy disk. The correction is expected to be stronger for edge-on galaxies and is estimated as the multiplicative factor $\kappa=(a/b)^{0.12}$ \citep{Giovanelli1994, Springob2005, Cicone2017}, where $a$ and $b$ are the optical major and minor axes of our galaxies, taken from HyperLeda. We verified that the correction due to HI self-absorption is negligible. It is not greater than 28$\%$ for the 245 sources in our sample, while on average it is at the level of a few percent, given the median value of $\kappa=1.07^{+0.09}_{-0.05}$. Table~\ref{tab:summary_CO_HI_properties_all_galaxies} summarizes the cold gas (both H$_2$ and HI) properties of our sample.


\section{Comparison samples}\label{sec:comparison_samples}

\subsection{Virgo cluster members}
There is a rich ensemble of data available for the galaxies inside the Virgo cluster. For our comparison we therefore considered the sample of Virgo cluster members from \citet{Boselli2014}, requiring both CO and HI observations. HI observations are reported in \citet{Boselli2014} and were taken from the literature, mostly from  the HI survey ALFALFA \citep{Giovanelli2005,Haynes2011}. We then updated the HI dataset using the most recent release of the ALFALFA survey by \citet{Haynes2018}.  Values for CO(1$\rightarrow$0) were obtained with several radio telescopes (e.g., Kitt Peak, IRAM-30m, FCRAO, SEST, Onsala, and BELL) and are also reported in \citet{Boselli2014}. 
For the sake of homogeneity, we converted CO(1$\rightarrow$0) and HI fluxes into gas masses, as described in Sect.~\ref{sec:gas_masses}. { Similarly to \citet{Boselli2014}, to estimate gas masses we assumed a distance of 23~Mpc for galaxies within the Virgo~B cloud, and of 17~Mpc for all other Virgo cluster members.} We also adopted a Galactic conversion factor $\alpha_{\rm CO}=4.3~M_\odot$~(K~km~s$^{-1}$~pc$^{2}$)$^{-1}$, and used the CO fluxes as reported by \cite{Boselli2014}. In the case of single-beam observations, the authors adopted a 3D aperture correction very similar to that used for our observations of filament galaxies (see also Sect.~\ref{sec:aperture_correction}).

To enable a homogeneous comparison in terms of stellar masses, $\log(M_\star/M_\odot)\gtrsim9$, and SFRs we then cross-matched the sample of Virgo sources with the \citet{Leroy2019} catalog, similarly to our analysis of the filament galaxies (see Sect.~\ref{sec:Mstar_SFR_our_sample}). This selection yields 109 galaxies.  

As our full catalog covers the Virgo cluster (see Fig.~\ref{fig:Virgo_field}), and in analogy to what { we} did for our sample of filament galaxies, we assigned morphological classifications from HyperLeda and local densities to all Virgo galaxies considered for our comparison.

\subsection{The field: AMIGA isolated galaxies}
The comparison of our sample of filament galaxies with field sources is also essential to understand the effect of the filamentary structures on galaxy evolution. To this aim, we considered the multi-wavelength Analysis of the interstellar Medium in Isolated GAlaxies (AMIGA) survey,\footnote{\url{http://amiga.iaa.es}} 
which comprises a sample of about $1000 $ galaxies in the local Universe peaking at distances of $\sim70$~Mpc \citep{VerdesMontenegro2005}. { AMIGA sources were primarily selected to be in isolation, by inspection of their local environment, using an isolation criterion whereby neighbors of comparable size of each AMIGA galaxy are at a projected distance of greater than 20 times their size \citep{Karachentseva1980,Verley2007a,Verley2007b}.}


This sample thus contains pure field galaxies and is optimal for our comparison because it has minimum contamination from field galaxies belonging to poor or moderately rich groups. 
{ As further outlined in the following,} the AMIGA sample also has a wealth of ancillary data, including { estimates of local densities, SFR, and both stellar and gas masses,} which allow us to perform a homogeneous comparison with respect to our filament galaxies.

\subsubsection{Local densities}\label{sec:AMIGA_loc_density}
{ We considered a subsample of 200 AMIGA sources with stellar mass and SFR estimates, as well as cold gas observations, as outlined below (Sect.~\ref{sec:mstar_sfr_comparison_sample}). }
{ Among these} 200 AMIGA sources, 45 (23\%) are in our full sample of galaxies around Virgo (Fig.~\ref{fig:Virgo_field}). We therefore assigned local density estimates  to these
sources, which were computed as described in Sect.~\ref{sec:environmental_properties}. For the remaining 155 galaxies, we assigned $n_5$ local densities using projected densities estimated by \citet{Verley2007a} by means of the $k$-th nearest neighbor, with $k=5$. These densities are estimated in projection, but we converted them into 3D densities, assuming statistically spherical symmetry around AMIGA sources and using the list of neighbors reported by \citet{Verley2007b}. { Local} densities for the 155 AMIGA sources were then increased by a factor of 4.3, which takes into account the mean logarithmic offset of $0.63\pm0.74$ that is found for the densities of the AMIGA sources that also belong to our catalog of sources around Virgo.

Finally, seven out of the 200 AMIGA sources, namely PGC~022100, UGC~04659, PGC~02731, NGC~5016, NGC~5375, UGC~09556, and NGC~6012, are also in our sample of filament galaxies and we removed them from our comparison. { This is consistent with the fact that our sample of filament galaxies span a broad range of local densities.} Our selection yields a sample of 193 AMIGA sources, all with local density estimates.

\subsubsection{Stellar mass, SFR, and gas masses}\label{sec:mstar_sfr_comparison_sample}

We assigned stellar mass and SFR estimates to AMIGA sources using the \citet{Leroy2019} catalog in order to make a homogeneous  comparison with our sample of filament galaxies and Virgo cluster members. Among the full sample of AMIGA galaxies, we then selected those with cold gas observations, both in  HI and CO. This selection yields 200 sources within AMIGA.  

HI observations of AMIGA sources were carried out 
with Arecibo, Effelsberg, Green Bank, and Nan\c{c}ay radio telescopes \citep{Jones2018}. CO(1$\rightarrow$0) observations are reported in \citet{Lisenfeld2011} and were mostly carried out with IRAM~30m and FCRAO telescopes, or were alternatively gathered from the literature by the authors. 
To convert the CO fluxes into gas masses, we followed the procedure described in Sect.~\ref{sec:gas_masses}. In particular, we adopted a Galactic conversion factor $\alpha_{\rm CO}=4.3~M_\odot$~(K~km~s$^{-1}$~pc$^{2}$)$^{-1}$. We also note that \citet{Lisenfeld2011} used the same CO aperture correction as the one adopted in this work (Eq.~\ref{eq:aperture_correction}). \\

For both AMIGA and Virgo cluster galaxies, in case of nondetection, { we derived} $3\sigma$ upper limits to gas masses at a resolution of 300~km/s for both CO and HI, analogously to what has been done for our sources of filament galaxies. { Among the 193 AMIGA sources, 61 and 11 have upper limits to the H$_2$ and HI masses, respectively, while out of the 109 Virgo cluster galaxies upper limits are used for 82  and 27 galaxies, respectively.
Cold gas observations for the sources in the comparison samples have typical rms values of $\sim3$~mJy for CO \citep{Lisenfeld2011,Boselli2014} and $\sim(0.4-0.5)$~mJy for HI \citep{Jones2018,Haynes2018}, evaluated at 300~km~s$^{-1}$ resolution, and are thus similar to the rms of our CO and HI observations.} {For further discussion, we refer to Appendix~\ref{sec:SFR_gas_diagnostic}, where diagnostic plots between gas content, stellar mass, and star formation are reported. }



 \begin{figure*}[h!]\centering
\captionsetup[subfigure]{labelformat=empty}
\subfloat[]{\hspace{0.cm}
\includegraphics[trim={0cm 0cm 2.3cm 
0cm},clip,width=0.5\textwidth,clip=true]{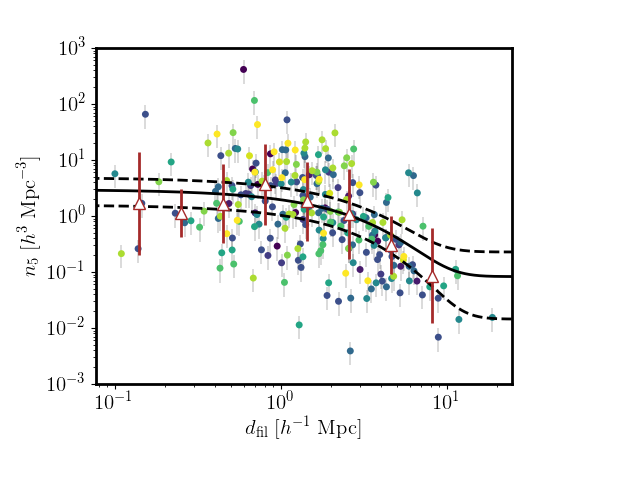}}
\subfloat[]{\hspace{-0.4cm}\includegraphics[trim={2.3cm 0cm 0cm 
0cm},clip,width=0.5\textwidth,clip=true]{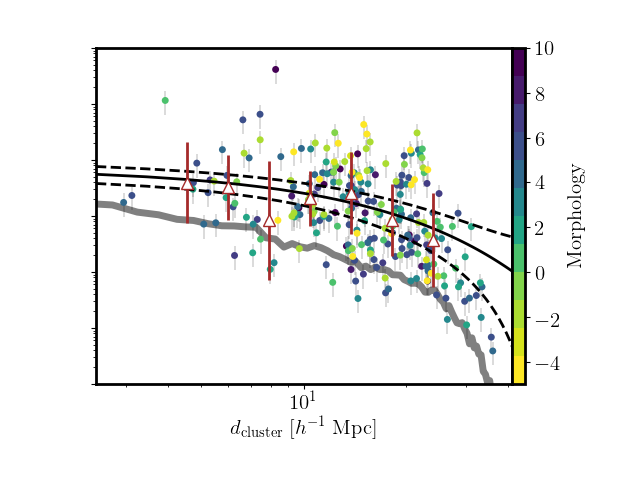}}
\caption{Density profile for filament sources in our sample as a function of the distance from the filament spine (left) and from the Virgo cluster center (right). { Sources are color-coded according to their morphological de Vaucouleurs classification.} The solid lines show the best fit to the data
performed with the exponential model described in the text and estimated at the median distances $\langle d_{\rm cluster} \rangle$ (left) and $\langle d_{\rm fil} \rangle$ (right). Dashed lines denote the $\pm1\sigma$ uncertainties to the fit. 
Triangles show the { binned median values}, while their error bars correspond to  the rms dispersion around the median; { equally spaced bins, in log scale, with at least five points each} have been considered.  In the right panel, the gray solid line shows the average field density estimated in{ consecutive spherical} shells, concentric with Virgo. { Each shell has a radius that is 0.5~$h^{-1}$~Mpc higher than the previous one.}
}\label{fig:density_profiles}
\end{figure*}

\begin{table}\centering
\begin{tabular}{ccc}
\hline\hline
&  \\
exponential scale length  \\
$d_{\rm 0,fil}$ & 2.29$\pm$0.55 & $h^{-1}$~Mpc\\
$d_{\rm 0,cluster}$ &  9.69$\pm$3.14 & $h^{-1}$~Mpc \\
&  \\
\hline
&  \\
 central density \\
$n(d_{\rm fil}\ll d_{\rm 0,fil},\langle d_{\rm  cluster}\rangle )$ & $2.94^{+1.86}_{-1.38}$  & $h^{3}$~Mpc$^{-3}$ \\
$n(\langle d_{\rm fil}\rangle,d_{\rm  cluster}\ll d_{\rm 0,cluster} )$ & $5.56^{+2.10}_{-1.75}$  & $h^{3}$~Mpc$^{-3}$ \\
&  \\
\hline
&  \\
 large scale density (field) \\
$n(d_{\rm fil}\gg d_{\rm 0,fil},\langle d_{\rm  cluster}\rangle )$ & $\sim0.08$  & $h^{3}$~Mpc$^{-3}$ \\
$n(\langle d_{\rm fil}\rangle,d_{\rm  cluster}\gg d_{\rm 0,cluster})$ & $\sim0.07$  & $h^{3}$~Mpc$^{-3}$ \\
&  \\
\hline
\end{tabular}
\caption{Density profile parameters{ as in Eqs.~(\ref{eq:density_formula}, \ref{eq:central_density},  \ref{eq:large_scale_density}). See text for further detail.}}
\label{tab:density_profiles}
\end{table}

%
%




\section{Environment}\label{sec:density_profiles}
In this section, we investigate the clustering properties of our sample of filament galaxies around Virgo.  To this aim, in Fig.~\ref{fig:density_profiles} we report the local densities $n_{5}$ estimated for the 245 sources in our sample  using the fifth-nearest neighbor. Local densities are plotted against the distances to the filament spines $d_{\rm fil}$ (left) and to the Virgo cluster $d_{\rm cluster}$ (right). The reported densities and distances are estimated in the 3D Super Galactic Coordinate frame. 
We refer to Sect.~\ref{sec:environmental_properties} for details.


{ In this work, we refer to the sample of filament galaxies and consider the filamentary structures altogether, independently of the associated environmental properties.} 
However, as seen in Fig.~\ref{fig:density_profiles} the sources in our sample span a broad range of distances from the spine, up to $d_{\rm fil}\sim20~h^{-1}$~Mpc for  a few sources, where the filamentary structures are less dense and more similar to the field in terms of their local environment. 
{ The filamentary structures considered in this work also} have different  structural properties (e.g., filament, cloud, sheet), as well as richness, { radius}, and length \citep[e.g.,][]{Kim2016}.

\subsection{Filament profiles}\label{sec:filament_profiles}
We fitted  the local densities of our 245 filament galaxies altogether  with an exponential model, as a function of their position relative to the filament spine and Virgo cluster. We assume that the density along filaments depends on two parameters, the distances  $d_{\rm fil}$ and $d_{\rm cluster}$, and  that these two dependencies are separable. We used exponential functions to fit the data points. According to this simple model, the 3D local density $n_{5}$ can be expressed as follows: 
\begin{equation}
\label{eq:n5_fit}
 n_5(d_{\rm fil},d_{\rm cluster}) = \Phi_{\rm fil}(d_{\rm fil})\times\Phi_{\rm cluster}(d_{\rm cluster}),
\end{equation}
where
\begin{equation}
\label{eq:density_formula}
\Phi_i(d_i) = a_i\exp(-d_i/d_{0,i})+b_i\,.
\end{equation}
The suffix $i=$~fil or $i=$~cluster corresponds to filaments or Virgo cluster, respectively. { The adopted model assumes that the local densities exponentially decline with increasing distance from the cluster and the filament spines, with scale length parameters  $d_{\rm 0,cluster}$ and $d_{\rm 0,fil}$, respectively. The remaining four best-fit parameters $a_{\rm cluster}$, $b_{\rm cluster}$, $a_{\rm fil}$, and $b_{\rm fil}$ can be related to central and large-scale densities evaluated for the 245 sources at the median distances  ${\langle d_{\rm cluster}\rangle=15.2~h^{-1}}$~Mpc and  ${\langle d_{\rm fil}\rangle=1.63~h^{-1}}$~Mpc, as follows.}
\begin{align}
\label{eq:central_density}
\hspace{3cm}\textrm{--- central densities ---}
\end{align}
\begin{align*}
n(d_{\rm fil}\ll d_{\rm 0,fil},\langle d_{\rm  cluster}\rangle ) = (a_{\rm fil}+b_{\rm fil})\cdot(a_{\rm cluster}\exp\frac{-\langle d_{\rm cluster} \rangle}{d_{\rm 0,cluster}}+b_{\rm cluster})\\
n(\langle d_{\rm fil}\rangle,d_{\rm  cluster}\ll d_{\rm 0,cluster} ) = (a_{\rm cluster}+b_{\rm cluster})\cdot(a_{\rm fil}\exp{\frac{- \langle d_{\rm fil} \rangle}{d_{\rm 0,fil}}}+b_{\rm fil}), 
\end{align*}
\begin{align}
\label{eq:large_scale_density}
\hspace{2.5cm}\textrm{--- large-scale densities ---}
\end{align}
\begin{align*}
n(d_{\rm fil}\gg d_{\rm 0,fil},\langle d_{\rm  cluster}\rangle ) = b_{\rm fil}(a_{\rm cluster}\exp\frac{- \langle d_{\rm cluster} \rangle}{d_{\rm 0,cluster}}+b_{\rm cluster})\\
n(\langle d_{\rm fil}\rangle,d_{\rm  cluster}\gg d_{\rm 0,cluster}) = b_{\rm cluster}(a_{\rm fil}\exp\frac{- \langle d_{\rm fil} \rangle}{d_{\rm 0,fil}}+b_{\rm fil}). 
\end{align*}

{ We stress that the fit to the local densities of the 245 sources has been done as a function of both $d_{\rm fil}$ and $d_{\rm cluster}$, simultaneously, as expressed in Eq.~\ref{eq:n5_fit}.} In the left and right panels of Fig.~\ref{fig:density_profiles}, we report instead the best-fit curves ${n_5(d_{\rm fil},\langle d_{\rm cluster} \rangle)}$ { and } ${n_5(\langle d_{\rm fil} \rangle,d_{\rm cluster})}$ evaluated separately at the median distances $\langle d_{\rm cluster}\rangle$ and  $\langle d_{\rm fil}\rangle$, respectively.
With 6 best-fit parameters and 245 data points, we have 239 degrees of freedom { (dof).} The best fit to the 245 data points yields a  $\chi^2/{\rm dof}=2730/239=11.4$. The reduced $\chi^2$ is thus higher than unity, which reflects the large scatter in the data points.
In Table~\ref{tab:density_profiles} we report the best-fit exponential scale{ lengths $d_{\rm 0,fil}$ and $d_{\rm 0,cluster}$ (Eq.~\ref{eq:density_formula}),} as well as the  central and large-scale densities that result from our fit{ (Eqs.~\ref{eq:central_density}, \ref{eq:large_scale_density}).}

A typical filament { radius} $d_{\rm 0, fil}=(2.29\pm0.55)~h^{-1}$~Mpc is inferred from our fit, which is within the range of values found in the literature. Our estimate is in fact lower than that of $(5.1\pm0.1)~h^{-1}$~Mpc found by \citet{Bonjean2020} by fitting an exponential profile to cosmological filaments at $0.1<z<0.3$, but is higher than both $\sim(0.2-0.9)~h^{-1}$~Mpc, found by \citet{Lee2021} by fitting the individual profiles of some major filaments around Virgo, and the values of $\sim(0.7-1.0)~h^{-1}$~Mpc  recently found in hydrodynamic simulations \citep{Kuchner2020,Rost2021}. These different values for the filament{ radius}, which is of the order of $\sim1.0~h^{-1}$~Mpc, are not surprising. Indeed, the specific smoothing scale associated  with different studies{ as well as the different filament lengths considered} may have an impact on the structural properties of the recovered filaments \citep[e.g.,][for a discussion]{Kuutma2020,Kraljic2018,GalarragaEspinosa2020}.

The local density declines less rapidly as a function of the distance to the cluster (Fig.~\ref{fig:density_profiles}, right), with a typical scale $d_{\rm 0,cluster}=(9.69\pm3.14)~h^{-1}$~Mpc, { which is about  four times the average filament radius $d_{\rm 0,fil}$.} We are indeed probing filaments over several virial radii between $\sim2$ and $36~h^{-1}$~Mpc in 3D but also in projection on the (SGX, SGZ) plane.
{ The fact that the considered filaments span the same range of Virgo clustercentric distances both in 3D and 2D shows} that the extension of the considered filaments is, on average, mostly along the projected sky, { as also appreciated in Fig.~\ref{fig:Virgo_field}.} This aspect minimizes the impact of line-of-sight uncertainties on our analysis.


As reported in Table~\ref{tab:density_profiles}, our fit also yields central densities of between $\sim3$ and $6~h^{3}$~Mpc$^{-3}$, which are more than an order of magnitude higher than those of $\sim0.1~h^{3}$~Mpc$^{-3}$ found at large distances from both cluster center and the filament spines, and nicely resemble those typical of our comparison field sample (i.e., AMIGA, see Sect.~\ref{sec:gas_content_tdep_environment}).  A similar density contrast of about an order of magnitude up to $d_{\rm cluster}\sim30~h^{-1}$~Mpc is also found in Fig.~\ref{fig:density_profiles} (right)
by comparing the best fit to the local density $n_5$  with the average field value (gray solid line) estimated in shells centered on Virgo. 

\subsection{Groups within filaments}

We further characterize the  environmental properties of our sample of filament galaxies using the group catalog of \citet{Kourkchi_Tully2017}. This catalog has a recession velocity cut at 3 500~km~s$^{-1}$, safely higher than that used for our sample. We indeed found all sources in our sample, except two (i.e., UGC~6455, KUG~1128+358).  The vast majority of our sources, namely $\sim72\%$  (177/245), are distributed in 81 groups, each having at least two members.  We report the group mass versus richness scatter plot for these groups in Fig.~\ref{fig:scatter_plot_ML_vs_richness}. The two quantities are nicely correlated.{ Here the richness refers to the number of galaxies potentially associated with a given group, as provided by \citet{Kourkchi_Tully2017}.}

\begin{figure}[h]\centering
\captionsetup[subfigure]{labelformat=empty}
\subfloat[]{\hspace{0.cm}\includegraphics[trim={0cm 0cm 0cm 
0cm},clip,width=0.5\textwidth,clip=true]{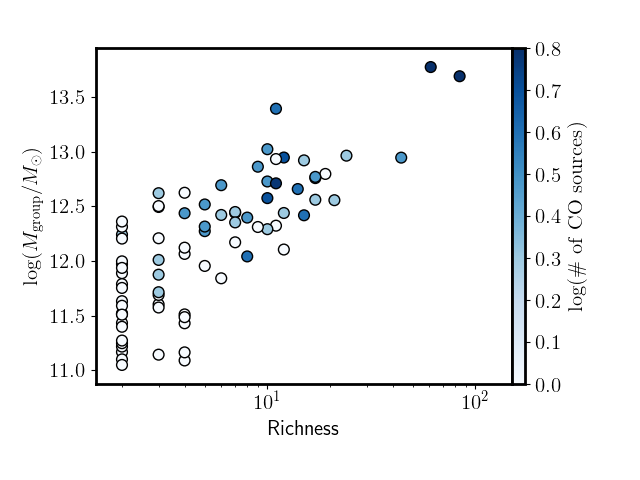}}
\caption{Group mass vs. richness for the 81 groups in filaments with at least two galaxies \citep{Kourkchi_Tully2017}. Points are color coded according to the number of group members that are part of our sample of filament galaxies.}\label{fig:scatter_plot_ML_vs_richness}
\end{figure}

Figure~\ref{fig:scatter_plot_ML_vs_richness} { (see color bar)} also indicates that the number of sources in our CO sample increases as a function of group richness. 
{ The} Spearman's test gives a probability of the null-hypothesis (no correlation), namely $p$-value~$=1.87\times10^{-11}$ (i.e., 6.7-$\sigma$).
Furthermore, about $12\%$ (30/245) of our sources belong { to two rich groups with masses}  $\gtrsim10^{13}~M_\odot$. In practice, a total of 12 and 18 of our filament galaxies belong to the well-known NGC~5846 and NGC~5353/4 groups, with a total of 84 and 61 members, respectively. These are the end points (i.e., knots) of the rich VirgoIII and NGC~5353/4 filaments.
{ We are thus sampling a wide range of galaxy environments in the  filamentary structures around Virgo,} from the poorest to the richest groups. We refer to Table~\ref{tab:environmental_prop}, where group properties for our sample, taken by \citet{Kourkchi_Tully2017} are reported.


 \begin{figure}[]\centering
\captionsetup[subfigure]{labelformat=empty}
\subfloat[]{\hspace{0.cm}\includegraphics[trim={0cm 0cm 0cm 
0cm},clip,width=0.5\textwidth,clip=true]{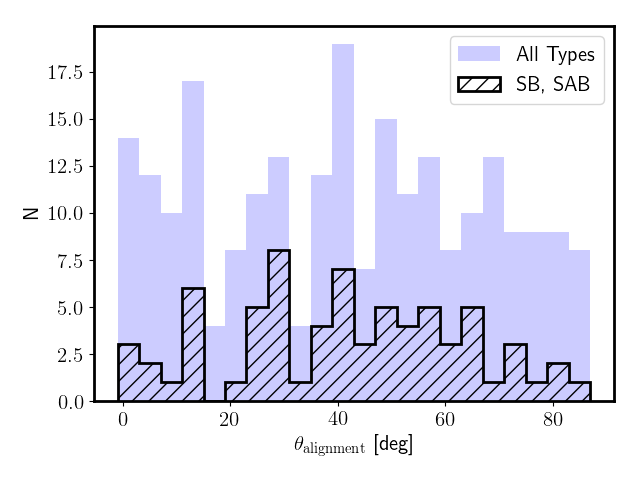}}
\caption{Distribution of the projected orientation $\theta_{\rm alignment}$ of the galaxy major axis with respect to the filament spine for the sources in our sample.\\
}\label{fig:scatter_plot_alignment}
\end{figure}

\begin{figure*}[]\centering
\captionsetup[subfigure]{labelformat=empty}
\subfloat[]{\hspace{0.cm}\includegraphics[trim={0.5cm 0cm 2.8cm 
0cm},clip,width=0.33\textwidth,clip=true]{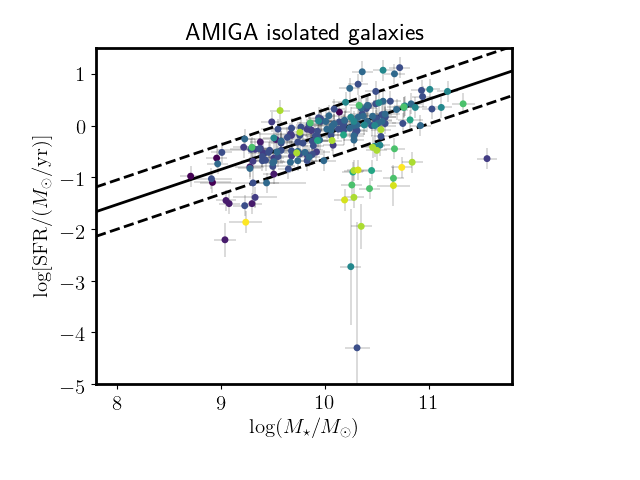}}
\subfloat[]{\hspace{-0.1cm}\includegraphics[trim={2.3cm 0cm 1cm 
0cm},clip,width=0.33\textwidth,clip=true]{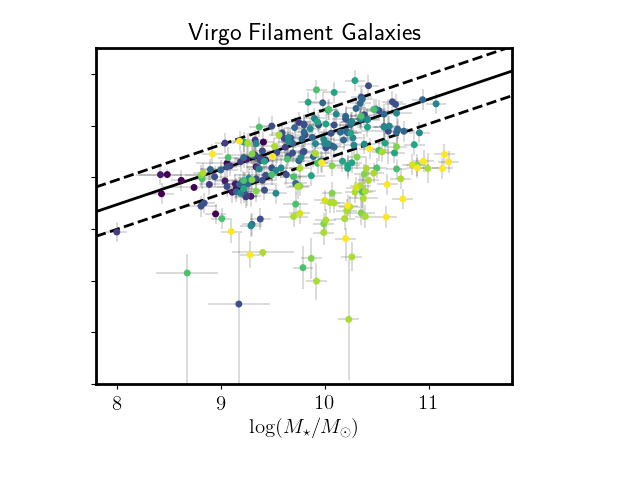}}
\subfloat[]{\hspace{-0.9cm}\includegraphics[trim={2.3cm 0cm 1cm 
0cm},clip,width=0.33\textwidth,clip=true]{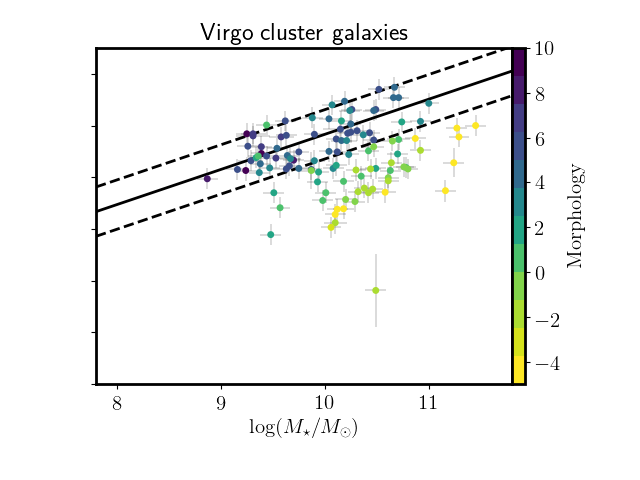}}
\caption{SFR vs. stellar mass scatter plots for the Virgo filament sources in our sample (center), AMIGA isolated galaxies (left), and Virgo cluster galaxies (right). Sources are color-coded according to their morphological de Vaucouleurs classification. The solid dashed line show the local MS relation by \citet{Leroy2019}, while the dashed lines correspond to $\pm\log(3)=\pm0.48$~dex uncertainty.}\label{fig:SFR_vs_Mstar}
\end{figure*}

 \begin{figure*}[]\centering
\captionsetup[subfigure]{labelformat=empty}
\subfloat[]{\hspace{0.4cm}\includegraphics[trim={0.8cm 0.5cm 2cm 
0.5cm},clip,width=0.5\textwidth,clip=true]{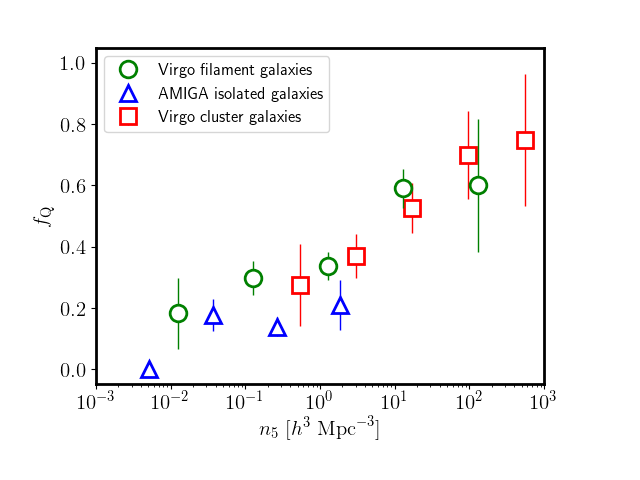}}
\subfloat[]{\hspace{-0.cm}\includegraphics[trim={2.3cm 0.5cm 0.5cm 
0.5cm},clip,width=0.5\textwidth,clip=true]{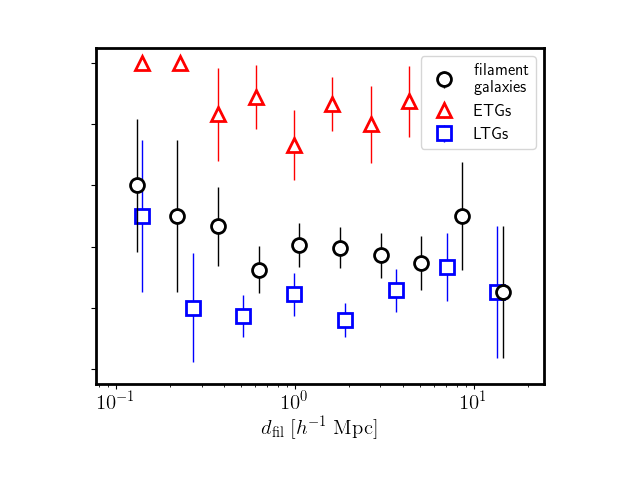}}
\caption{Quenching fraction $f_{\rm Q}$ vs. local density (left) and distance to the filament (right).}\label{fig:scatter_plot_fQ_vs_density_dfil}
\end{figure*}

\subsection{Galaxy alignments with filaments}
As part of our environmental analysis, we also measured the  orientation $\theta_{\rm alignment}$ between the major axis of the source and the direction of the filament spine, which has been estimated locally.  { These are the galaxy position angles,  $0~{\rm deg}\leq\theta_{\rm alignment}\leq90~{\rm deg}$, with respect to the projected orientation of the filament in the plane of the sky.} The values of $\theta_{\rm alignment}$ are reported in Table~\ref{tab:environmental_prop} for each galaxy. 

In Fig.~\ref{fig:scatter_plot_alignment} we show the distribution of $\theta_{\rm alignment}$ for our sources, where barred galaxies (SB, SAB) are highlighted. Interestingly, our sources  span all possible alignments quite homogeneously, which implies that there is no preferred direction with respect to the filament spine. This applies also when barred galaxies are considered{ separately.} Similarly, we find no clear trend when cross-correlating $\theta_{\rm alignment}$ with (i) local densities, (ii) distances from the filament spine, or (iii) galaxy morphology, separately. 

Interestingly, the absence of correlation of $\theta_{\rm alignment}$ with respect to galaxy morphology is at odds with the findings of some previous studies based on both cosmological simulations and wide field surveys \citep{Tempel2013,Tempel_Libeskind2013,Hirv2017,Codis2018,Chen2019,Welker2020}. These latter studies show that the spin axis or major axis of low-mass galaxies is preferentially aligned with their local environment or the nearest filament,  while  higher  mass  galaxies more  likely  display  an  orthogonal orientation. However, this effect is found to be small in all these studies. The absence of any correlation for our sample could be explained by our small sample size, which is particularly small if we limit ourselves to the more massive early-type galaxies, or by the fact that{ we are looking at galaxies up to large distances of ${\sim20~h^{-1}}$~Mpc from the filaments where the effect of filament environment is weaker.}  Uncertainties in the position angles and filament directions, as well as the fact that  $\theta_{\rm alignment}$  is estimated in projection could contribute to weaken any possible correlation. Similarly to what we found for our sample, we also note that \citet{Krolewski2019} recently found no evidence for alignment between galaxy spin and filament direction for nearby galaxies within the MANGA survey.

\section{Results}\label{sec:results}

{ In the following, we discuss filament galaxies and their properties in the general cosmic context presented above. In Sect.~\ref{sec:Mstar_SFR_environment} we present the stellar, star formation, and environmental properties, while in Sect.~\ref{sec:gas_content_morphology} we focus on the HI and H$_2$ gas. We also compare filament galaxies with sources in the field (AMIGA isolated galaxies) and Virgo cluster.} { In our analysis, we consider upper/lower limits as true detections.}

\subsection{ Stellar, star formation, and environmental properties}\label{sec:Mstar_SFR_environment}

We start by investigating the distribution of the galaxy morphologies, stellar masses, and star formation rates in connection with their { large-scale environment} as traced by the field (isolation), filaments, and { the Virgo cluster.}

Figure~\ref{fig:SFR_vs_Mstar} presents the distribution of the different samples in the star-formation-rate-versus-stellar-mass  diagram, with our sample of filament galaxies in the central panel, the AMIGA sources in the left panel, and the Virgo cluster galaxies on the right. For the three samples, galaxies{ sample} the same stellar mass range of $\log(M_\star/M_\odot)\sim9-11$. The star formation rate at the MS, i.e., ${\rm SFR}_{\rm MS}$, as derived by \citet{Leroy2019}, and the associated scatter are displayed with solid and dashed lines in the figure, respectively. Galaxies with ${\rm SFR}>3~{\rm SFR}_{\rm MS}$ (located above the upper dashed line) are star forming, while sources with  ${\rm SFR}<1/3~{\rm SFR}_{\rm MS}$ (below the lower dashed line) have low levels of star formation activity. Galaxies with intermediate SFR values are instead MS galaxies. For the analysis presented in the following, we explicitly define the quenching fraction $f_{\rm Q}$ as that of sources with suppressed star formation, i.e., ${\rm SFR}<1/3~{\rm SFR}_{\rm MS}$. According to the MS prescription of \citet{Leroy2019}, these are galaxies with specific star formation rate sSFR$<9\times10^{-11}$~yr$^{-1}$ in the stellar mass range of interest here. }



\subsubsection{{ Main sequence and morphology}}\label{sec:main_sequence_and_morphology}

The bulk of the galaxy population in the three samples under investigation is composed of late-type galaxies (LTGs, $T\geq0$). Indeed the fraction of early-type galaxies (ETGs, $T<0$) is only about $\sim27\%$ for both filament (65/245) and Virgo cluster (30/109) galaxies, while it drops significantly to $\sim8\%$ (i.e., 16/193) for the AMIGA sources.  


We further investigate the location of galaxies with respect to the MS, distinguishing between LTGs and ETGs.
For AMIGA, filaments, and Virgo, the bulk of the galaxy population lies on the MS and is essentially composed of LTGs, as mentioned above. 


{ From Fig.~\ref{fig:SFR_vs_Mstar} we can see that the quenching fraction $f_{\rm Q}$ of both ETGs and LTGs  increases from the field (AMIGA) to filaments. For the LTGs of the AMIGA sample,} $f_{\rm Q}=11\%\pm2\%$\footnote{Hereafter, rms uncertainties in the fractions are estimated using binomial statistics \citep[see e.g.,][]{Castignani2014}.} compared to $f_{\rm Q}=23\%\pm3\%$ (41/180) in the filaments, and $f_{\rm Q}=25\%\pm5\%$ (20/79) in the Virgo cluster.  { For ETGs, the quenching fraction is higher than for LTGs in all three environments, and is equal to} $f_{\rm Q}=69\%\pm12\%$ (11/16)  for AMIGA sources, $f_{\rm Q}=83\%\pm5\%$ (54/65) for filament galaxies, and 100\% (30/30) for the Virgo cluster galaxies.

{ The fact that the fraction of galaxies with low levels of star formation increases from the field, to filaments, to the Virgo cluster implies that galaxy star formation activity is suppressed  in  filaments more effectively than the field, but less strongly than in the Virgo cluster.} Therefore, the processes that lead to star formation quenching are already at play in filaments for LTGs { and are even more effective in ETGs.}


 \begin{figure*}[]\centering
\captionsetup[subfigure]{labelformat=empty}
\subfloat[]{\hspace{0.cm}\includegraphics[trim={0.5cm 0.5cm 0.5cm 
0.5cm},clip,width=0.5\textwidth,clip=true]{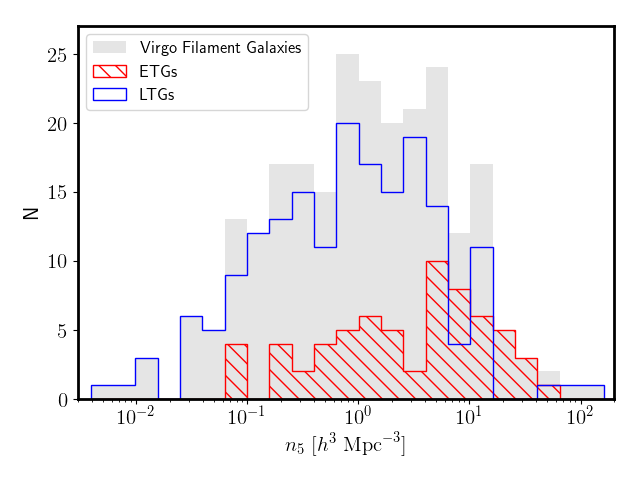}}
\subfloat[]{\hspace{0.cm}\includegraphics[trim={0.5cm 0.5cm 0.5cm 
0.5cm},clip,width=0.5\textwidth,clip=true]{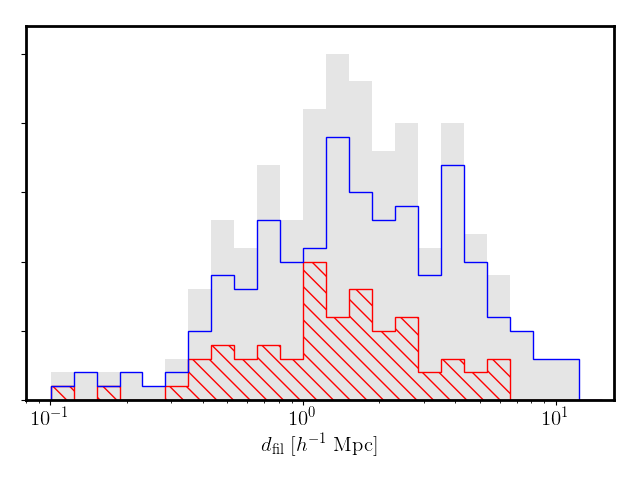}}
\caption{Distribution in local density and distance to the filament spine for our sample of filament galaxies (gray) as well as for the subsamples of ETGs and LTGs.}\label{fig:histo_density_dfil}
\end{figure*}

\subsubsection{Fraction of galaxies with suppressed star formation as a function of their environment}\label{sec:fQ}
We now quantify the impact of filamentary structures on the star formation activity of galaxies. Figure~\ref{fig:scatter_plot_fQ_vs_density_dfil} displays{ the quenching fraction} $f_{\rm Q}$ as a function of the environment, which we parameterize using the local density $n_5$ (left) and distance to the filament (right). 
The AMIGA and Virgo galaxies are associated with the lowest ($n_5=(0.19^{+0.41}_{-0.13})~h^3~{\rm Mpc}^{-3}$)
and highest ($n_5=(6.29^{+27.47}_{-4.54})~h^3~{\rm Mpc}^{-3}$) densities, respectively.\footnote{ We remind that local densities of AMIGA sources have been increased by a factor of 4.3 to correct for the offset with respect to our $n_5$ estimates (see Sect.~\ref{sec:AMIGA_loc_density}).} Filament galaxies are found at intermediate densities, with a median of $n_5=(1.12^{+5.64}_{-0.98})~h^3~{\rm Mpc}^{-3}$. They also span a broader range of $n_5$ with an overlap in density with field and cluster galaxies at the low and high ends of the distribution, respectively, similarly to what has been found for example in simulations \citep{Cautun2014}.

The left panel of Fig.~\ref{fig:scatter_plot_fQ_vs_density_dfil} reveals  a monotonic increase of the quenching fraction from the lowest to the highest densities for field, filament, and cluster galaxies.
The AMIGA sources, which are found at the lowest densities $n_5\lesssim1~h^{3}$~Mpc$^{-3}$, exhibit the lowest values $f_{\rm Q}\lesssim0.2$. On the other extreme, the Virgo cluster shows an increasing fraction ranging from $f_{\rm Q}\sim0.3$ to $f_{\rm Q}\sim0.8$ in  the highest density regions with $n_5\sim10^3~h^{3}$~Mpc$^{-3}$. 
For the filaments specifically, the quenching fraction increases from $f_{\rm Q}\sim0.2$ to $0.6$ within the broad range of densities that they cover, namely $n_5\sim(10^{-2}$ to $10^2)~h^{3}$~Mpc$^{-3}$.  At comparable densities, filaments have similar  $f_{\rm Q}$ values to the field and the Virgo cluster.  { We verified that the observed trend of increasing $f_{\rm Q}$ with $n_5$ for filament and Virgo galaxies is mostly driven by massive $\log(M_\star/M_\odot)>10$ galaxies. For less massive sources, $f_{\rm Q}$ shows a flatter and more noisy behavior when plotted against $n_5$, averaging around $f_{\rm Q}\lesssim0.4$ for each of the three cosmic environments considered.}


Considering the filament galaxies altogether, we find $f_{\rm Q}=(39\pm3)\%$, i.e., 95/245. This value is fairly consistent with  the quiescent fraction of $\sim50\%$ found by \citet{Bonjean2020} for higher redshift $0.1<z<0.3$ cosmological filaments. Interestingly, this may indicate that the average star formation properties of filament galaxies have not  dramatically evolved over the last $\sim2$~Gyr.

The right panel of Fig.~\ref{fig:scatter_plot_fQ_vs_density_dfil} shows the variation of $f_{\rm Q}$ with distance to the filament spine.  The fraction of filament galaxies with suppressed star formation declines from $f_{\rm Q}\sim0.6$ in the central regions down to $f_{\rm Q}\sim0.2$ at $d_{\rm fil}\lesssim1~h^{-1}$~Mpc. This suggests that filaments do not start affecting star formation until galaxies are within $\sim1~h^{-1}$~Mpc. This radius at which quenching{ appears to} start is fairly consistent with that of $(3.0\pm1.1)~h^{-1}$~Mpc found by \citet{Bonjean2020} in filaments at higher redshifts $0.1<z<0.3$ and with the average filament{ radius}  $d_{0,fil}\sim2~h^{-1}$~Mpc for the density profile reported in Sect.~\ref{sec:density_profiles}.

The $f_{\rm Q}$ of ETGs and LTGs is dramatically different at all distances from the filament spine. In the central regions ($d_{\rm fil}\lesssim0.2~h^{-1}$~Mpc), half of the LTGs have suppressed star formation ($f_{\rm Q}\sim0.5$), and this fraction declines with increasing distance to the filament.  In contrast, nearly all the ETGs show suppressed star formation regardless of separation. At larger distances, both LTGs and ETGs see their $f_{\rm Q}$ decreasing but the ETGs stay at much higher values than LTGs ($f_{\rm Q}\sim0.8$ vs. 0.2). 


Figure~\ref{fig:scatter_plot_fQ_vs_density_dfil} therefore suggests that the suppression of star formation in filaments is strongly regulated by the local density, and remains at a level that is intermediate between that of  the field and that of the Virgo cluster. The distance to the filament spine appears to be a secondary parameter, but the onset of the passive population in filaments emerges once a separation{ of} $\sim1~h^{-1}$~Mpc is reached.
{ Overall, our analysis suggests that,} at the  lookback time of Virgo, nearly all ETGs have already seen their star formation activity impacted, while LTGs  are still sensitive to their position relative to the filament spine. The following section further clarifies the relationship between local density and distance to the filament spines.

 \begin{figure*}[]\centering
\captionsetup[subfigure]{labelformat=empty}
\subfloat[]{\hspace{0.cm}\includegraphics[trim={0.2cm 2.2cm 2.3cm 
0.5cm},clip,width=0.33\textwidth,clip=true]{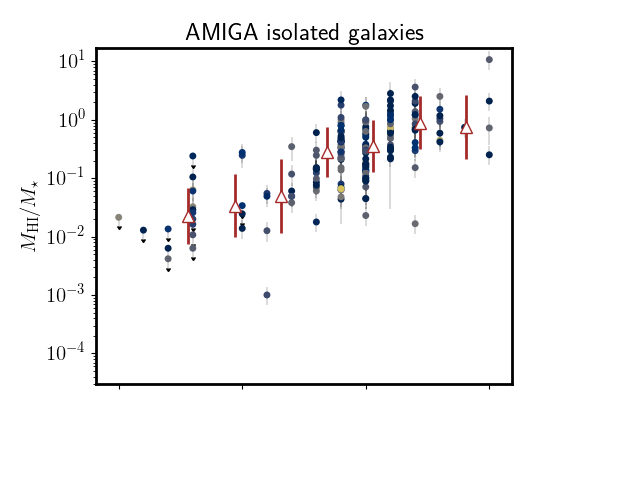}}
\subfloat[]{\hspace{-0.3cm}\includegraphics[trim={1.5cm 2.2cm 1cm 0.5cm},clip,width=0.33\textwidth,clip=true]{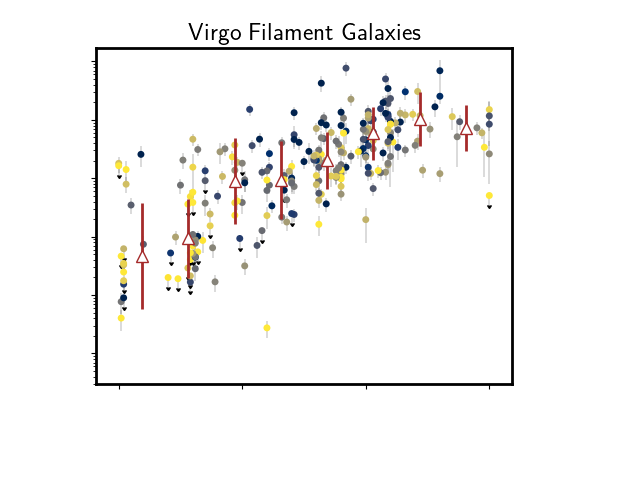}}
\subfloat[]{\hspace{-0.6cm}\includegraphics[trim={2.3cm 2.2cm 0.2cm 
0.5cm},clip,width=0.33\textwidth,clip=true]{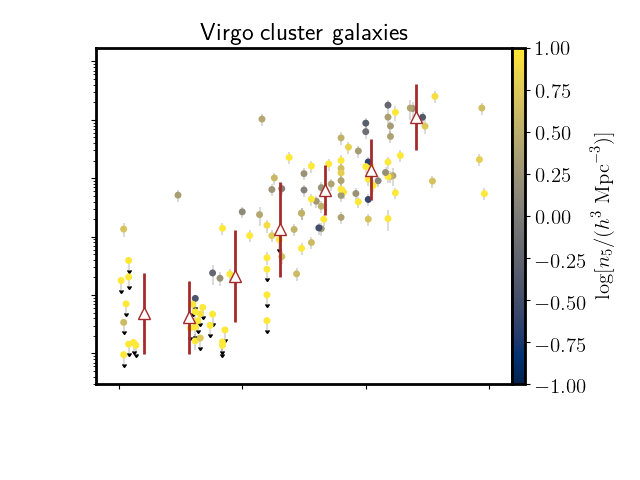}}\\
\vspace{-0.8cm}
\subfloat[]{\hspace{0.cm}\includegraphics[trim={0.2cm 2.3cm 2.3cm 
1.2cm},clip,width=0.33\textwidth,clip=true]{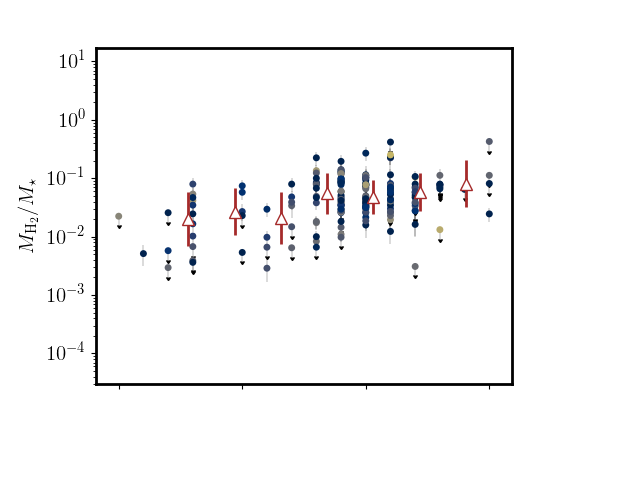}}
\subfloat[]{\hspace{-0.3cm}\includegraphics[trim={1.5cm 2.3cm 1cm 1.2cm},clip,width=0.33\textwidth,clip=true]{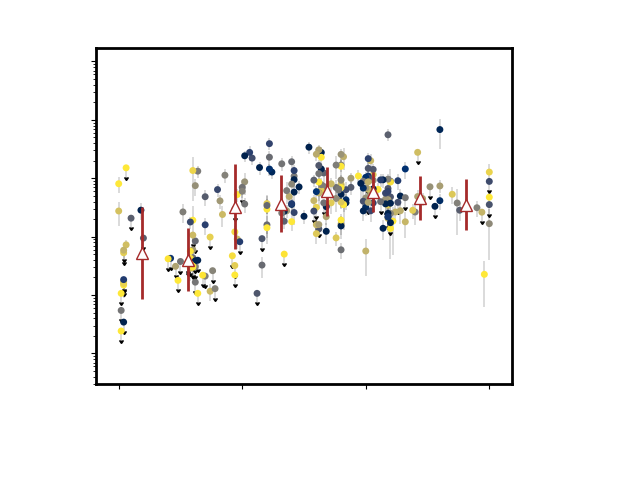}}
\subfloat[]{\hspace{-0.6cm}\includegraphics[trim={2.3cm 2.3cm 0.2cm 
1.2cm},clip,width=0.33\textwidth,clip=true]{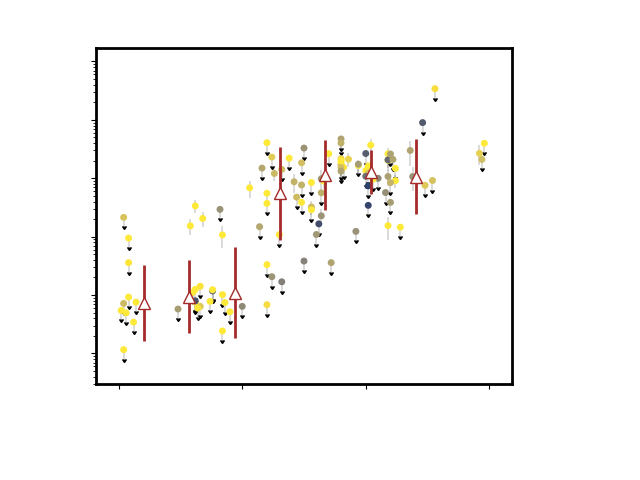}}\\
\vspace{-0.8cm}
\subfloat[]{\hspace{0.cm}\includegraphics[trim={0.2cm 0.6cm 2.3cm 1.2cm},clip,width=0.33\textwidth,clip=true]{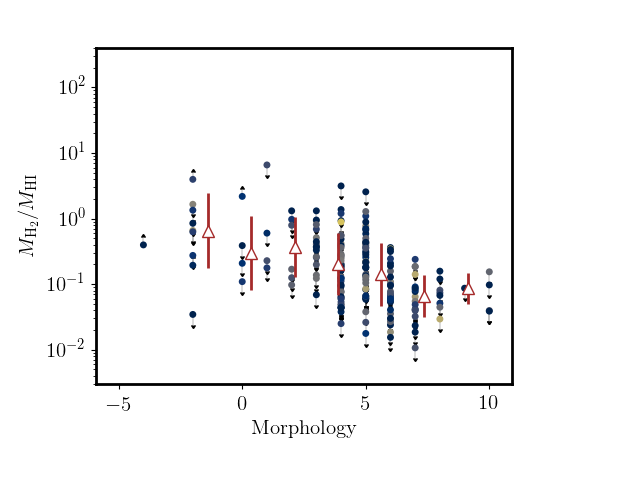}}
\subfloat[]{\hspace{-0.3cm}\includegraphics[trim={1.5cm 0.6cm 1cm 1.2cm},clip,width=0.33\textwidth,clip=true]{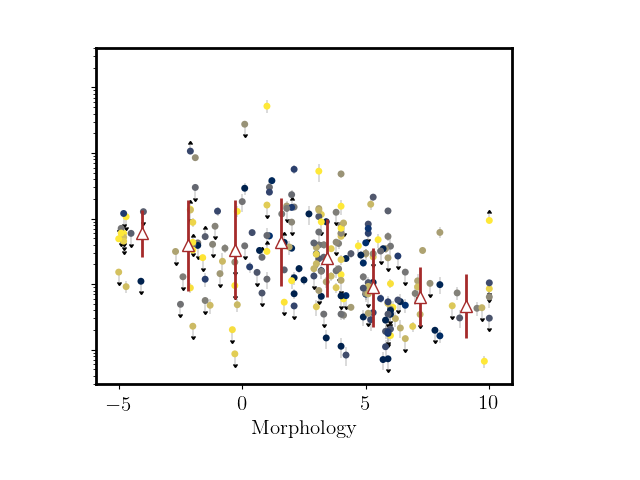}}
\subfloat[]{\hspace{-0.6cm}\includegraphics[trim={2.3cm 0.6cm 0.2cm 1.2cm},clip,width=0.33\textwidth,clip=true]{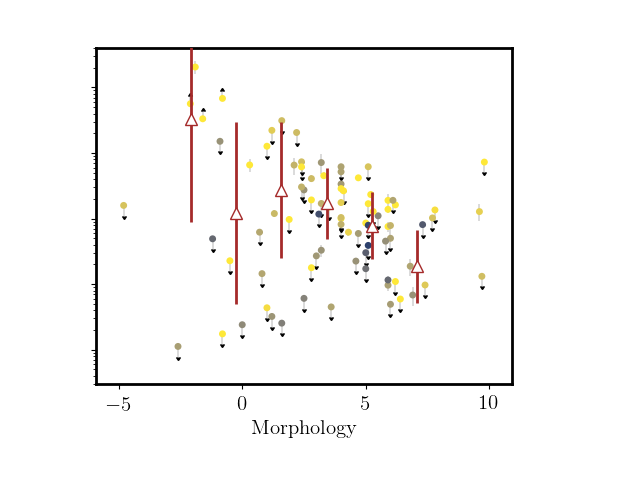}}\\
\caption{From top to bottom: $M_{\rm HI}/M_\star$, $M_{\rm H_2}/M_\star$,  and $M_{\rm H_2}/M_{\rm HI}$ as a function of morphology for Virgo filament galaxies (center), AMIGA isolated galaxies (left), and Virgo cluster galaxies (right). Points are color coded according to local density $n_5$.
Triangles show the{ binned median values}, while their error bars correspond to  the rms dispersion around the median; { equally spaced bins with at least five points each} have been considered. { In the bottom row, we do not plot a source if both the HI and H$_2$ masses are upper limits.}}\label{fig:scatter_plot_MHI_MH2_vs_morphology}
\end{figure*}

 \begin{figure*}[]\centering
\captionsetup[subfigure]{labelformat=empty}
\subfloat[]{\hspace{0.cm}\includegraphics[trim={0.2cm 0cm 2.8cm 
0cm},clip,width=0.33\textwidth,clip=true]{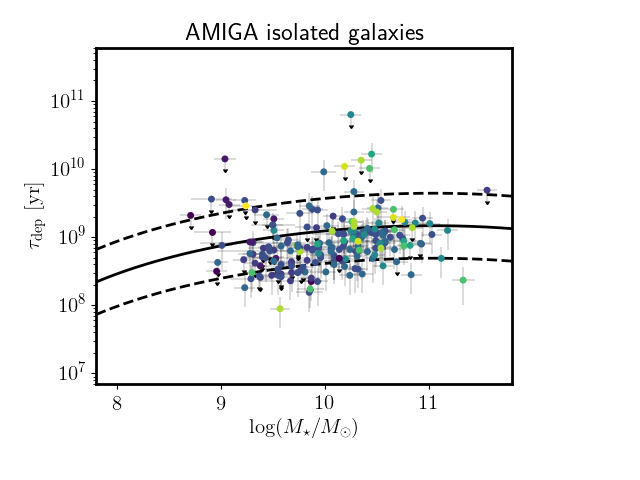}}
\subfloat[]{\hspace{-0.1cm}\includegraphics[trim={2.cm 0cm 1cm 
0cm},clip,width=0.33\textwidth,clip=true]{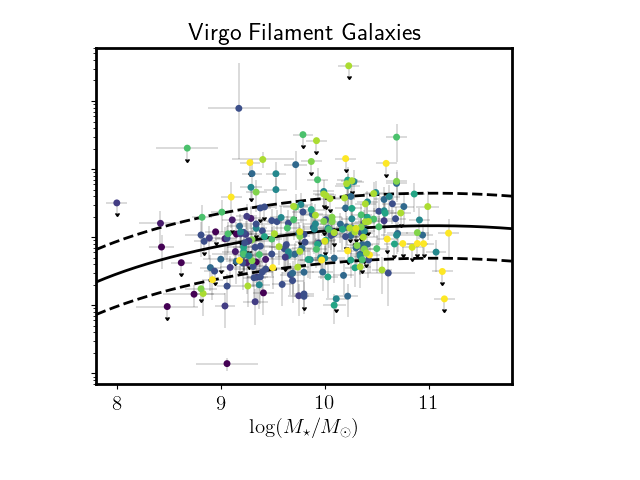}}
\subfloat[]{\hspace{-1.cm}\includegraphics[trim={2.cm 0cm 1cm 
0cm},clip,width=0.33\textwidth,clip=true]{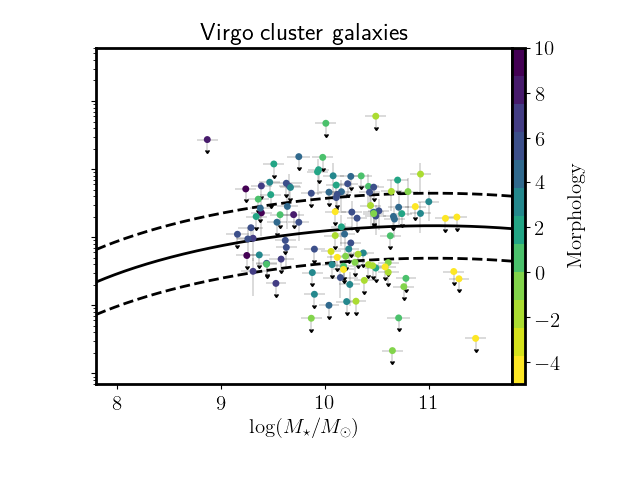}}
\caption{Depletion time vs. stellar mass. Different panels refer to Virgo filament galaxies (center), AMIGA isolated galaxies (left), and Virgo cluster galaxies (right).
{ Sources are color coded according to their morphological de Vaucouleurs classification.}
Solid and dashed lines correspond to the local prescription and model uncertainties by \citet{Tacconi2018} for MS galaxies, calibrated using a Galactic conversion factor $\alpha_{\rm CO}=4.3~M_\odot$~(K~km~s$^{-1}$~pc$^{2}$)$^{-1}$.
}\label{fig:scatter_plot_tdep}
\end{figure*}

\subsubsection{Local densities and distances to the filaments}\label{sec:result_local_densities_distances}

In Fig.~\ref{fig:histo_density_dfil} we show the distribution of our filament galaxies as a function of local density (left panel) and distance to the filament spine (right panel). The distribution of ETGs and LTGs are shown with the hatched red and blue histograms, respectively, while the filled gray histogram shows the combined populations.
As shown in the left panel of Fig. \ref{fig:histo_density_dfil}, LTGs are spread over a wide range of local densities and distances to the filaments, with a moderate median  density of $n_5=(0.9^{+4.4}_{-0.7})~h^{3}~{\rm Mpc}^{-3}$. ETGs are preferentially found in regions of higher local densities, with a median value of $n_5=(4.1^{+10.8}_{-3.6})~h^{3}~{\rm Mpc}^{-3}$.
While the ETGs also span a wide range of local densities (a factor of 1000 in density), they do not populate the lowest density regions. 
The right panel of Fig.~ \ref{fig:histo_density_dfil} shows that ETGs are also preferentially located
closer to the filament spines than LTGs; the median distance for ETGs is $\langle d_{\rm fil}\rangle =(1.4^{+1.3}_{- 0.8})~h^{-1}$~Mpc, whereas the median distance for LTGs is $\langle d_{\rm fil } \rangle =(1.7^{+2.6}_{- 1.1})~h^{-1}$~Mpc.
The Mann–Whitney–Wilcoxon test indicates that the difference in the median values between the two morphological classes is statistically significant for both local densities ($p$-value=$5.61\times10^{-6}$, { i.e., 4.5-$\sigma$} ) and distances $d_{\rm fil}$ ($p$-value=$3.89\times10^{-2}$, { i.e., 2.1-$\sigma$} ). { The $p$-values are those of the null hypothesis that both classes are drawn from the same parent distribution.} 
{ The reported significance is higher for the local density than for the distance to the filament, which suggests that the latter may be a secondary parameter, as also found for the quenching fraction in Sect.~\ref{sec:fQ}.}

{ Overall, our results suggest} that massive galaxies{ tend to preferentially reside in} the densest central regions of filaments in the local Universe, possibly as a result of previous mergers of lower mass sources. This{ result} is in agreement with similar findings found at higher redshifts for a number of  spectroscopic surveys \citep{Tempel_Libeskind2013,Malavasi2017,Laigle2018,Kraljic2018,Welker2020,Kuutma2020}.
Our finding is also consistent with those of \citet{Rost2020}, who compared three catalogs of cosmological filaments identified in the { SDSS} and found that the overdensity profile of red galaxies is systematically higher than that of blue galaxies.
Figure~\ref{fig:density_profiles} shows that there is a wide range of local densities at fixed distance from the filament spines, except for the largest distances for which densities are only low. Therefore, while distances and densities are correlated, we show in the following sections
(e.g., Sects~\ref{sec:gas_content_tdep_environment}, \ref{sec:gasdef_density}) that local density seems to be a stronger driver of morphological evolution than distance from the filament.



\subsection{Atomic and molecular gas}\label{sec:gas_content_morphology}
We now investigate the general properties of the gas reservoirs of filament galaxies in comparison with those of AMIGA and Virgo cluster galaxies. 

\subsubsection{Relation between gas fraction and morphology} \label{sec:result_gas_fraction_vs_morphology}
Figure~\ref{fig:scatter_plot_MHI_MH2_vs_morphology} presents the $M_{\rm HI}/M_\star$ ratio as a function of galaxy morphology for the three samples considered, as well as the molecular-gas-to-stellar-mass ratio, $M_{\rm H_2}/M_\star$, and the $M_{\rm H_2}/M_{\rm HI}$ ratio. { The reported y-axis uncertainties in the data points were estimated by propagating in quadrature those of $M_{\rm HI}$, $M_{\rm H_2}$, and $M_\star$.}
Both $M_{\rm HI}/M_\star$ and $M_{\rm H_2}/M_\star$ ratios clearly increase from ETGs to LTGs. The slope of this function is increasingly steep{ when comparing galaxies in isolation to those in the cluster,} as a consequence of the fact that while LTGs have very similar gas fractions in all cosmic structures, ETGs are more strongly gas depleted in very dense {environments} such as clusters.  This observed trend of gas fraction with morphology is  stronger for $M_{\rm HI}/M_\star$  than for $M_{\rm H_2}/M_\star$, again driven by the ETGs being increasingly gas poor with increasing environmental density. 


The median gas-to-stellar-mass ratios for ETGs decrease in the three environments as follows: {  $M_{\rm HI}/M_\star=0.02^{+0.04}_{-0.01}$ (AMIGA), $0.01^{+0.15}_{-0.01}$ (filaments), and $0.0005^{+0.0018}_{-0.0003}$ (Virgo cluster), while $M_{\rm H_2}/M_\star=0.02^{+0.03}_{-0.02}$ (AMIGA),  $0.004^{+0.025}_{-0.003}$ (filaments), and $0.0009^{+0.0117}_{-0.0003}$ (Virgo cluster).}
For LTGs, the median ratios are higher than those of ETGs and are fairly similar { in} the three different environments, within the dispersions: { $M_{\rm HI}/M_\star=0.34^{+0.71}_{-0.25}$ (AMIGA), $0.30^{+0.85}_{-0.21}$ (filaments), and $0.08^{+0.51}_{-0.07}$ (Virgo cluster), while $M_{\rm H_2}/M_\star=0.05^{+0.05}_{-0.02}$ (AMIGA), $0.05^{+0.09}_{-0.03}$ (filaments), and $0.11^{+0.15}_{-0.09}$ (Virgo cluster).}



{ For both ETGs and LTGs,} the scatter around the median values increases{ from the field, to filaments, and then to clusters}, with an increased fraction of upper limits{ in HI (H$_2$) equal to 6\% (32\%) for AMIGA galaxies, 12\% (27\%) for filament sources, and 25\% (75\%) for Virgo galaxies.} This result is consistent with those of \citet{Bok2020}, who found that the scatter in the gas content is significantly higher for galaxies in pairs than for isolated AMIGA galaxies.

\subsubsection{Relation between H$_2$-to-HI mass ratio and morphology} \label{sec:result_gas_ratio_vs_morphology}
The bottom row of Figure~\ref{fig:scatter_plot_MHI_MH2_vs_morphology} shows the molecular-to-atomic-mass ratio of $M_{\rm H_2}/M_{\rm HI}$ as a function of galaxy morphology, where sources with upper limits in  both H$_2$ and HI have been conservatively excluded. The ratios of $M_{\rm H_2}/M_{\rm HI}$ are  essentially $\lesssim1$  regardless of the considered environment of the galaxy: isolation, in filament, or in the Virgo cluster.

Both the isolated AMIGA and filament galaxies show a decrease in $M_{\rm H_2}/M_{\rm HI}$ with increasing $T$ type. This is mostly driven by{ the} increase  in HI mass in galaxies moving towards late-type morphology, which is steeper than for $M_{\rm H_2}$. { This correlation was also discussed by \citet{Obreschkow_Rawlings2009}. Indeed, from LTGs to ETGs, galaxies have less gas and higher metallicities.  While ETGs are deficient in HI relative to LTGs, they are less deficient in H$_2$. This explains the observed trend in $M_{\rm H_2}/M_{\rm HI}$ going from LTGs to ETGs.}


The median H$_2$-to-HI mass ratios for the three considered environments are: { $M_{\rm H_2}/M_{\rm HI}=0.16^{+0.42}_{-0.11}$ (AMIGA), $0.24^{+0.84}_{-0.19}$ (filaments), and $1.10^{+5.33}_{-0.99}$ (Virgo cluster).}   As we move from the field, to filaments, and then to cluster galaxies, the ratio $M_{\rm H_2}/M_{\rm HI}$, plotted against the morphology, shows increased scatter, of the order of $\sim$1~dex. This value is higher than that observed for $M_{\rm H_2}$ and $M_{\rm HI}$ separately, but is similar to that found in the literature for field galaxies which have, on average, $M_{\rm H_2}/M_{\rm HI}\sim{0.3}$ \citep[][]{Saintonge2011}.

 \begin{figure*}[]\centering
\captionsetup[subfigure]{labelformat=empty}
\subfloat[]{\includegraphics[trim={0.2cm 0.3cm 2.8cm 1cm},clip,width=0.4\textwidth,clip=true]{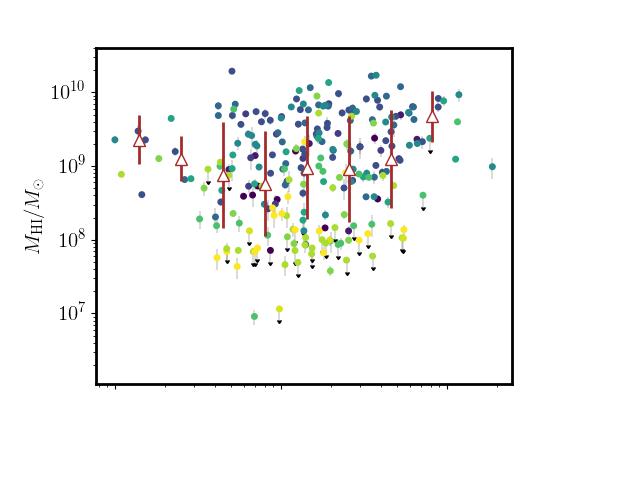}}
\subfloat[]{\includegraphics[trim={2cm 0.3cm 1cm 
1cm},clip,width=0.4\textwidth,clip=true]{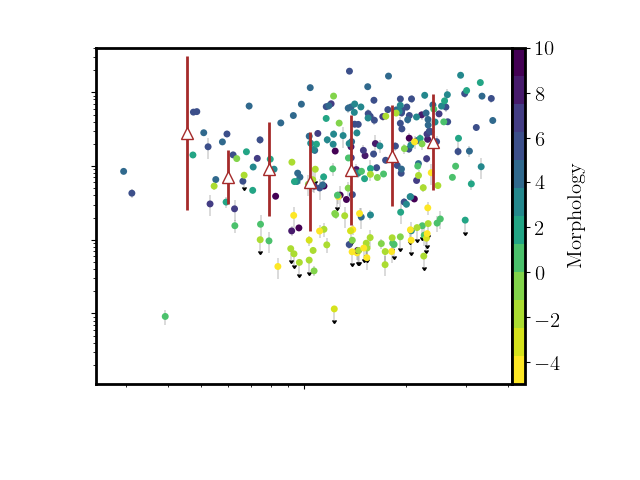}}\\
\vspace{-2.cm}
\subfloat[]{\includegraphics[trim={0.2cm 0.3cm 2.8cm 1cm},clip,width=0.4\textwidth,clip=true]{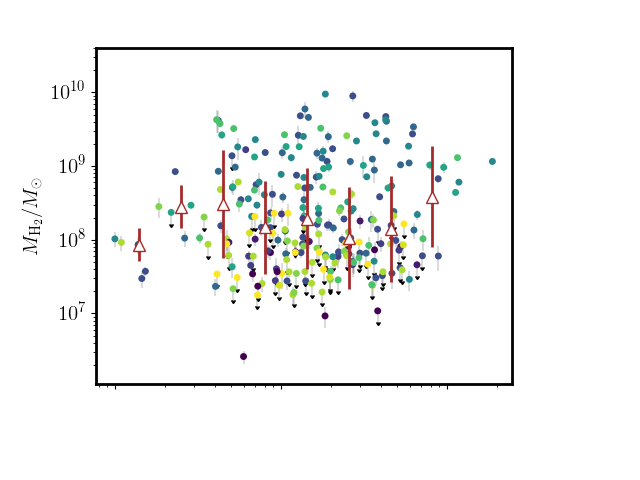}}
\subfloat[]{\includegraphics[trim={2cm 0.3cm 1cm 
1cm},clip,width=0.4\textwidth,clip=true]{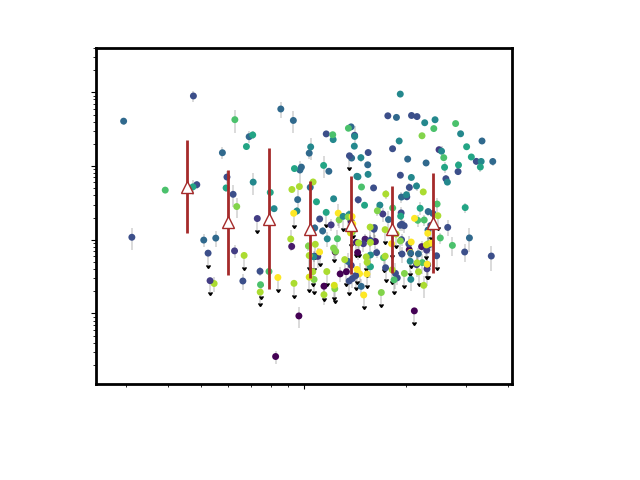}}\\
\vspace{-2.cm}
\subfloat[]{\includegraphics[trim={0.2cm 0.5cm 2.8cm 1cm},clip,width=0.4\textwidth,clip=true]{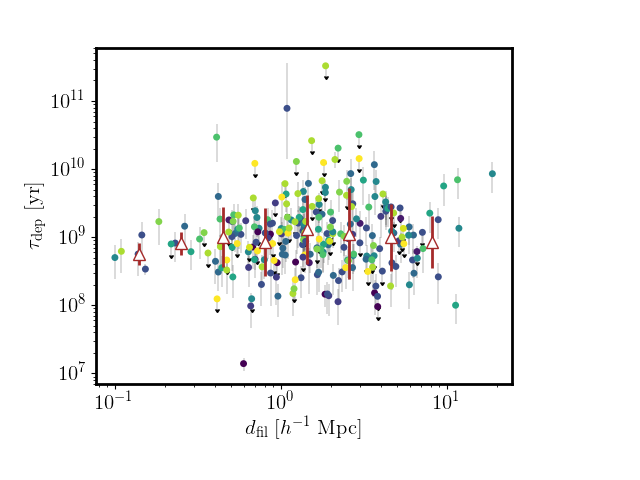}}
\subfloat[]{\includegraphics[trim={2cm 0.5cm 1cm 
1cm},clip,width=0.4\textwidth,clip=true]{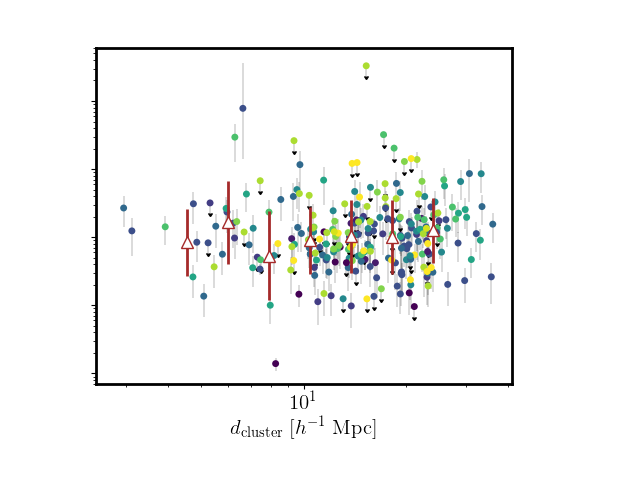}}
\caption{$M_{\rm HI}$ { (top row)}, $M_{\rm H_2}$ { (center row)}, and depletion time { (bottom row)} plotted against  the distance from the filament spine { (left column)} and the distance to the Virgo cluster { (right column)} for our sample of filament galaxies.
Triangles show the { binned median values}, while their error bars correspond to  the rms dispersion around the median; { equally spaced bins, in log scale, with at least five points each} have been considered. { Sources are color coded according to their morphological de Vaucouleurs classification.}} \label{fig:dist_vs_MHI_and_MH2}
\end{figure*}

 \begin{figure*}[]\centering
\captionsetup[subfigure]{labelformat=empty}
\subfloat[]{\hspace{0.cm}\includegraphics[trim={0.2cm 2.3cm 2.8cm 0.2cm},clip,width=0.33\textwidth,clip=true]{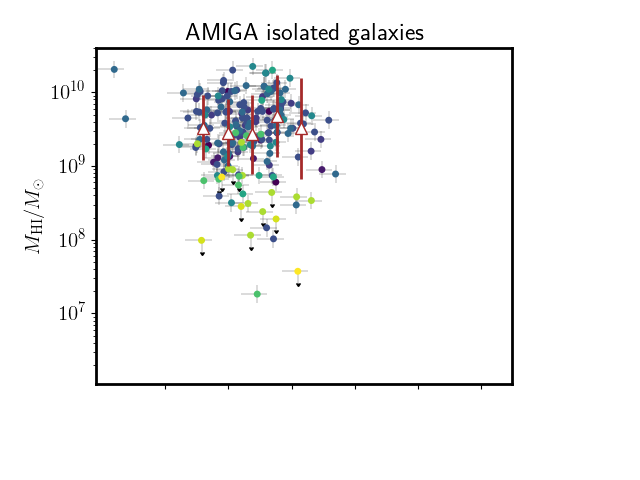}}
\subfloat[]{\hspace{-0.cm}\includegraphics[trim={2cm 2.3cm 1cm 
0.2cm},clip,width=0.33\textwidth,clip=true]{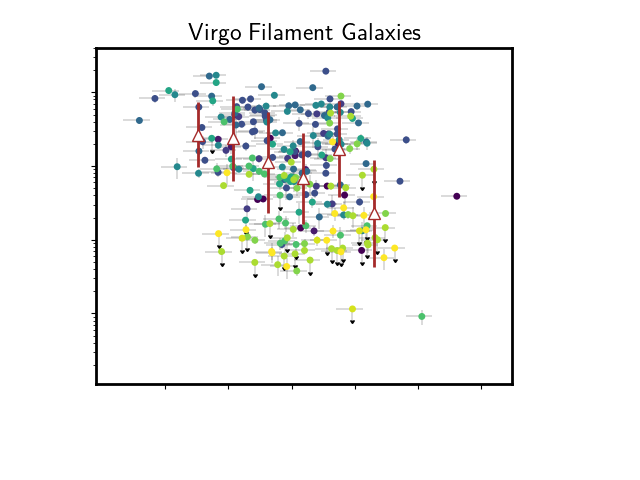}}
\subfloat[]{\hspace{-0.9cm}\includegraphics[trim={2cm 2.3cm 1cm 
0.2cm},clip,width=0.33\textwidth,clip=true]{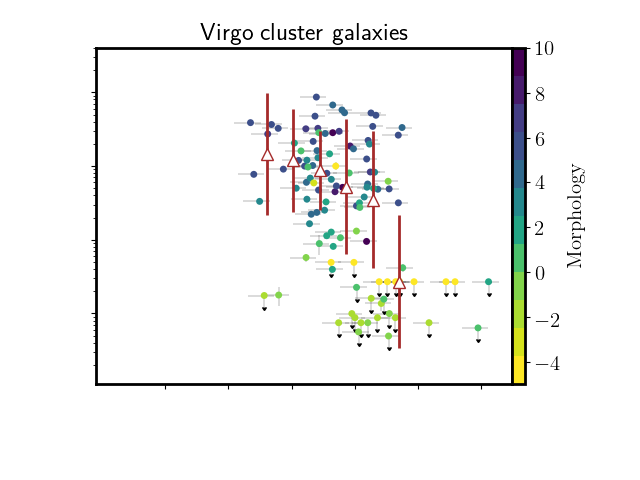}}\\
\vspace{-0.8cm}
\subfloat[]{\hspace{0.cm}\includegraphics[trim={0.2cm 2.3cm 2.8cm 1.2cm},clip,width=0.33\textwidth,clip=True]{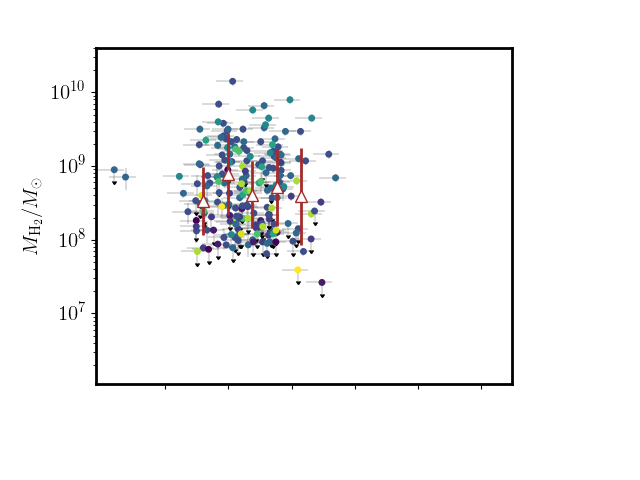}}
\subfloat[]{\hspace{-0.cm}\includegraphics[trim={2cm 2.3cm 1cm 
1.2cm},clip,width=0.33\textwidth,clip=true]{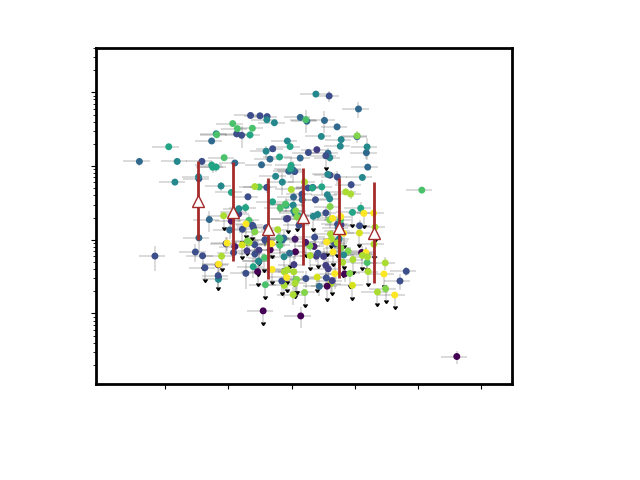}}
\subfloat[]{\hspace{-0.9cm}\includegraphics[trim={2cm 2.3cm 1cm 
1.2cm},clip,width=0.33\textwidth,clip=true]{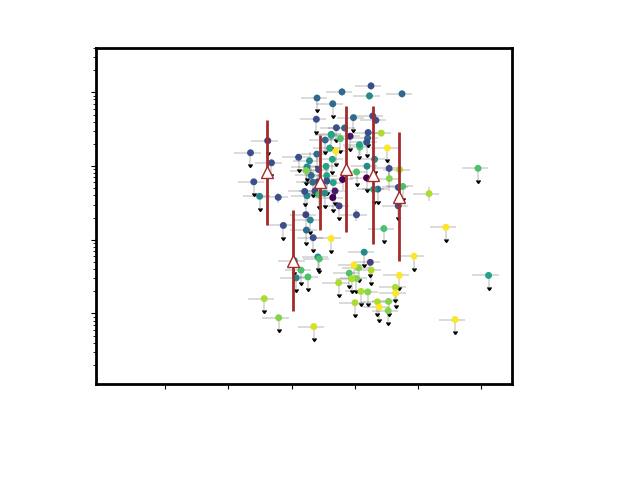}}\\
\vspace{-0.8cm}
\subfloat[]{\hspace{0.cm}\includegraphics[trim={0.2cm 2.3cm 2.8cm 1.2cm},clip,width=0.33\textwidth,clip=true]{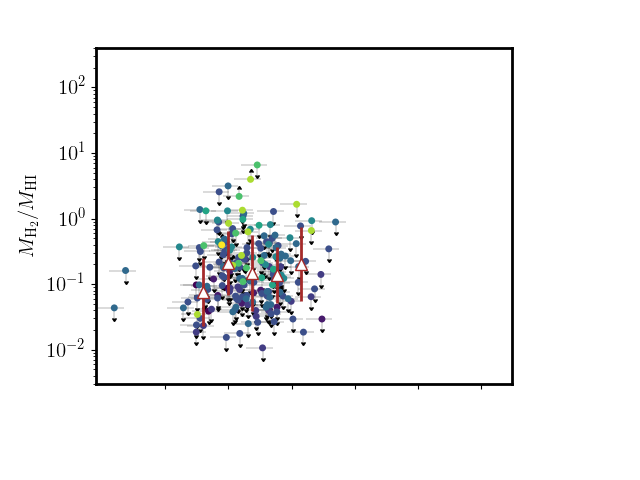}}
\subfloat[]{\hspace{-0.cm}\includegraphics[trim={2cm 2.3cm 1cm 
1.2cm},clip,width=0.33\textwidth,clip=true]{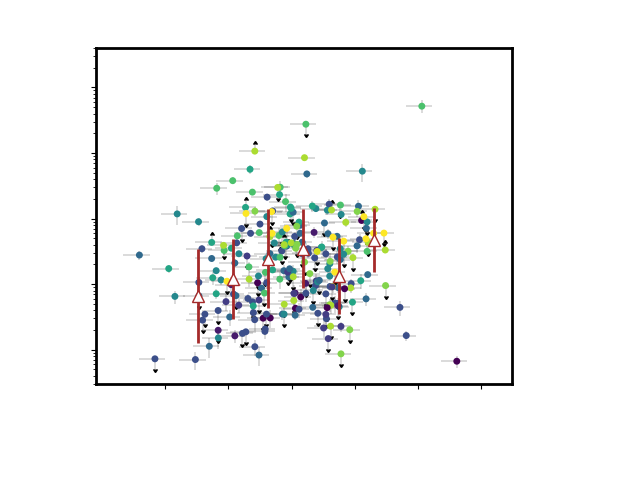}}
\subfloat[]{\hspace{-0.9cm}\includegraphics[trim={2cm 2.3cm 1cm 1.2cm},clip,width=0.33\textwidth,clip=true]{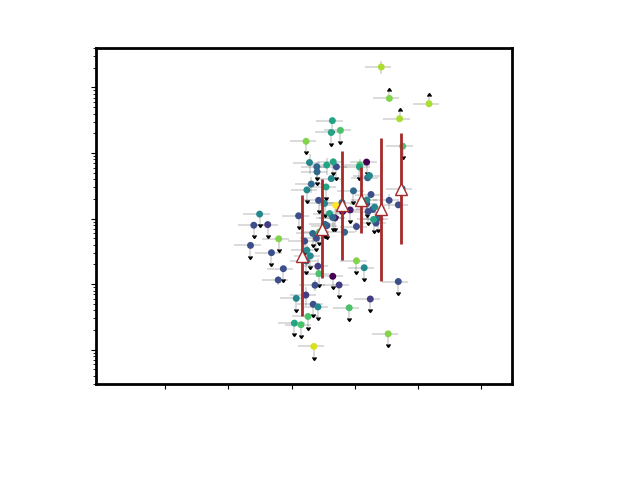}}\\
\vspace{-0.8cm}
\subfloat[]{\hspace{0.cm}\includegraphics[trim={0.2cm 0.6cm 2.8cm 1.2cm},clip,width=0.33\textwidth,clip=true]{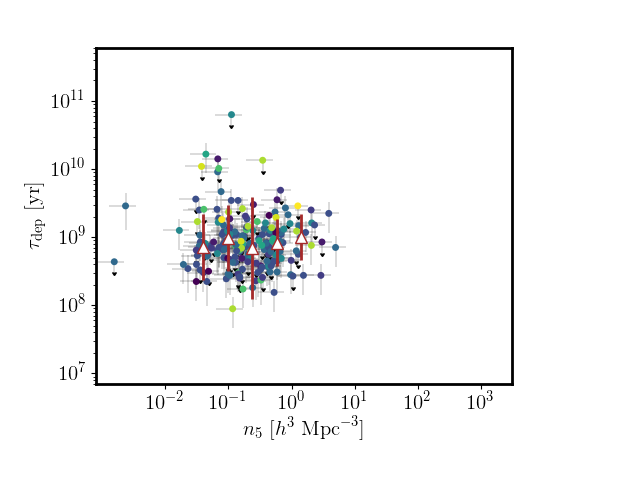}}
\subfloat[]{\hspace{-0.cm}\includegraphics[trim={2cm 0.6cm 1cm 
1.2cm},clip,width=0.33\textwidth,clip=true]{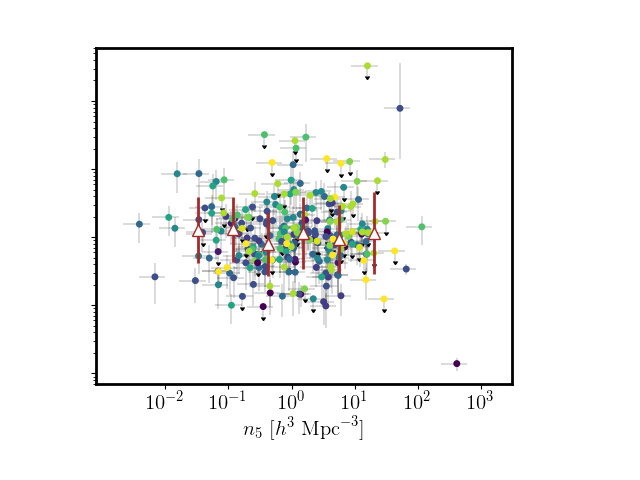}}
\subfloat[]{\hspace{-0.9cm}\includegraphics[trim={2cm 0.6cm 1cm 
1.2cm},clip,width=0.33\textwidth,clip=true]{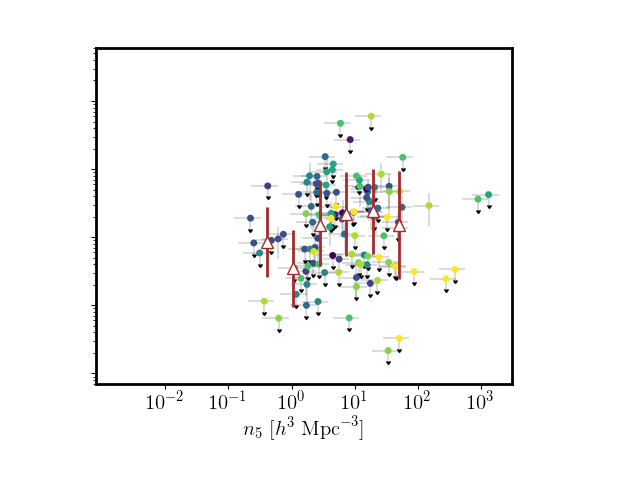}}
\caption{From top to bottom: $M_{\rm HI}$, $M_{\rm H_2}$, $M_{\rm H_2}/M_{\rm HI}$, and $\tau_{\rm dep}$  plotted against  local densities for Virgo filament galaxies (center),  AMIGA isolated galaxies (left), and Virgo cluster galaxies (right). 
Triangles show the{ binned median values}, while their error bars correspond to  the rms dispersion around the median; { equally spaced bins, in log scale, with at least five points each} have been considered. { Sources are color coded according to their morphological de Vaucouleurs classification.}}\label{fig:density_vs_MHI_MH2_tdep_comparison}
\end{figure*}

 \begin{figure*}[]\centering
\captionsetup[subfigure]{labelformat=empty}
\subfloat[]{\hspace{0.cm}\includegraphics[trim={0.2cm 0.6cm 2.8cm 0.5cm},clip,width=0.33\textwidth,clip=true]{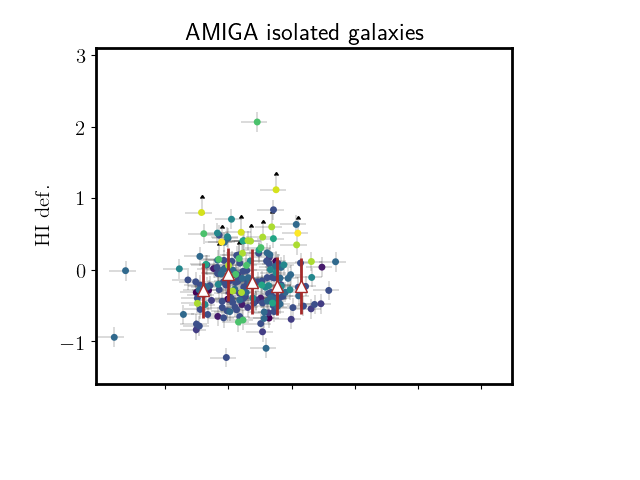}}
\subfloat[]{\hspace{-0.cm}\includegraphics[trim={2cm 0.6cm 1cm 
0.5cm},clip,width=0.33\textwidth,clip=true]{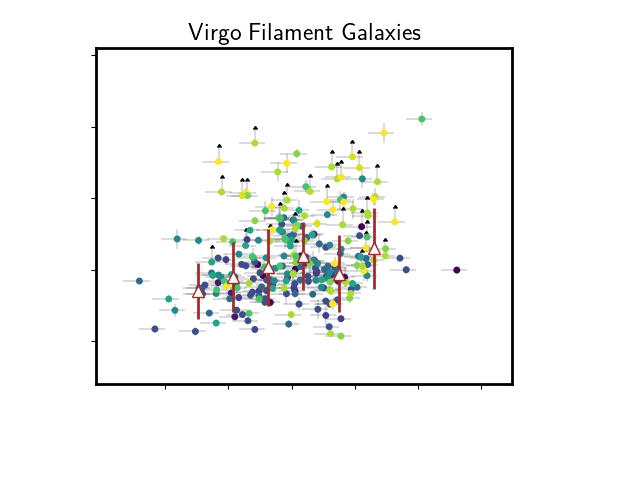}}
\subfloat[]{\hspace{-0.9cm}\includegraphics[trim={2cm 0.6cm 1cm 
0.5cm},clip,width=0.33\textwidth,clip=true]{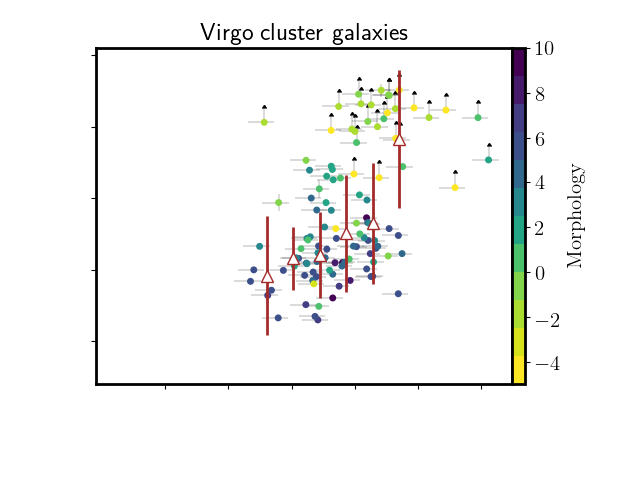}}\\
\vspace{-1.6cm}
\subfloat[]{\hspace{0.cm}\includegraphics[trim={0.2cm 0.6cm 2.8cm 1.2cm},clip,width=0.33\textwidth,clip=true]{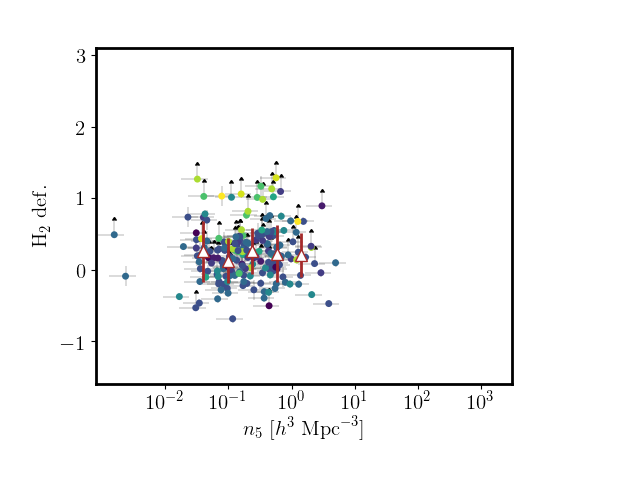}}
\subfloat[]{\hspace{-0.cm}\includegraphics[trim={2cm 0.6cm 1cm 
1.2cm},clip,width=0.33\textwidth,clip=true]{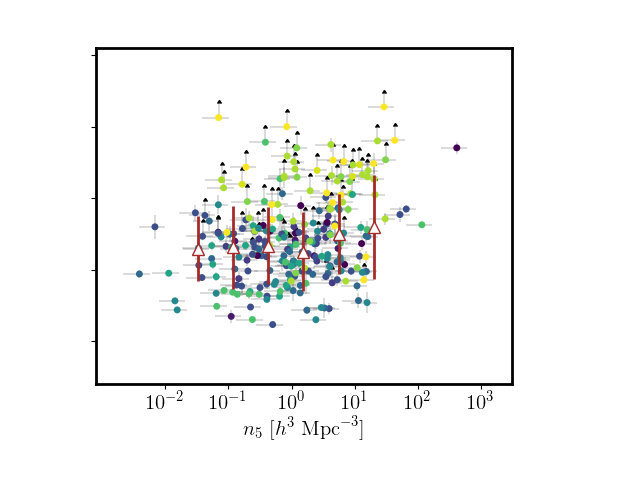}}
\subfloat[]{\hspace{-0.9cm}\includegraphics[trim={2cm 0.6cm 1cm 
1.2cm},clip,width=0.33\textwidth,clip=true]{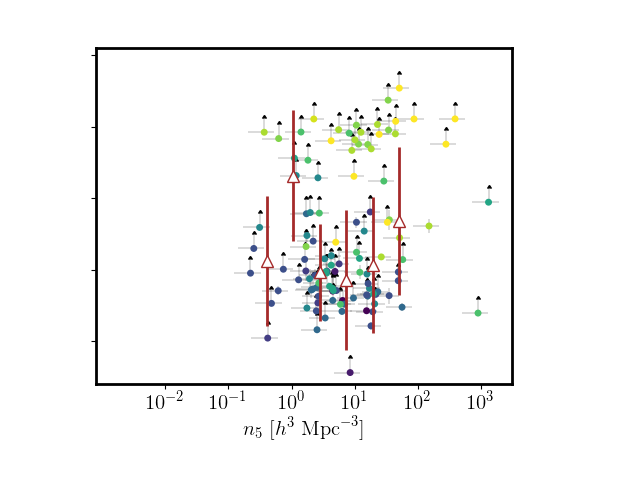}}\\
\caption{HI deficiency (top row) and H$_2$ deficiency (bottom row) plotted against  local densities for Virgo filament galaxies (center), AMIGA isolated galaxies (left),  and Virgo cluster galaxies (right). 
Triangles show the binned median values, while their error bars correspond to  the rms dispersion around the median; equally, { equally spaced bins, in log scale, with at least five points each} have been considered. { Sources are color coded according to their morphological de Vaucouleurs classification.}}\label{fig:HI_H2_def_plots_comparison}
\end{figure*}

\subsubsection{Relation between depletion timescale and stellar mass}\label{sec:depletion_time}

Figure~\ref{fig:scatter_plot_tdep} presents the depletion timescale, $\tau_{\rm dep}=M_{\rm H_2}/{\rm SFR}$, i.e., the time over which the galaxy molecular gas reservoir will be consumed by star formation, as a function of stellar mass. { The uncertainty in  $\tau_{\rm dep}$ was estimated by propagating in quadrature those of both $M_{\rm H_2}$ and the SFR.}
{ The solid line in Fig.~\ref{fig:scatter_plot_tdep} shows the prescription of $\tau_{\rm dep}$ versus $M_\star$ for MS galaxies from \citet{Tacconi2018}, and the dashed lines show the uncertainty in the relation.}{ The vast majority of AMIGA isolated galaxies have low depletion times, which are scattered between the MS relation represented by the solid line and the lower dashed line.
On the other hand, Virgo filament galaxies show average MS depletion times, as the sources fully sample the region between the dashed lines, with only a fraction of them outside this region.
Virgo cluster galaxies are instead more scattered, reaching higher and lower depletion times outside the region delimited by the dashed lines.}

{ The scatter of the data points thus increases between isolated, filament, and cluster samples,} to a point where  the relation between $\tau_{\rm dep}$ and $M_\star$ breaks down for Virgo galaxies.  As can be seen in Appendix~\ref{sec:SFR_gas_diagnostic}, the number of galaxies with upper limits in H$_2$ and nonnegligible SFR, often at the level of the MS, increases at stellar masses $\gtrsim 10^{9.5}~M_\odot$. This suggests that gas depletion precedes star formation quenching for these sources, the majority of which are ETGs.  In the Virgo cluster, this effect is also observed in LTGs.

{ Following the MS scaling relation,} in isolation and in filaments, more massive galaxies tend to have longer depletion timescales than less massive galaxies. This{ trend} is mainly driven by LTGs { on the MS, for which the H$_2$ mass increases by a factor of approximately 100 within $\log(M_\star/M_\odot)\simeq9-11$ (see Fig.~\ref{fig:MHI_MH2_vs_Mstar} in the Appendix~\ref{sec:SFR_gas_diagnostic}), while the increase in SFR is less strong within the same stellar mass range (Fig.~\ref{fig:SFR_vs_Mstar}).}

{ However, the fact that AMIGA sources are shifted to low $\tau_{\rm dep}$ with respect to the relation for MS galaxies may be explained by noting that $M_{\rm H_2}/M_{\rm HI}$ of AMIGA galaxies is slightly lower than for the other two environments (Fig.~\ref{fig:scatter_plot_MHI_MH2_vs_morphology}). In fact, many AMIGA sources have lower $M_{\rm H_2}$ and higher $M_{\rm HI}$ than that predicted for MS galaxies at any given stellar mass (Fig.~\ref{fig:MHI_MH2_vs_Mstar}). This suggests that, being in isolation, the disks of AMIGA sources are undisturbed and the molecular gas reservoirs are consumed rapidly, without being effectively replenished \citep[e.g.,][]{Braine_Combes1992}, while at the same time the HI-to-H$_2$ conversion is not efficient.}

\subsubsection{Relation between gas masses and environmental parameters}\label{sec:gas_content_tdep_environment}

In Fig.~\ref{fig:dist_vs_MHI_and_MH2}, we show the HI and H$_2$ gas masses and the gas-depletion timescale as a function of distance to the filament (left column) and distance to Virgo (right column) for the 245 filament galaxies in our sample.
The HI gas mass increases with increasing distance to the filaments ($p$-value~=~$7.88\times10^{-4}$, { i.e., 3.4-$\sigma$}) and to the cluster core ($p$-value=$1.56\times10^{-4}$, { i.e., 3.8-$\sigma$}).\footnote{The $p$-values reported in Sects~\ref{sec:gas_content_tdep_environment} and \ref{sec:gasdef_density} are those of the Spearman's test for{ the null hypothesis (no correlation)}.  A lower significance{ is found} when LTGs and ETGs are considered separately.  This is because of the poorer statistics of the two individual subsamples.}

In contrast, the $M_{\rm H_2}$ gas mass shows no trend with distance to either the filaments or Virgo.  Finally, the depletion time $\tau_{\rm dep}$ shows no correlation with distance from the Virgo cluster, but a tentative increase with increasing distance from the filament spine is observed by means of the increasing $\tau_{\rm dep}$ binned median with $d_{\rm fil}$ in Fig.~\ref{fig:dist_vs_MHI_and_MH2}.

In Fig.~\ref{fig:density_vs_MHI_MH2_tdep_comparison} we show the gas mass, molecular-to-atomic
gas ratio, and depletion time as a function of local density{ for our sample of filament galaxies (center column), AMIGA sources (left column), and Virgo cluster galaxies (right column).}  
For the filament galaxies, we see that the HI content decreases with increasing density ($p$-value~$=5.25\times10^{-5}$, { i.e., 4.0$\sigma$}). The H$_2$ mass decreases with local density as well, but less strongly ($p$-value=0.017, { i.e., 2.4$\sigma$}) than $M_{\rm HI}$. For both HI and H$_2$ masses, the decrease is also more clear above a threshold density of $\sim0.2~h^{3}$~Mpc$^{-3}$, { which corresponds to the median density of the field sources in the AMIGA comparison sample.} The decrease becomes extreme at the highest densities $\gtrsim10~h^{3}$~Mpc$^{-3}$ in filaments and in Virgo cluster, where the fraction of sources with only gas mass upper limits significantly increases.

As seen in the top row{ of Fig.~\ref{fig:density_vs_MHI_MH2_tdep_comparison} and illustrated by the color coding,} ETGs in all environments have the lowest HI masses,{ which are roughly} an order of magnitude lower than those of LTGs.  In filaments, the early-type and late-type filament galaxies have indeed a median HI mass of $M_{\rm HI}=(1.38^{+6.71}_{-0.73})10^8~M_\odot$ and  $M_{\rm HI}=(1.89^{+4.34}_{-1.49})10^9~M_\odot$, respectively. 
Moreover, moving from the status of an isolated system to a cluster member, the HI content of the ETGs drops dramatically.  
Inspection of the upper range of HI masses shows that the fraction of galaxies with the highest values of $M_{\rm HI}\gtrsim 10^{10}~M_\odot$ (these are all LTGs) depends on the global environment, steadily decreasing from the AMIGA sample of isolated galaxies, to that of filaments, and then to that of the Virgo cluster. This suggest that these late-type galaxies indeed experience a certain degree of HI processing already in filaments. 

Early- and late-type galaxies can also be distinguished in terms of their molecular gas content, as seen in the
second row of Figure~\ref{fig:density_vs_MHI_MH2_tdep_comparison}. For example, in filaments, ETGs never reach $M_{\rm H_2}$ values as high as those of LTGs and have a low median value of $M_{\rm H_2}=(0.08^{+0.13}_{-0.05})10^9~M_\odot$.
However, the median does not tell the full story, as a large number of the LTGs have as low a reservoir of molecular gas as the ETGs, and this is true in all three environments.
In the filaments, these LTGs with a low molecular reservoir ($M_{\rm H_2}\lesssim10^8~M_\odot$) have, on average, normal HI content, with a median value of  $M_{\rm HI}=(0.94^{+1.80}_{-0.69})10^9~M_\odot$.
They therefore belong to the subsample of LTGs with low $M_{\rm H_2}/M_{\rm HI}$ (already seen in Fig.~\ref{fig:scatter_plot_MHI_MH2_vs_morphology}) and a relatively high median morphological index of $T=6.0^{+2.6}_{-2.8}$; they are for the most part only seen in moderately dense regions in filaments ($n_5=(0.77^{+4.26}_{-0.65})~h^3~{\rm Mpc}^{-3}$).

The third and fourth rows of Fig.~\ref{fig:density_vs_MHI_MH2_tdep_comparison} show the ratio of molecular-to-atomic gas mass and the depletion time, respectively.
As $M_{\rm HI}$ depends on the local density to a greater extent    than $M_{\rm H_2}$, the $M_{\rm H_2}/M_{\rm HI}$ ratio  in filaments increases with increasing local density. The dispersion  in the ratio also increases significantly as the environmental density increases.
However, the mean value of $M_{\rm H_2}/M_{\rm HI}$ does not vary considerably from isolation to the cluster core.

The dichotomy between the two broad classes of galaxy morphologies is not seen in the depletion timescales:  both ETGs and LTGs in filaments scatter around the median value of $\tau_{\rm dep}=(1.08^{+2.05}_{-0.72})$~Gyr. At the highest local densities  $\gtrsim10~h^{3}$~Mpc$^{-3}$ in filaments and Virgo cluster, the depletion times are strongly suppressed, with
about $\sim40\%$ and $\sim70\%$ of the sources, respectively,  only having upper limits in $M_{\rm H_2}$ and $\tau_{\rm dep}$.
This suggests a rapid environment-driven exhaustion of the H$_2$ reservoirs in these dense regions.

Taken together, the results from Fig.~\ref{fig:dist_vs_MHI_and_MH2} and Fig.~\ref{fig:density_vs_MHI_MH2_tdep_comparison} show that a galaxy's gas content varies with location within the cosmic web (isolation, filament, cluster). While local density plays a stronger role than distance to the filament spines or to cluster core, it appears to be a secondary factor, contributing to amplify the gas depletion. The impact of the filaments is clear, leading to a rise of the fraction of ETGs and the removal of HI gas for even the most gaseous systems, { which imply some similarities} with the galaxy population that is commonly witnessed in clusters.

The HI gas is essentially distributed in the outer regions of galaxies, and is therefore easily stripped { \citep[e.g.,][]{Gavazzi2018,Kenney2004,Yoon2017}}, while the dense and molecular gas is shielded in the inner parts of galaxies. As a consequence, H$_2$ is therefore less impacted by the environment, except for the densest regions in filaments and the Virgo cluster { \citep[e.g.,][]{Verdugo2015,Lee2017}}. The clear emergence of ETGs in filament environments, with increasingly low gas content and low depletion times, is a striking feature of Fig. \ref{fig:density_vs_MHI_MH2_tdep_comparison}, as we move from the field, to filaments, and then to Virgo. This raises the question of the timescale and relative contribution of the different physical mechanisms that can play a role, namely gas exhaustion during mergers and consecutive change in morphologies or lack of gas supply (e.g., starvation). In particular, for ETGs in filaments that show low gas masses, the HI stripping likely cuts off the supply of cold gas. This then favors the suppression of star formation and quenching, possibly via starvation or strangulation in group environments \citep{Kawata_Mulchaey2008}.
Some of the cluster members may experience ---{in situ}--- the same transformations as in the filaments, but a fraction of the infalling population is certainly pre-processed in the filaments before reaching the cluster core.

\begin{table*}
\centering
\hspace{-0.6cm}
\begin{tabular}{lccc}
{ HI-def.}  & { All galaxies} & { LTGs} & { ETGs} \\ 
  &  &  &   \\ 
\hline
  &  &  &   \\ 
AMIGA & $(5.7\pm1.7)\%$ & $(3.4\pm1.4)\%$ & $(31.2\pm11.6)\%$ \\
      &   11/193 & 6/177& 5/16 \\
  &  &  &   \\ 
Filaments & $(23.3\pm2.7)\%$ & $(11.7\pm2.4)\%$ & $(55.4\pm6.2)\%$ \\
      &   57/245 & 21/180 & 36/65 \\
  &  &  &   \\ 
Virgo Cluster& $(47.7\pm4.8)\%$ & $(30.4\pm5.2)\%$ & $(93.3\pm4.6)\%$ \\
      &   52/109 & 24/79  & 28/30 \\
\hline\hline
&    & & \\
 { H$_2$-def.}  & { All galaxies} & { LTGs} & { ETGs} \\   &  &  &   \\ 
\hline
  &  &  &   \\ 
AMIGA & $(18.7\pm2.8)\%$ & $(15.3\pm2.7)\%$ & $(56.2\pm12.4)\%$ \\
      &   36/193 & 27/177& 9/16 \\
  &  &  &   \\ 
Filaments & $(41.6\pm3.1)\%$ & $(26.7\pm3.3)\%$ & $(83.1\pm4.1)\%$ \\
  & 102/245 & 48/180 & 54/65  \\ 
  &  &  &   \\ 
Virgo Cluster& $(37.6\pm4.6)\%$ & $(19.0\pm4.4)\%$ & $(86.7\pm6.2)\%$ \\
&   41/109 & 15/79  & 26/30 \\
\hline\hline
&    & & \\
 { HI-def. \& H$_2$-def.}  & { All galaxies} & { LTGs} & { ETGs} \\   &  &  &   \\ 
\hline
  &  &  &   \\ 
AMIGA & $(4.1\pm1.4)\%$ & $(2.3\pm1.1)\%$ & $(25.0\pm10.8)\%$ \\
      &   8/193 & 4/177& 4/16 \\
  &  &  &   \\ 
Filaments & $(15.5\pm2.3)\%$ & $(2.2\pm1.1)\%$ & $(52.3\pm6.2)\%$ \\
  & 38/245 & 4/180 & 34/65  \\ 
  &  &  &   \\ 
Virgo Cluster & $(24.8\pm4.1)\%$ & $(3.8\pm2.2)\%$ & $(80.0\pm7.3)\%$ \\
&   27/109 & 3/79  & 24/30 \\
\end{tabular}
\caption{Fractions of deficient galaxies in HI (top), H$_2$ (center), and simultaneously both in HI and H$_2$ (bottom).\label{tab:fraction_gas_def}}
\end{table*}

\subsubsection{Gas deficiency}\label{sec:gasdef_density}

The HI-deficiency parameter \HIdef\  is defined as the logarithmic difference between{ the average HI mass of a reference sample of  isolated  galaxies with the same morphological type and size as the galaxy in question and the observed HI mass of that galaxy:
\begin{equation}
{\rm HI-def.}~=~\log(M_{\rm HI,\,ref})-\log(M_{\rm HI})\,,
\end{equation}
{ where, according to \citet{Haynes_Giovanelli1984},}
\begin{equation}
\log\bigg(\frac{h^2~M_{\rm HI,\,ref}}{M_\odot}\bigg)= c+2d\,\log\bigg(\frac{h~D_{25}}{\rm kpc}\bigg)\,.
\end{equation}
{ In this latter formula, we adopt $h=0.74$ and we use the diameter $D_{25}$ as found in HyperLeda (Table~\ref{tab:general_properties_all_galaxies}). The parameters $c\sim6.9-7.8$ and $d\sim0.6-0.9$ depend instead on the morphology and we adopt the values reported in Table~3 of \citet{Boselli_Gavazzi2009}.}

{ Similarly, we use the definition of the H$_2$-deficiency parameter (\Htwodef) by \citet{Boselli2014b}, as a function of the stellar mass,  calibrated to $\alpha_{\rm CO}=4.3~M_\odot$~(K~km~s$^{-1}$~pc$^{2}$)$^{-1}$:}
\begin{equation}
{\rm H_2-def.}~=~\log(M_{\rm H_2,ref})-\log(M_{\rm H_2})\,,
\end{equation}
{ where}
\begin{equation}
\log(M_{\rm H_2,ref})=0.81\log(M_\star/M_\odot)-0.79\,.
\end{equation}
Galaxies with \HIdef$>0.5$ { are} considered as deficient in HI, and { sources} with \Htwodef$>0.5$ are considered as deficient in H$_2$.

Figure~\ref{fig:HI_H2_def_plots_comparison} presents the relation between  \HIdef\  and \Htwodef\ and the local density for the isolated, filament, and cluster galaxies.
The uncertainties on the gas deficiency parameters  reported in the figure were estimated{ taking only the errors on the fluxes into account, but not those on the distances or those on the deficiency parameters. For the galaxies with upper limits in HI or H$_2$, we report lower limits in the corresponding gas deficiency parameters.} 
In both the filament and cluster environments, both the number of HI and H$_2$-deficient galaxies and the median deficiency parameters increase with increasing local density. In addition, the number of both HI- and H$_2$-deficient galaxies at a fixed local density increases going from isolated, to filament, to cluster environments. On average, AMIGA galaxies have a normal gas content, with median \HIdef, \Htwodef$<0.5$, while the average values steadily increase with density in the filaments and reach the highest values $\gtrsim2$ in the Virgo cluster.

For Virgo filament galaxies, \HIdef\ and \Htwodef\ correlate well ($p$-value$\simeq(5-8)\times10^{-5}$, { i.e., ${4\sigma}$)} with the local density. This correlation is driven by the rise in the number of gas-deficient galaxies in denser environments mainly composed of ETGs.
Table~\ref{tab:fraction_gas_def} reports the fractions of gas-deficient galaxies. There is a strong dichotomy in the gas deficiencies between LTGs and ETGs, similarly to what has been discussed in Sect.~\ref{sec:gas_content_tdep_environment}. ETGs form the majority of the gas-deficient galaxies in any cosmic environment; they  indeed represent $\sim90\%$ of the sources that are simultaneously
deficient in both HI and H$_2$. 

\HIdef\ rises with decreasing  distance to the filament ($p$-value$~=1.59\times10^{-3}$, { i.e., 3.2$\sigma$)} and to the Virgo cluster core (p-value$~=7.65\times10^{-4}$, { i.e., 3.4$\sigma$)}, unlike for \Htwodef\  which is insensitive to these parameters. This echoes what is found for the HI and H$_2$ gas masses in Fig.~\ref{fig:dist_vs_MHI_and_MH2}.  No similar study exists on the H$_2$ content of galaxies in filaments. 

{ Some studies have instead investigated the HI content of galaxies in the filamentary structures in the local Universe. Interestingly,} \citet{Lee2021} did not detect the same correlation between distance and HI deficiency as highlighted here, but our results on \HIdef\ are in agreement with those of \citet{CroneOdekon2018} based on data from the ALFALFA HI survey, who found that the \HIdef\ of their galaxies{ decreases} with{ increasing} distance from the filament spine, and suggest that galaxies{ in dense regions} are cut off from their supply of cold gas. 



\section{Summary and Conclusions}\label{sec:summary_conclusions}


We present the first large observational effort to gather the gas status ---in both molecular and atomic gas phases--- of large-scale structures linked to a central galaxy cluster. This  comprehensive study was undertaken in order to evaluate the impact of cosmological filaments in processing of the cold gas reservoirs of galaxies as they move within the cosmic web and before they fall into the cluster itself. 
To this end, we built a homogeneous sample of 245 galaxies with stellar masses $\log(M_\star/M_\odot)\sim9-11$  located in major filaments surrounding Virgo. Stellar masses and SFR estimates were gathered from the literature. H$_2$ and HI masses were estimated thanks to our CO and HI campaigns at the IRAM-30m and Nan\c{c}ay telescopes, respectively, or were taken from previous published observations. 

Environmental parameters such as alignments with respect to the filament spines, local densities, and distances from the filament spines and from Virgo cluster were calculated following { a rigorous}  3D characterization of the cosmic web around Virgo.
We compared the properties of these filament galaxies with those of two samples of galaxies of similar sizes: (i) isolated field galaxies (AMIGA) and (ii) galaxies belonging to the Virgo cluster itself. Stellar and gas masses, SFRs, depletion times, quiescent fractions, and environmental properties have been considered for all three samples, with a homogeneous treatment of the data.

\subsection{Summary of the results}


\begin{itemize}
 \item[$\bullet$] The filamentary structures around Virgo contain a large number of groups, from poor groups to rich ones. As such, they exhibit a broad range of local densities. { The filaments} are fairly well described by exponentially decaying profiles (Sect.~\ref{sec:density_profiles}, { Fig.~\ref{fig:density_profiles}}).

\smallskip\smallskip

\item[$\bullet$] { Our} filament galaxy sample with CO and HI observations is primarily composed of spiral galaxies, { similarly to the comparison samples of AMIGA isolated field galaxies and Virgo cluster sources.} The large majority of these LTGs fall  on the MS of star forming galaxies. ETGs appear in significant numbers in the filaments, with the majority in the quenching phase,  while they are hardly present in isolation  (Sects.~\ref{sec:main_sequence_and_morphology}, { Fig.~\ref{fig:SFR_vs_Mstar}}).

 \smallskip\smallskip

\item[$\bullet$] The fraction of galaxies with suppressed star formation ($f_Q$), that is, well below the MS, monotonically increases from the field,  to filaments, and to the Virgo cluster.  At comparable local density, filaments have similar quenching fractions to the field and Virgo cluster (Sect.~\ref{sec:fQ}, { Fig.~\ref{fig:scatter_plot_fQ_vs_density_dfil} left}).

\smallskip\smallskip

\item[$\bullet$] The $f_Q$ fraction  significantly differs between ETGs and LTGs. { For ETGs, it is} high, $f_Q\gtrsim80\%$, at all distances from the filament spine. For LTGs, it reaches $f_Q\sim$50\% close to the filament spines, but is otherwise of the order of { $\sim$20\%} (Sect.~\ref{sec:fQ}, { Fig.~\ref{fig:scatter_plot_fQ_vs_density_dfil} right}).


\smallskip\smallskip

 \item[$\bullet$] { ETGs not only} tend to populate the densest regions within filaments, but they are also preferentially located closer to the filament spines than LTGs (Sect.~\ref{sec:result_local_densities_distances}, { Fig.~\ref{fig:histo_density_dfil}}).\smallskip\smallskip
 

 \smallskip\smallskip

 \item[$\bullet$] The atomic and molecular gas fractions, $M_{\rm HI}/M_\star$ and $M_{\rm H_2}/M_\star$, increase from early-type to late-type morphologies, with a slope that steepens when passing from isolation, to filaments, and then to Virgo.  This is a consequence of the fact that while LTGs have very similar gas fractions in all cosmic structures, ETGs are more strongly gas-depleted in very dense regions such as clusters.  This effect is stronger for HI than for H$_2$ (Sects.~\ref{sec:result_gas_fraction_vs_morphology}, \ref{sec:result_gas_ratio_vs_morphology}, { Fig.~\ref{fig:scatter_plot_MHI_MH2_vs_morphology}}).
 
 \smallskip\smallskip

 


\smallskip\smallskip
 \item[$\bullet$]  In isolation and in filaments, the  average depletion timescale increases with stellar mass as a consequence of massive galaxies having larger H$_2$ gas reservoirs.  The scatter in the relation between depletion timescale and galaxy stellar mass increases steadily from the field, to the filaments, and then to Virgo, up to a point where  the relation between $\tau_{\rm dep}$ and $M_\star$ breaks down. { This suggests that gas depletion precedes star-formation quenching (Sect.~\ref{sec:depletion_time}, Fig.~\ref{fig:scatter_plot_tdep}).}
 
\smallskip\smallskip


 \item[$\bullet$] Both the H$_2$ and HI  mass of the galaxies in filaments decrease with increasing local density, the latter more steeply than the former.
 The HI mass also increases  with increasing distance from the filament spine ($d_{\rm fil}$) and Virgo cluster ($d_{\rm cluster}$). The average depletion time tentatively increases with $d_{\rm fil}$.  In contrast, the galaxy H$_2$ mass shows no clear trend as a function of distance (Sect.~\ref{sec:gas_content_tdep_environment}, { Figs.~\ref{fig:dist_vs_MHI_and_MH2}, \ref{fig:density_vs_MHI_MH2_tdep_comparison}}). 
 \smallskip\smallskip
 
  \item[$\bullet$] The number of LTGs  with very large HI reservoirs ($M_{\rm HI}\gtrsim 10^{10}~M_\odot$) steadily decreases { as a function of the global cosmic structure in which galaxies reside, i.e., from isolation, to filaments, and then to the Virgo cluster.} 
 Filament ETGs have average HI and H$_2$ reservoirs of $\sim10^8~M_\odot$ and never reach the highest values found for LTGs, in all three considered environments (Sect.~\ref{sec:gas_content_tdep_environment}, { Fig.~\ref{fig:density_vs_MHI_MH2_tdep_comparison}}). 
 \smallskip\smallskip

 \item[$\bullet$] The average \HIdef\  and \Htwodef\  parameters increase with increasing local density. Also, HI-deficiency is anti-correlated with distances $d_{\rm fil}$ and $d_{\rm cluster}$. The fraction of gas-deficient galaxies, which are mostly ETGs, as well as the average deficiency parameters, both in HI and in H$_2$, significantly increase from the field, to filaments, to the Virgo cluster (Sect.~\ref{sec:gasdef_density}, { Fig.~\ref{fig:HI_H2_def_plots_comparison}}).
 
\end{itemize}

\subsection{Conclusions}

Our survey highlights the importance of cosmic filaments in modifying galaxy properties.
Our study reveals indeed that the specific environment in which galaxies are located (field, filaments, cluster) acts as the primary driver of galaxy transformation. The local density of the cosmic web, distance to the filament spines, and distance to the cluster center, in that order, are secondary parameters, but significant dependencies of galaxy properties on these environmental parameters are nevertheless found.
Many of the gas deficiencies and the changes in morphological composition of the galaxy population that are  classically attributed to galaxy clusters  are already advanced in the medium-density environments { associated with the} filaments. Some of the properties of { cluster galaxies}  can therefore be acquired in the clusters themselves, the rest being the consequence of the infall of the filament galaxies onto the cluster cores.

{ As regards to the physical processes at play in the environments considered here, this work is unable to provide definitive answers, but} mergers, stripping, tidal interactions, and starvation are the most commonly invoked mechanisms. 
Overall, going from the field, to filaments, and then to the Virgo cluster, we observe the emergence of ETGs in particular, but also LTGs, with increasing levels of gas deficiency, both in HI and H$_2$. Indeed, a large fraction, 121/245 (i.e., $49\%\pm6\%$), of filament galaxies are deficient either in HI or H$_2$, or both. As these gas-deficient sources are observed preferentially in dense regions within filaments in groups, their gas reservoirs have likely experienced strong environment-driven pre-processing.


The HI envelopes of HI-deficient galaxies in filaments have likely been stripped, possibly via tidal interactions or ram pressure. 
For HI, our proposed scenario is in agreement with the findings of \citet{Dzudzar2021}, who investigated  environmental processing in late-type-dominated groups using high-resolution HI observations. 
These latter authors  further discussed the possibility that groups with the highest levels of processing are transitioning towards Hickson compact or fossil groups,  which are environments similar to those associated with the isolated ellipticals in our AMIGA comparison sample. Filament galaxies around Virgo that are HI deficient and live in the highest density regions within filaments may be experiencing a similar transition. 

Our scenario is also in line with previous studies  \citep[e.g., on the HIPASS survey,][]{Denes2014,Reynolds2020}, where it was found that HI sources living in denser environments show, on average, asymmetries and higher \HIdef~ than those in less dense environments. However, \citet{Reynolds2020} also find groups and clusters that are not HI poor, confirming the large dispersion between \HIdef\  and  local density that we report in the present study.


While the densest regions in the filaments are able to effectively remove or deplete the HI envelope of galaxies via stripping,{ cosmic starvation in HI \citep[i.e., the reduction of the gas supply from the cosmic web;][]{Feldmann_Mayer2015} is a less likely mechanism.} Indeed, filaments are quite rich in hot gas and baryons, as shown by both simulations  and observations \citep{Eckert2015,Martizzi2019,Libeskind2020,Tanimura2020}. 
Gas accretion from filaments \citep[][]{Bournaud2005}, ultimately feeding star formation, or hydrodynamical interactions with  the  intergalactic  medium \citep{Watts2020} could also be responsible for gas asymmetries, which we may ultimately observe in terms of different levels of gas deficiencies.

Concerning H$_2$, {exhaustion of the molecular gas reservoirs and inefficient HI-to-H$_2$ conversion} may explain the low molecular gas content associated with a fraction of our filament galaxies, in particular those LTGs in filaments{ (discussed in Sect.~\ref{sec:gas_content_tdep_environment})} with low H$_2$ gas reservoirs{ and normal HI content. As they live in relatively low-density environments of $\sim0.8~h^{3}$~Mpc$^{-3}$, their H$_2$ gas reservoirs may not be effectively replenished.} Starvation (e.g., strangulation in groups) or H$_2$ exhaustion induced by past mergers are possible scenarios to explain the H$_2$ gas deficiencies observed in the large fraction ($\sim84\%$) of filament ETGs, { which} preferentially live in denser regions than LTGs.
While ram pressure or tidal stripping in H$_2$ is still a viable mechanism to explain the H$_2$ gas deficiencies, it may be a less likely mechanism.
{ Molecular} gas reservoirs {are} less extended than the HI envelopes and are associated with higher gas densities. H$_2$ is therefore more difficult to strip than HI. Furthermore,
H$_2$ stripping would require higher densities and infall velocities more typical of galaxies in clusters than in filaments \citep[e.g.,][]{Jachym2014,Jachym2019}. Higher angular resolution observations in CO would provide further insight into the physical processes responsible for the processing of the cold gas reservoirs in filament galaxies.

\begin{acknowledgements}
{ We thank the anonymous referee for helpful comments which contributed to improve the paper significantly.}
This work is based on observations carried out with the Nan\c{c}ay  decimetric  radio telescope and the IRAM 30m telescope. IRAM is supported by INSU/CNRS (France), MPG (Germany) and IGN (Spain). { The Nan\c{c}ay radio Observatory is operated by the Paris Observatory, associated with the French Centre National de la Recherche Scientifique (CNRS) and with the University of Orl\'{e}ans. }
GC acknowledges financial support from the Swiss National Science Foundation (SNSF) and fruitful discussion with Amelie Saintonge concerning the CO aperture correction. The authors acknowledge  Mindy Townsend, Dara Norman, and Kim Conger for helpful discussion.  GHR acknowledges funding support from NSF AST-1716690.  BV acknowledges financial contribution from the grant PRIN MIUR 2017 n.20173ML3WW\_001 (PI Cimatti) and from the INAF main-stream funding program (PI Vulcani).
RAF gratefully acknowledges support from NSF grants AST-0847430 and AST-1716657.  The authors thank the hospitality of International Space Science Institute (ISSI) in Bern (Switzerland) and of the Lorentz Center in Leiden (Netherlands). Regular group meetings in these institutes allowed the authors to make substantial progress on the project and finalize the present work. We acknowledge the usage of the NASA Sloan Atlas, HyperLeda, NASA/IPAC Extragalactic, and DustPedia databases.
\end{acknowledgements}


\newpage

\label{lastpage}
\onecolumn

\begin{appendix}

\FloatBarrier

\section{Properties of Virgo filament galaxies}\label{sec:tables}
In this section we provide several tables with the properties of our sample of filament galaxies.

\onecolumn{
\LTcapwidth=0.8\textwidth
\captionsetup[longtable]{margin=0.6in}
\setlength{\LTleft}{-1.cm}
\setlength\tabcolsep{1.6pt} 
\begin{tiny}
\begin{longtable}{cccccccccccccccccc} 
Galaxy &   R.A.     & Dec.     &   SG (X;Y;Z) & Filament & $d_{\rm fil}$ & $d_{\rm cluster}$ & $\theta_{\rm alignment}$ & Group &  Richness  & $\log\big(\frac{M_{\rm group}}{M_\odot})$ &  $\log\big(\frac{n_5^{\rm 2D}}{h^2~{\rm Mpc}^{-2}}\big)$ & $\log\big(\frac{n_5^{\rm 3D}}{h^3~{\rm Mpc}^{-3}}\big)$  \\ 
    & (hh:mm:ss.s) & (dd:mm:ss.s)  &   ($h^{-1}$~Mpc)  & & ($h^{-1}$~Mpc)  & ($h^{-1}$~Mpc)  & (deg) & ID  &    & (Mpc) & &  &   &  \\
(1) & (2) & (3) & (4) & (5) & (6) & (7) & (8) & (9) & (10) & (11) & (12)  & (13) \\
\hline
PGC022100 & 07:53:45.27 & +21:02:57.90 & (9.48;10.63;-17.29) & LeoII A & 3.64 (3D) & 20.57 (3D) & 13 & 22100 & 1 & 11.46 & $0.68^{+0.16}_{-0.26}$ & $-0.52^{+0.16}_{-0.26}$ \\ 
  &   &   &   &   & 3.04 (2D)  & 20.55 (2D)  &   &   &   &   &   &   \\ 
  &   &   &   &   &   &   &   &   &  \\ 
IC2256 & 08:16:54.45 & +24:10:35.76 & (10.93;14.76;-18.61) & LeoII A & 5.14 (3D) & 23.00 (3D) & 38 & 23214 & 1 & 11.16 & $0.50^{+0.16}_{-0.26}$ & $-0.52^{+0.16}_{-0.26}$ \\ 
  &   &   &   &   & 4.69 (2D)  & 22.48 (2D)  &   &   &   &   &   &   \\
\end{longtable}
\end{tiny}}


\onecolumn{
\begin{tiny}
\LTcapwidth=0.8\textwidth
\captionsetup[longtable]{margin=0.6in}
\setlength{\LTleft}{-0.6cm}
\setlength\tabcolsep{5pt}
\begin{longtable}{ccccccccccccccc} 
\hline
Galaxy & R.A.          & Dec.         & $V_{\rm H}$  & $V_{\rm cosmic}$ & $\log(D_{25}/0.1^\prime)$ & $i$    & PA &$\log(M_\star/M_\odot)$ & $\log(\frac{\rm SFR}{M_\odot/{\rm yr}})$  & $\log(\frac{\rm SFR}{\rm SFR_{\rm MS}})$ & Morphology   \\
       & (hh:mm:ss.s)  & (dd:mm:ss.s) & (km/s) & (km/s)    &  & (degree) & (degree) & & & &  \\
(1) & (2) & (3) & (4) & (5) & (6) & (7) & (8) & (9) & (10) & (11) & (12) \\
\hline
  &   &   &   &   &   &   &   &   &   &  \\ 
PGC022100 & 07:53:45.27 & +21:02:57.90 & 2106 & 2240 & $0.92\pm0.05$ & 52.8 & 33 & $9.27\pm0.10$ & $-1.02\pm0.20$ & -0.35 & SABb (2.9) \\ 
  &   &   &   &   &   &   &   &   &   &  \\ 
IC2256 & 08:16:54.45 & +24:10:35.76 & 2208 & 2615$^{a}$ & $0.90\pm0.06$ & 61.4 & 8 & $9.27\pm0.10$ & $-1.09\pm0.20$ & -0.42 & SABc (6.3) \\ 
  &   &   &   &   &   &   &   &   &   &  \\
\end{longtable}
\end{tiny}}


\onecolumn{
\begin{tiny}
\LTcapwidth=0.8\textwidth
\captionsetup[longtable]{margin=0.6in}
\setlength{\LTleft}{-0.2cm}
\setlength\tabcolsep{5pt} 
\begin{longtable}{cccccccccc} 
\hline
 Galaxy &  Distance & $\log(D_{25}/0.1^\prime)$ & CO(${\rm J}\rightarrow {\rm J-1}$) &   $f_{ap}$ &  $\log\Big(\frac{L'_{\rm CO(J\rightarrow J-1)}}{{\rm K~km~s}^{-1}~{\rm pc}^2}\Big)$ & $r_{J1}$ & $\log\Big(\frac{M_{\rm H_2}}{M_\odot}\Big)$ & Telescope  & Reference \\ 
       & (Mpc)    &  & &   & &  &  & & \\
(1) & (2) & (3) & (4) & (5) & (6) & (7) & (8) & (9) & (10)  \\
\hline
PGC022100 & 30.27 & $0.92\pm0.05$ & $1\rightarrow 0$ & 0.66 & $<7.07$ & 1.0 & $<7.71$ & IRAM 30m & \citet{Lisenfeld2011} \\ 
  &   &   &   &   &   &   &   &   &  \\ 
NGC2577 & 35.62$^{a}$ & $1.19\pm0.04$ & $1\rightarrow 0$ & 0.52 & $<7.42$ & 1.0 & $<7.93$ & IRAM 30m & \citet{Young2011} \\ 
  &   &   & $2\rightarrow 1$ & 0.28 & $<7.20$ & --- &   & IRAM 30m & \citet{Young2011} \\ 
  &   &   &   &   &   &   &   &   &  \\
\end{longtable}
\end{tiny}}


\onecolumn{
\begin{small}
\LTcapwidth=0.8\textwidth
\captionsetup[longtable]{margin=0.6in}
\setlength{\LTleft}{0cm}
\begin{longtable}{cccccccccc} 
\hline
 Galaxy &  Distance & $\log(D_{25}/0.1^\prime)$ & CO(${\rm J}\rightarrow {\rm J-1}$) & $S_{\rm CO(J\rightarrow J-1)}\Delta\varv$  &  $f_{ap}$ & FWHM & $\log\Big(\frac{L'_{\rm CO(J\rightarrow J-1)}}{{\rm K~km~s}^{-1}~{\rm pc}^2}\Big)$ & $r_{21}$ & $\log\Big(\frac{M_{\rm H_2}}{M_\odot}\Big)$ \\ 
       & (Mpc)    &  & &  (Jy~km~s$^{-1}$) & & (km/s) &  &  \\
(1) & (2) & (3) & (4) & (5) & (6) & (7) & (8) & (9) & (10)   \\
\hline
IC2256 & 35.33$^{a}$ & $0.90\pm0.06$ & $1\rightarrow 0$ & $<10.02$ & $0.71$ & --- & $<7.63$ & --- & $<7.86$ \\ 
  &   &   & $2\rightarrow 1$ & $<7.80$ & $0.44$ & --- & $<7.13$ &  \\ 
  &   &   &   &   &   &   &   &   &  \\ 
UGC04375 & 40.34$^{a}$ & $1.23\pm0.04$ & $1\rightarrow 0$ & $14.79\pm1.86$ & $0.38$ & $156\pm21$ & $8.19^{+0.05}_{-0.06}$ & $0.30\pm0.07$ & $8.83^{+0.05}_{-0.06}$ \\ 
  &   &   & $2\rightarrow 1$ & $7.49\pm1.64$ & $0.16$ & $145\pm27$ & $7.67^{+0.09}_{-0.11}$ &  \\
\end{longtable}
\end{small}}


\onecolumn{
\begin{small}
\LTcapwidth=0.8\textwidth
\captionsetup[longtable]{margin=0.6in}
\setlength{\LTleft}{2cm}
\begin{longtable}{ccccccccccccc} 
\hline
 Galaxy &  Distance & $S_{\rm HI} \Delta\varv$ &  W$_{50}$ &   $\log\Big(\frac{M_{\rm HI}}{M_\odot}\Big)$ &   Reference \\
        & (Mpc)    & (Jy~km~s$^{-1}$) & (km/s) & &     \\
(1) & (2) & (3) & (4) & (5) & (6) \\
\hline
PGC022100 & 30.27 & $1.77\pm0.13$ & $150\pm11$ & $8.58\pm0.03$ & This work \\ 
  &   &   &   &  \\ 
IC2256 & 35.33$^{a}$ & $4.27\pm0.53$ & --- & $9.10^{+0.05}_{-0.06}$ & \citet{Springob2005} \\ 
  &   &   &   &  \\ 
NGC2577 & 35.62$^{a}$ & $<0.35$ & --- & $<8.02$ & \citet{Huchtmeier_Richter1989} \\
\end{longtable}
\end{small}}



\onecolumn{
\begin{tiny}
\LTcapwidth=0.8\textwidth
\captionsetup[longtable]{margin=0.6in}
\setlength{\LTleft}{-1cm}
\setlength\tabcolsep{5.5pt} 
\begin{longtable}{ccccccccccccccccc} 
\hline
 Galaxy &  Distance & $r_{21}$ & $\log\Big(\frac{M_{\rm H_2}}{M_\odot}\Big)$ &   $\log\Big(\frac{M_{\rm H_2}}{M_\star}\Big)$  &   $\log\Big(\frac{M_{\rm H_2}}{M_\star}\Big)_{\rm MS}$ &     $\log\Big(\frac{\tau_{\rm dep}}{\rm yr}\Big)$ & $\log\Big(\frac{\tau_{\rm dep, MS}}{\rm yr}\Big)$ & $\log\Big(\frac{M_{\rm HI}}{M_\odot}\Big)$ & $\log\Big(\frac{M_{\rm HI}}{M_\star}\Big)$ & $\log\Big(\frac{M_{\rm HI}}{M_\star}\Big)_{\rm MS}$ & Ref.  & Ref.  \\ 
       & (Mpc)    &  & &  & & & & & &   & CO &  HI  \\
(1) & (2) & (3) & (4) & (5) & (6) & (7) & (8) & (9) & (10) & (11) & (12) & (13)  \\
\hline
PGC022100 & 30.27 & --- & $<7.71$ & $<-1.56$ & $-0.99\pm0.53$ & $<8.73$ & $8.95\pm0.08$ & $8.58\pm0.03$ & $-0.69^{+0.09}_{-0.12}$ & $-0.37\pm0.25$ & 2 & 1 \\ 
  &   &   &   &   &   &   &   &   &  \\ 
IC2256 & 35.33$^{a}$ & --- & $<7.86$ & $<-1.41$ & $-0.99\pm0.53$ & $<8.95$ & $8.95\pm0.08$ & $9.10^{+0.05}_{-0.06}$ & $-0.17^{+0.10}_{-0.13}$ & $-0.37\pm0.25$ & 1 & 2 \\ 
  &   &   &   &   &   &   &   &   &  \\ 
NGC2577 & 35.62$^{a}$ & --- & $<7.93$ & $<-2.37$ & $-1.26\pm0.53$ & $<9.13$ & $9.13\pm0.05$ & $<8.02$ & $<-2.28$ & $-0.67\pm0.25$ & 2 & 2 \\
\end{longtable}
\end{tiny}}




\FloatBarrier


\section{CO and HI spectra}\label{sec:CO_HI_spectra}

In this section we show CO and HI spectra from our IRAM-30m and Nan\c{c}ay campaigns.


\begin{figure*}[!ht]\centering
\captionsetup[subfigure]{labelformat=empty}
\subfloat[]{\includegraphics[trim={3.cm 2.cm 3.cm 7.5cm},clip,page=1,scale=0.9,clip=true]{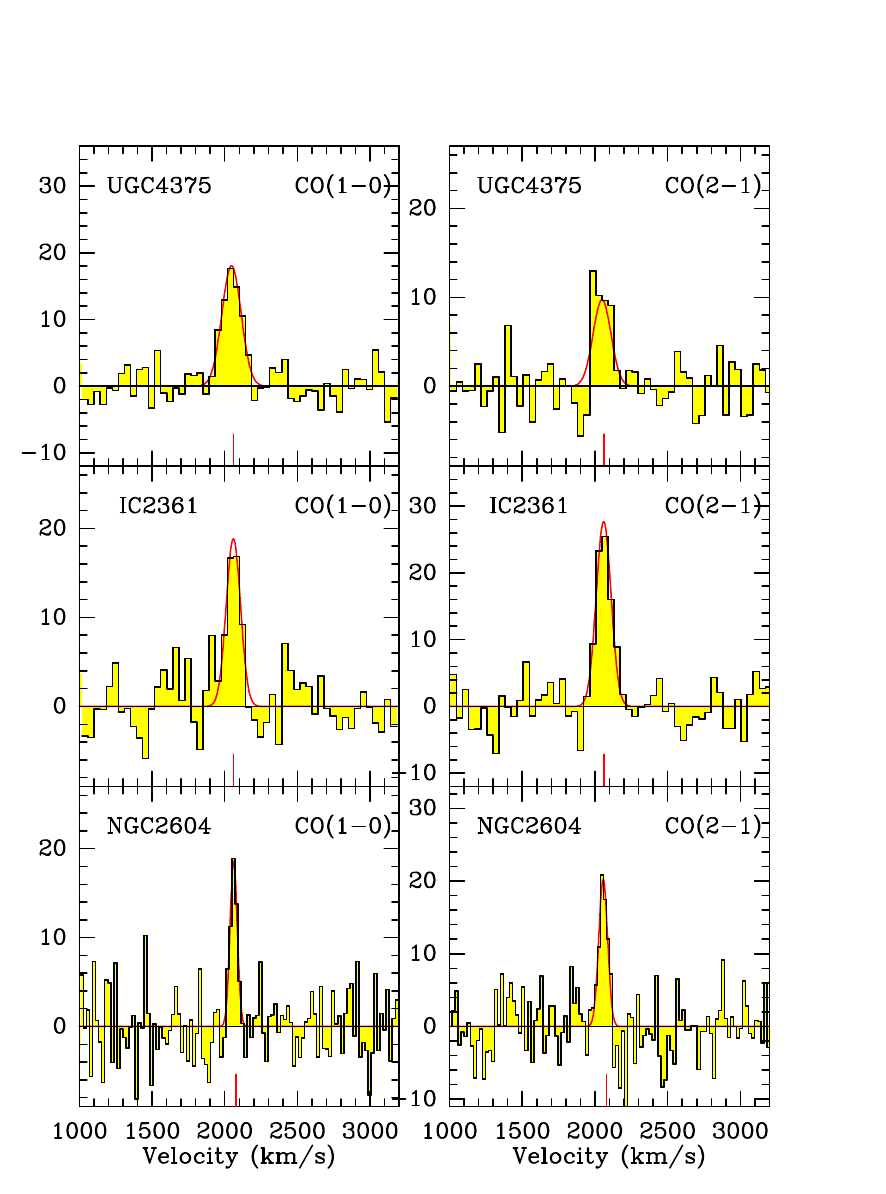}}
\caption{Baseline subtracted CO(1$\rightarrow$0) (left) and CO(2$\rightarrow$1) (right) spectra from our IRAM-30m campaign for the sources with secure or tentative detections in CO. For each spectrum, the x-axis displays the relative velocity, { where the vertical red segment shows the heliocentric velocity of the galaxy.} In the y-axis $T_{\rm mb}$ is shown in units of mK. Solid red curves show the single Gaussian fits to the CO(1$\rightarrow$0) and CO(2$\rightarrow$1) lines. { Three}  Gaussian components are adopted to fit the CO(1$\rightarrow$0) emission of NGC~5311. For the following sources two heliocentric velocities are shown with the vertical segments. The CO emission observed from NGC~2799 is consistent with the source recession velocity of $\sim$1870~km/s from the optical.  \citet{Monnier-Ragaigne2003} report instead a lower velocity of 1673~km/s inferred from{ their} HI spectrum taken at the Nançay telescope. However, this HI redshift might not be reliable because of confusion, as also noted by these latter authors. Both optical and radio velocities are nevertheless reported with vertical segments. Similarly, for PGC 214137, i.e., UGC 08656 NOTES01, we report both its recession velocity 2715~km/s and that 2865~km/s of its more massive companion UGC 08656.}\label{fig:30m_spectrum}
\end{figure*}
\begin{figure*} \centering
\captionsetup[subfigure]{labelformat=empty}
\ContinuedFloat
\subfloat[]{\includegraphics[trim={3.cm 2.cm 3.cm 2cm},clip,page=2,scale=0.9,clip=true]{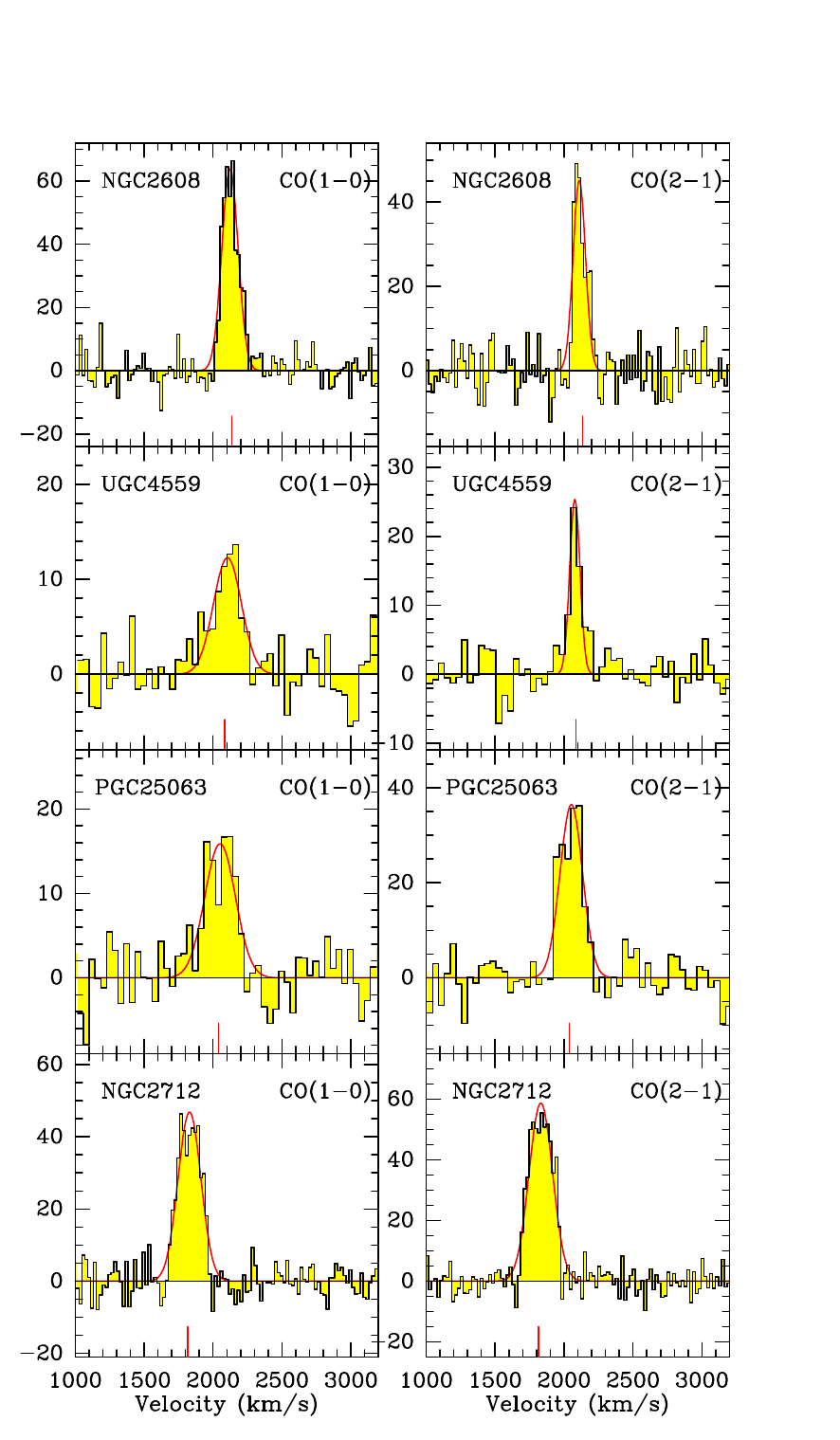}}
\caption{Continued.}
\end{figure*}
\begin{figure*} \centering
\captionsetup[subfigure]{labelformat=empty}
\ContinuedFloat
\subfloat[]{\includegraphics[trim={3.cm 2.cm 3.cm 2cm},clip,page=3,scale=0.9,clip=true]{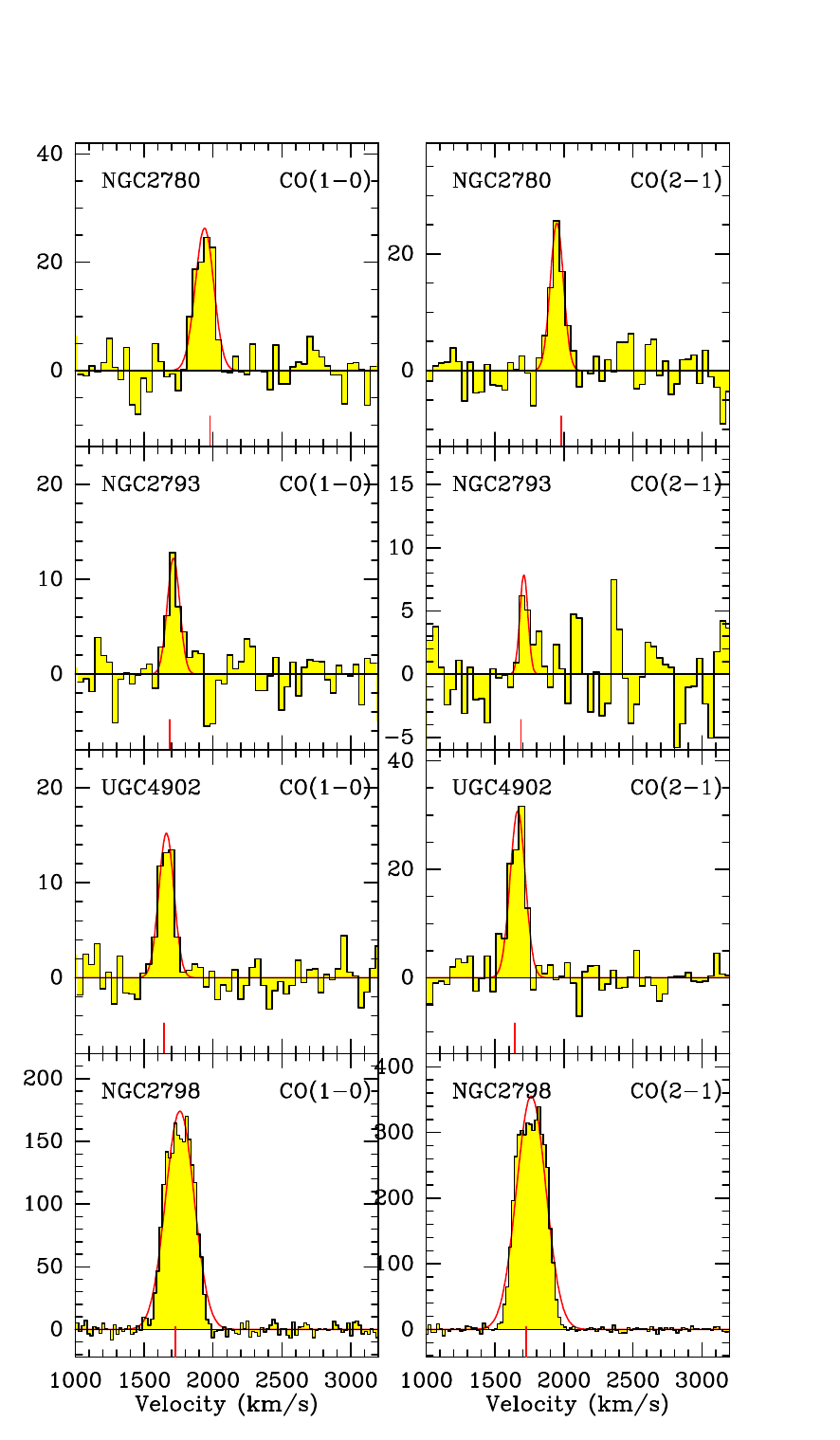}}
\caption{Continued.}
\end{figure*}
\begin{figure*} \centering
\captionsetup[subfigure]{labelformat=empty}
\ContinuedFloat
\subfloat[]{\includegraphics[trim={3.cm 2.cm 3.cm 2cm},clip,page=4,scale=0.9,clip=true]{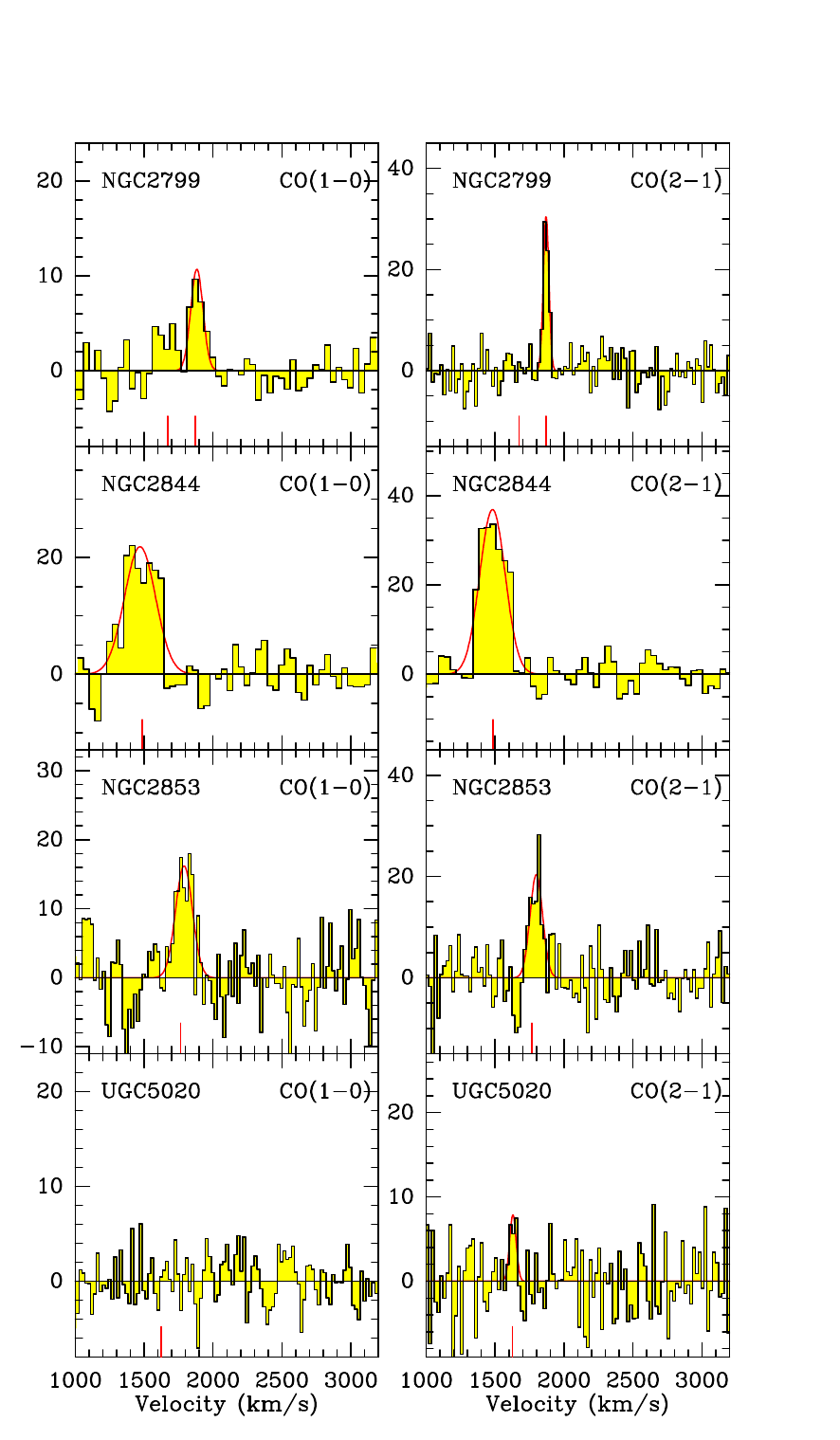}}
\caption{Continued.}
\end{figure*}
\begin{figure*} \centering
\captionsetup[subfigure]{labelformat=empty}
\ContinuedFloat
\subfloat[]{\includegraphics[trim={3.cm 2.cm 3.cm 2cm},clip,page=5,scale=0.9,clip=true]{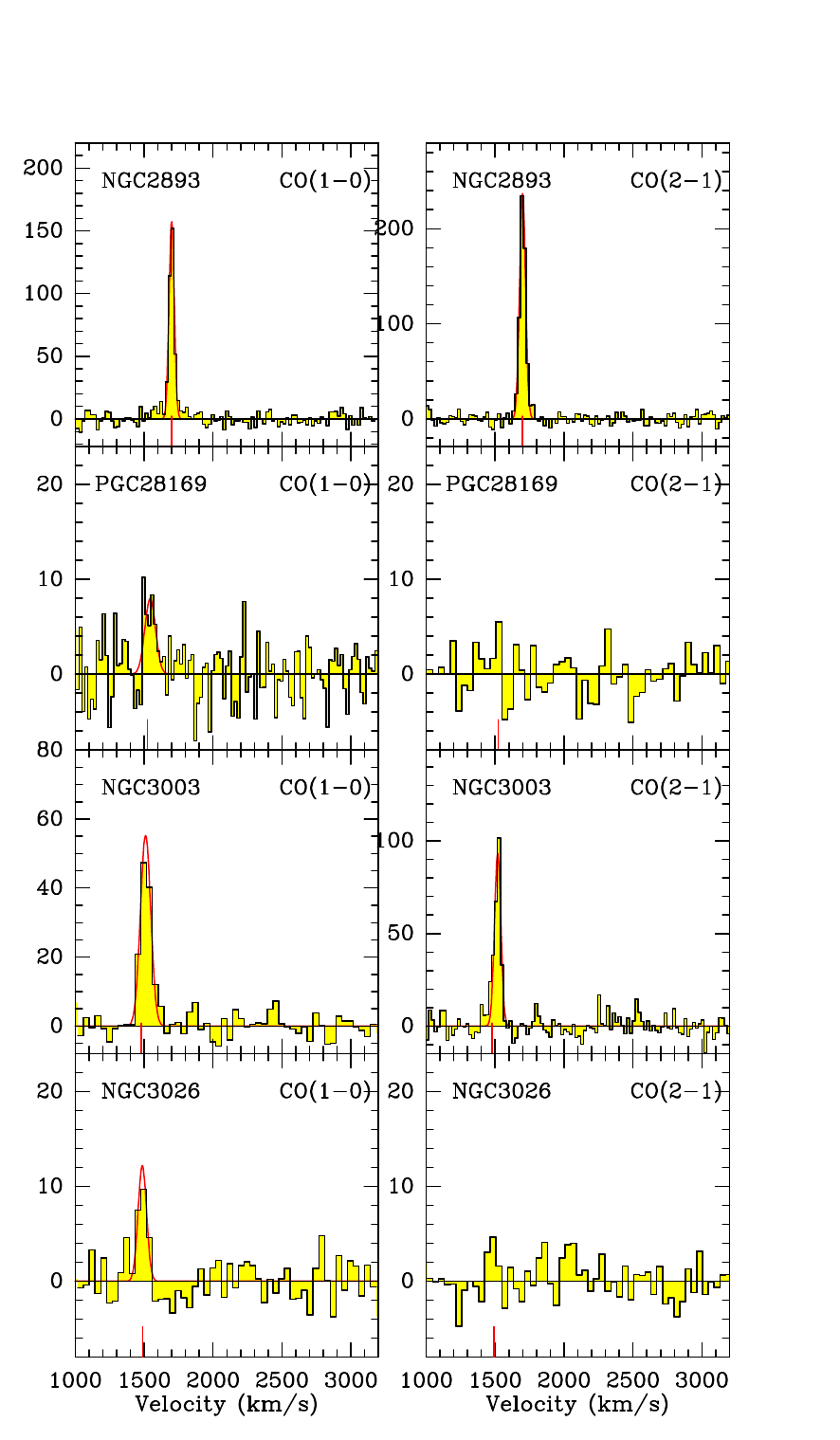}}
\caption{Continued.}
\end{figure*}
\begin{figure*} \centering
\captionsetup[subfigure]{labelformat=empty}
\ContinuedFloat
\subfloat[]{\includegraphics[trim={3.cm 2.cm 3.cm 2cm},clip,page=6,scale=0.9,clip=true]{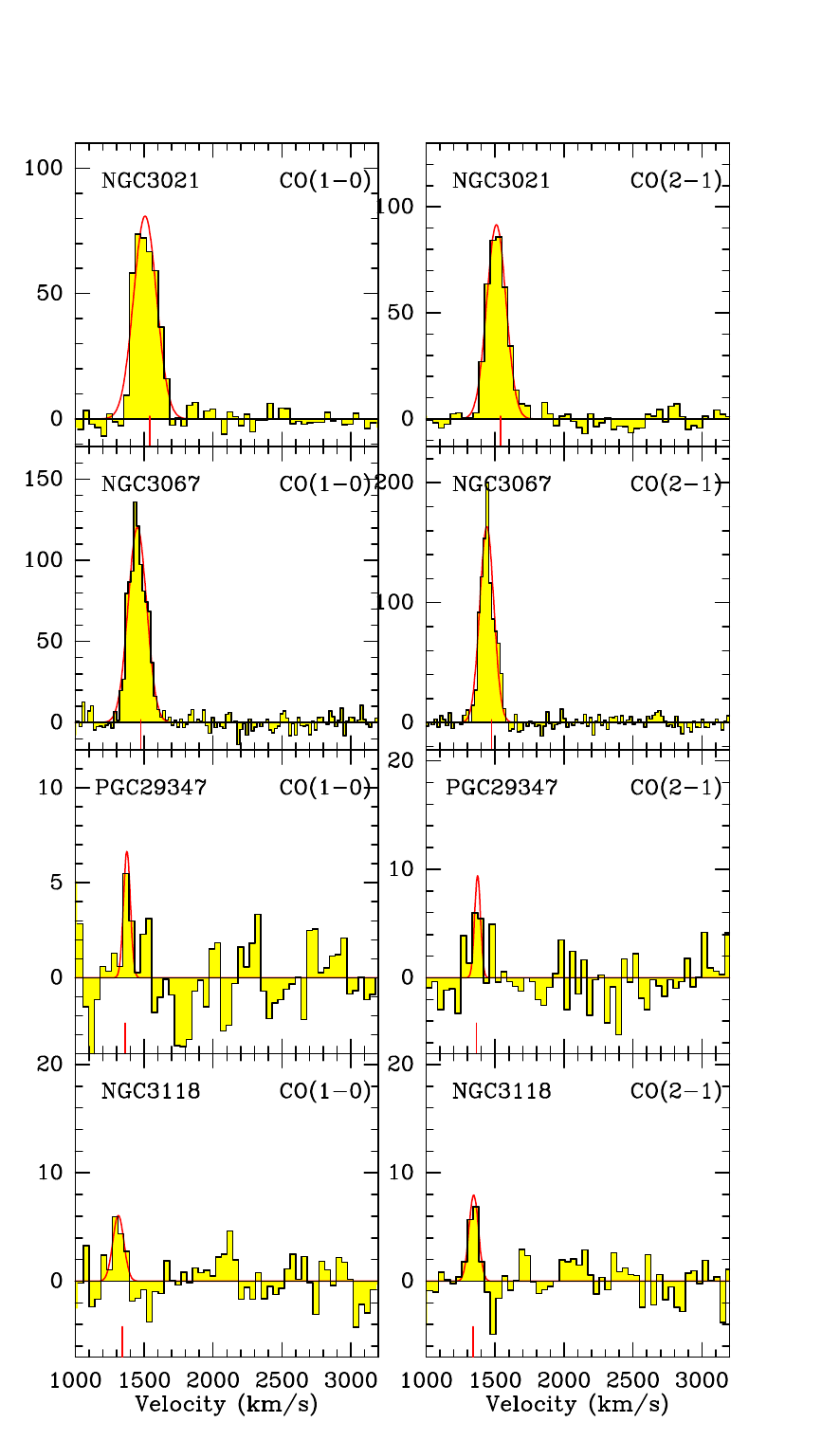}}
\caption{Continued.}
\end{figure*}
\begin{figure*} \centering
\captionsetup[subfigure]{labelformat=empty}
\ContinuedFloat
\subfloat[]{\includegraphics[trim={3.cm 2.cm 3.cm 2cm},clip,page=7,scale=0.9,clip=true]{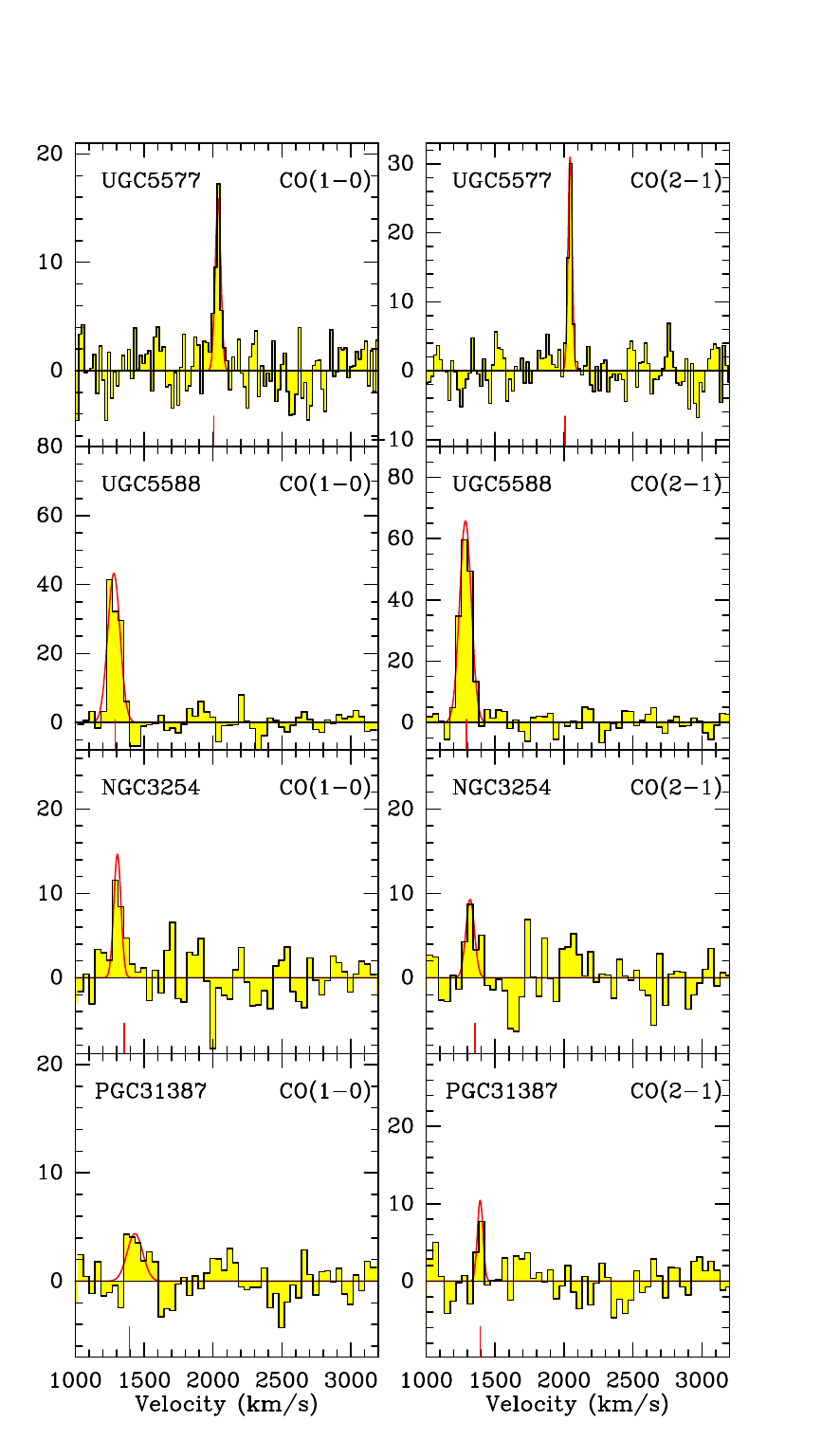}}
\caption{Continued.}
\end{figure*}
\begin{figure*} \centering
\captionsetup[subfigure]{labelformat=empty}
\ContinuedFloat
\subfloat[]{\includegraphics[trim={3.cm 2.cm 3.cm 2cm},clip,page=8,scale=0.9,clip=true]{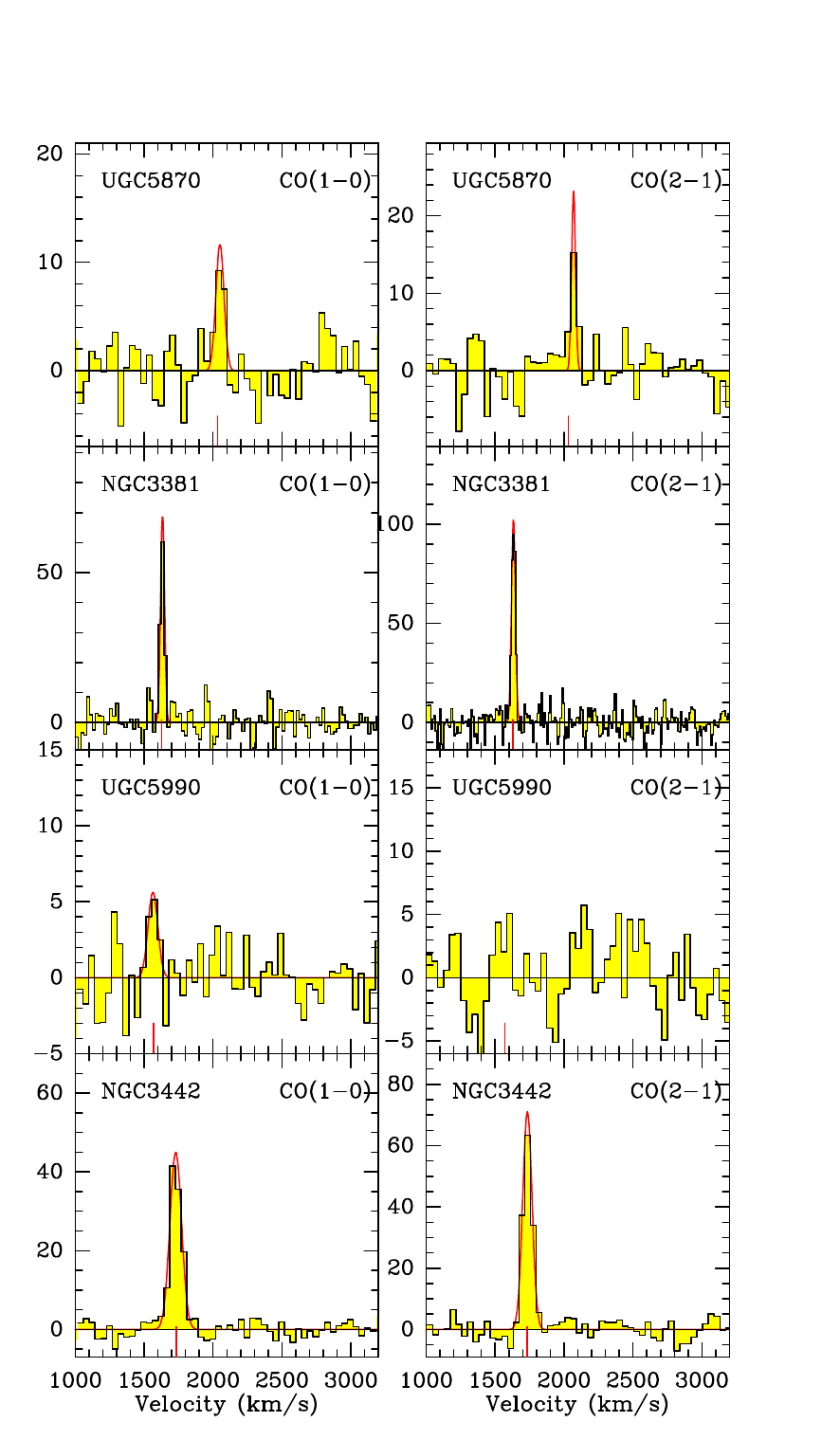}}
\caption{Continued.}
\end{figure*}
\begin{figure*} \centering
\captionsetup[subfigure]{labelformat=empty}
\ContinuedFloat
\subfloat[]{\includegraphics[trim={3.cm 2.cm 3.cm 2cm},clip,page=9,scale=0.9,clip=true]{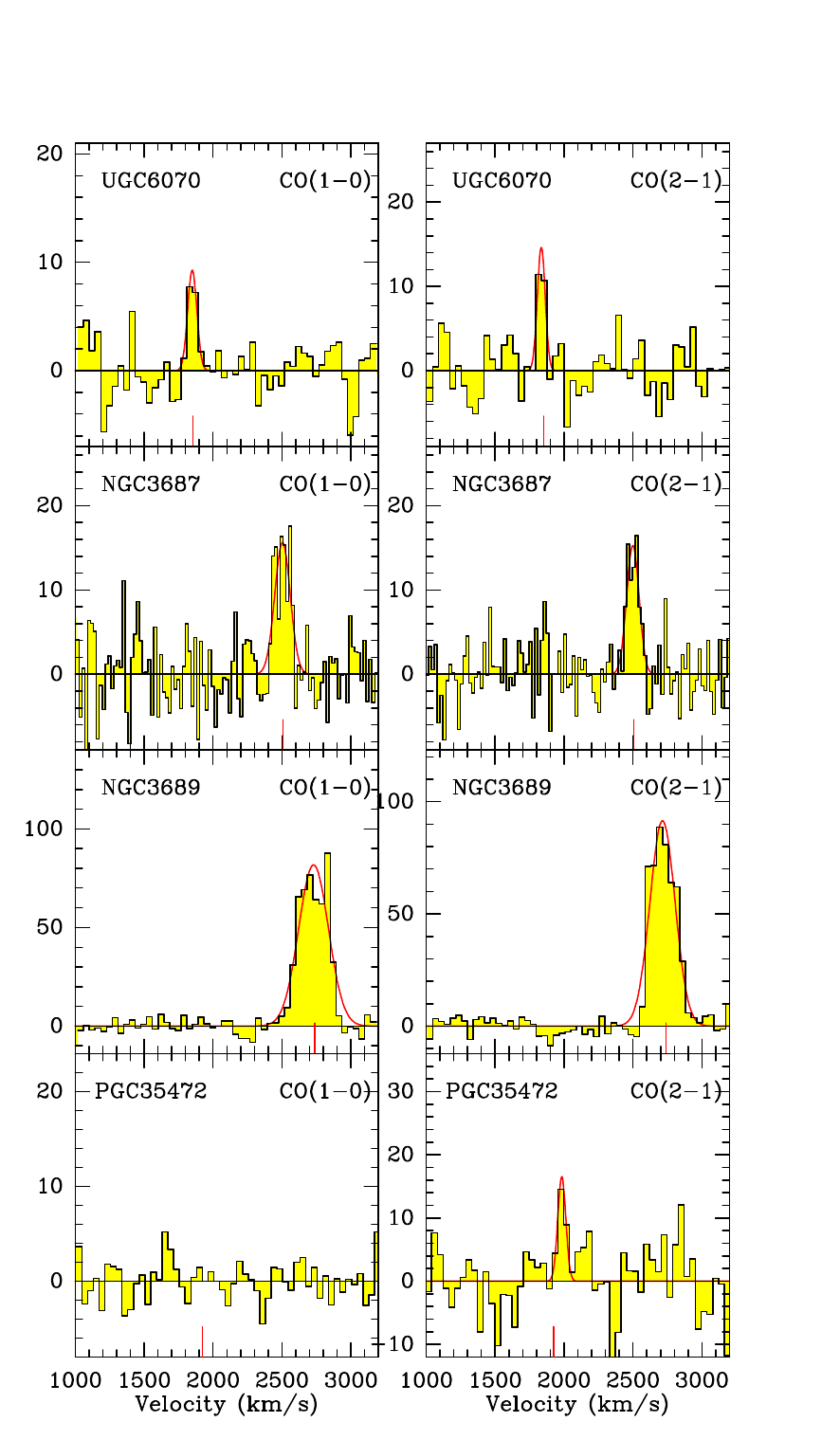}}
\caption{Continued.}
\end{figure*}
\begin{figure*} \centering
\captionsetup[subfigure]{labelformat=empty}
\ContinuedFloat
\subfloat[]{\includegraphics[trim={3.cm 2.cm 3.cm 2cm},clip,page=10,scale=0.9,clip=true]{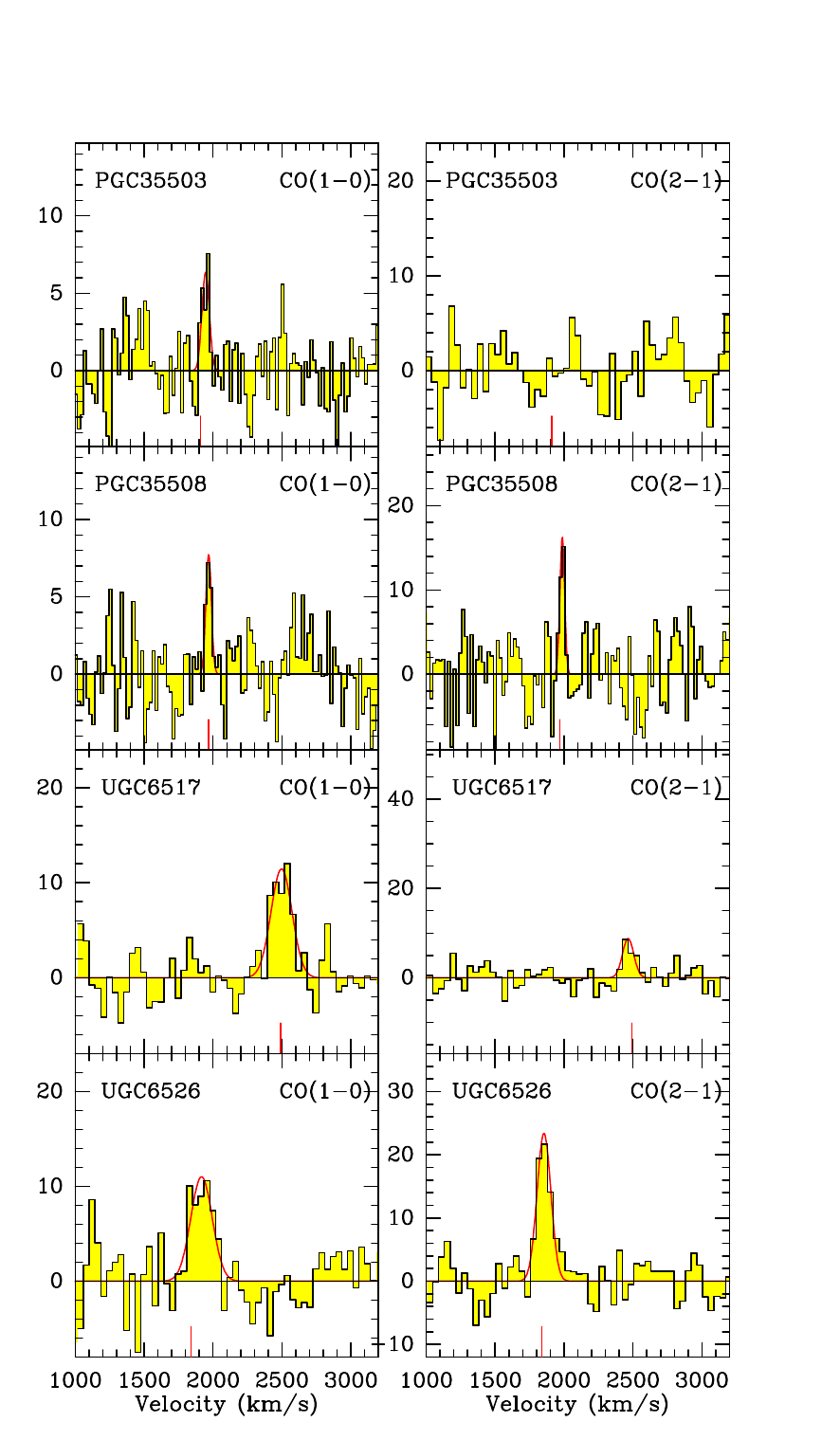}}
\caption{Continued.}
\end{figure*}
\begin{figure*} \centering
\captionsetup[subfigure]{labelformat=empty}
\ContinuedFloat
\subfloat[]{\includegraphics[trim={3.cm 2.cm 3.cm 2cm},clip,page=11,scale=0.9,clip=true]{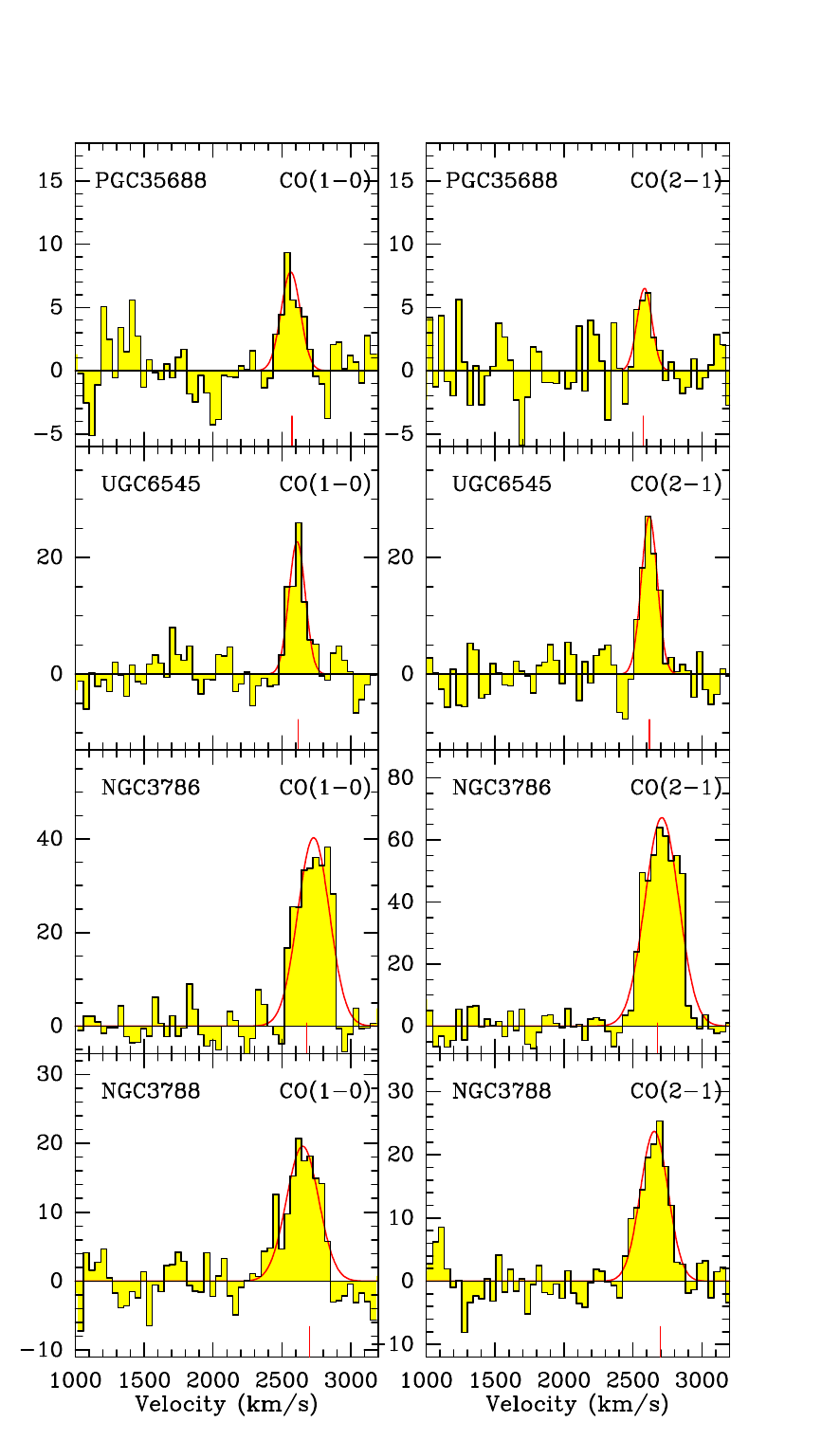}}
\caption{Continued.}
\end{figure*}
\begin{figure*} \centering
\captionsetup[subfigure]{labelformat=empty}
\ContinuedFloat
\subfloat[]{\includegraphics[trim={3.cm 2.cm 3.cm 2cm},clip,page=12,scale=0.9,clip=true]{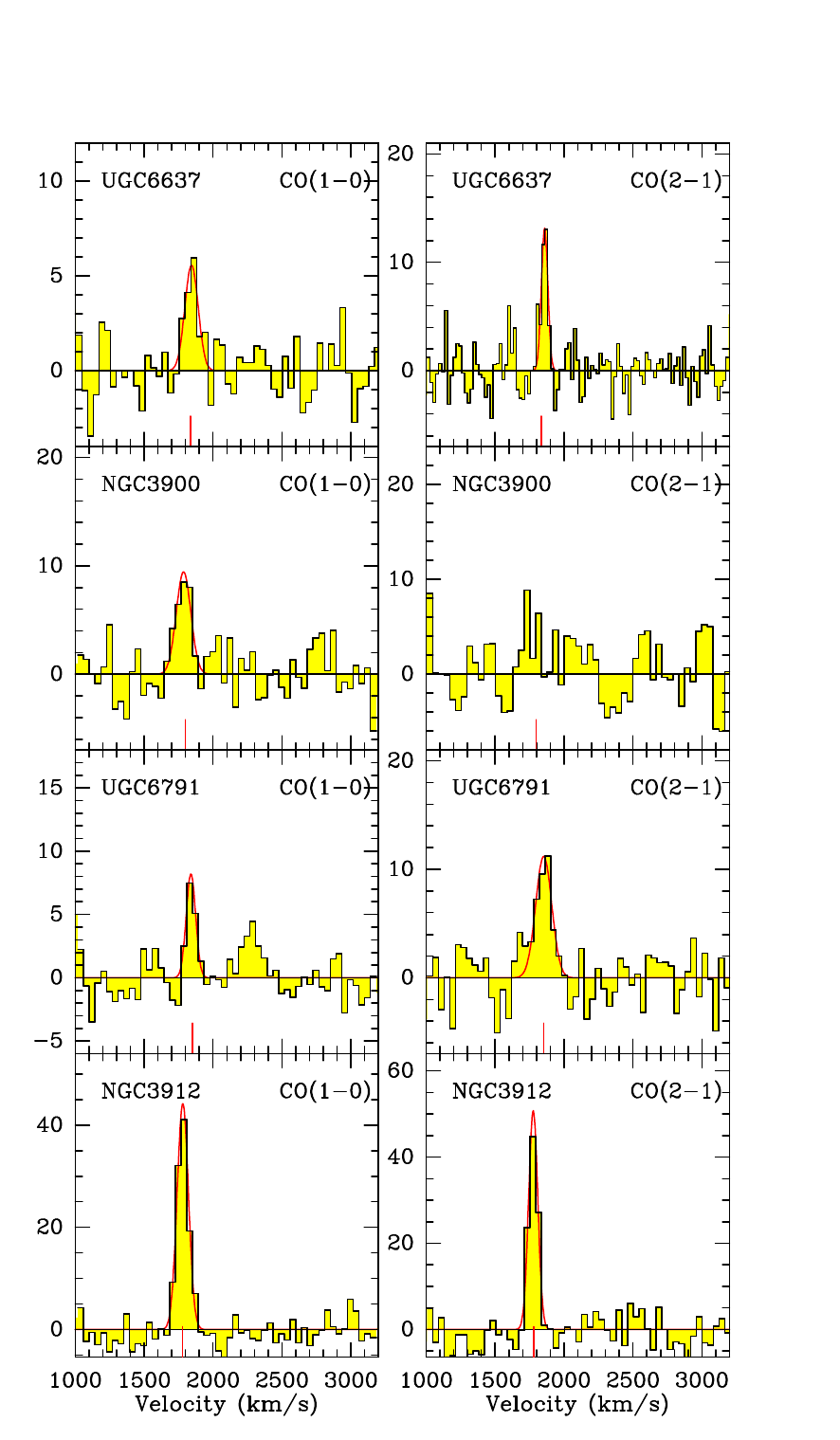}}
\caption{Continued.}
\end{figure*}
\begin{figure*} \centering
\captionsetup[subfigure]{labelformat=empty}
\ContinuedFloat
\subfloat[]{\includegraphics[trim={3.cm 2.cm 3.cm 2cm},clip,page=13,scale=0.9,clip=true]{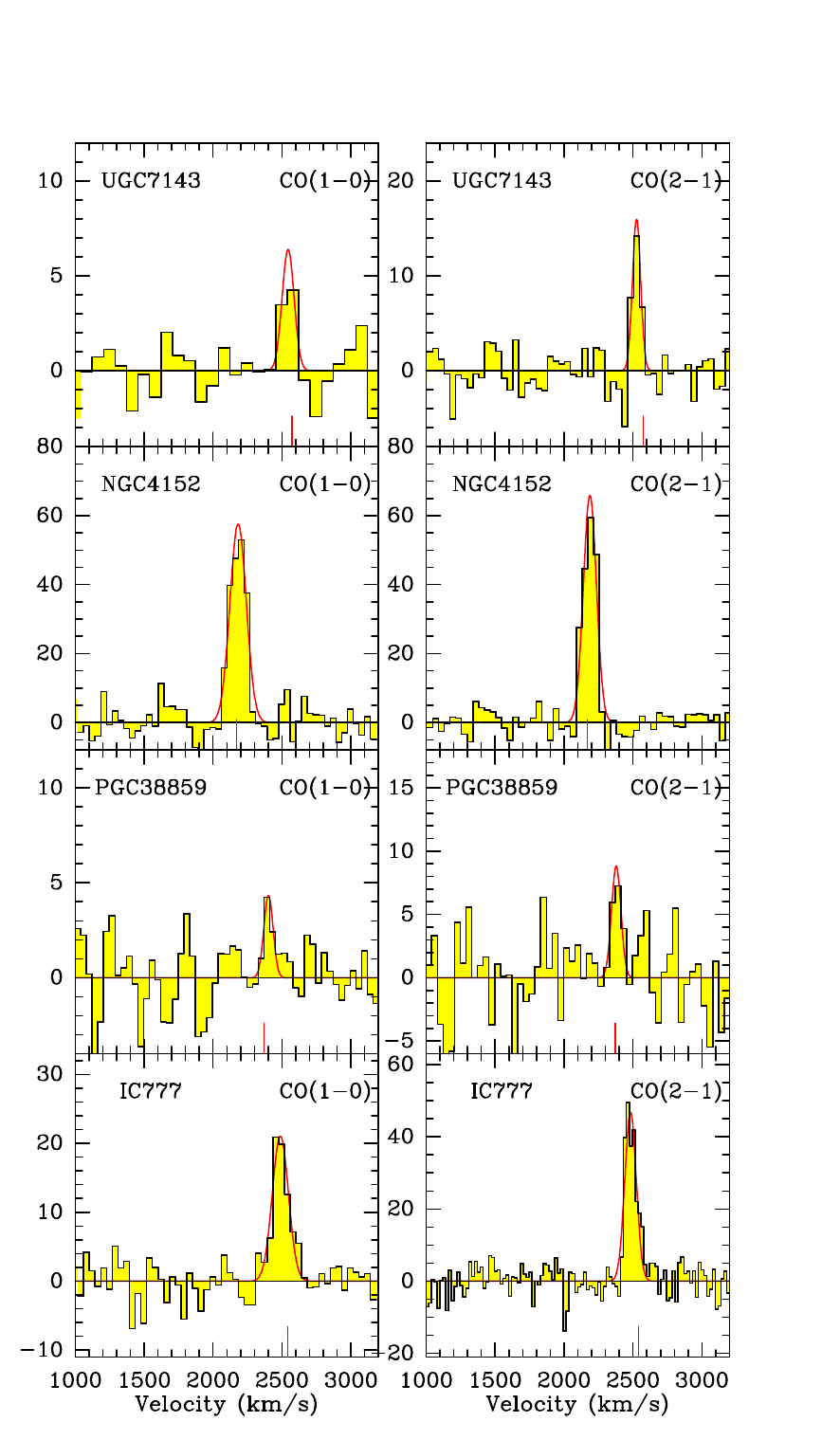}}
\caption{Continued.}
\end{figure*}
\begin{figure*} \centering
\captionsetup[subfigure]{labelformat=empty}
\ContinuedFloat
\subfloat[]{\includegraphics[trim={3.cm 2.cm 3.cm 2cm},clip,page=14,scale=0.9,clip=true]{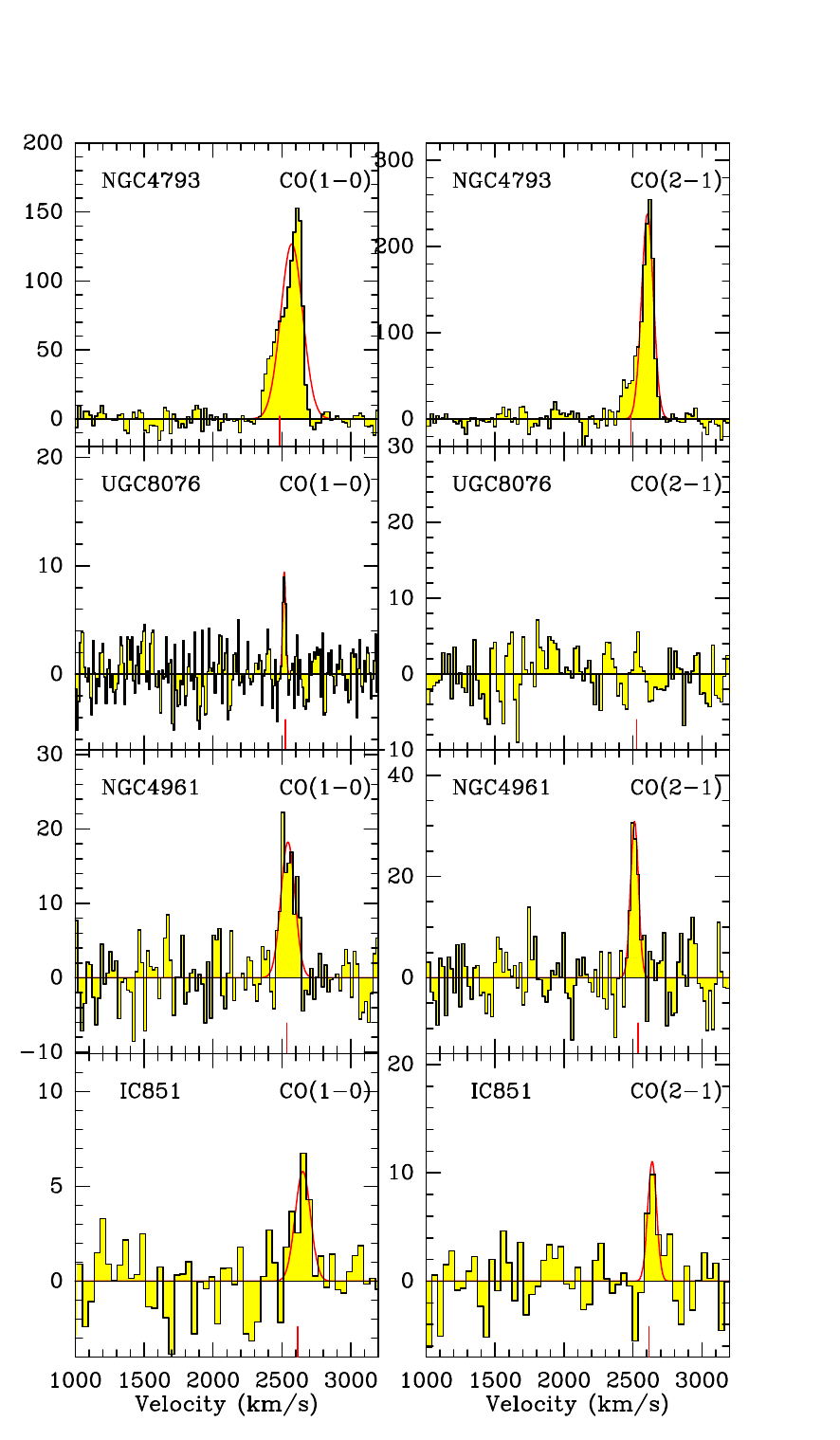}}
\caption{Continued.}
\end{figure*}
\begin{figure*} \centering
\captionsetup[subfigure]{labelformat=empty}
\ContinuedFloat
\subfloat[]{\includegraphics[trim={3.cm 2.cm 3.cm 2cm},clip,page=15,scale=0.9,clip=true]{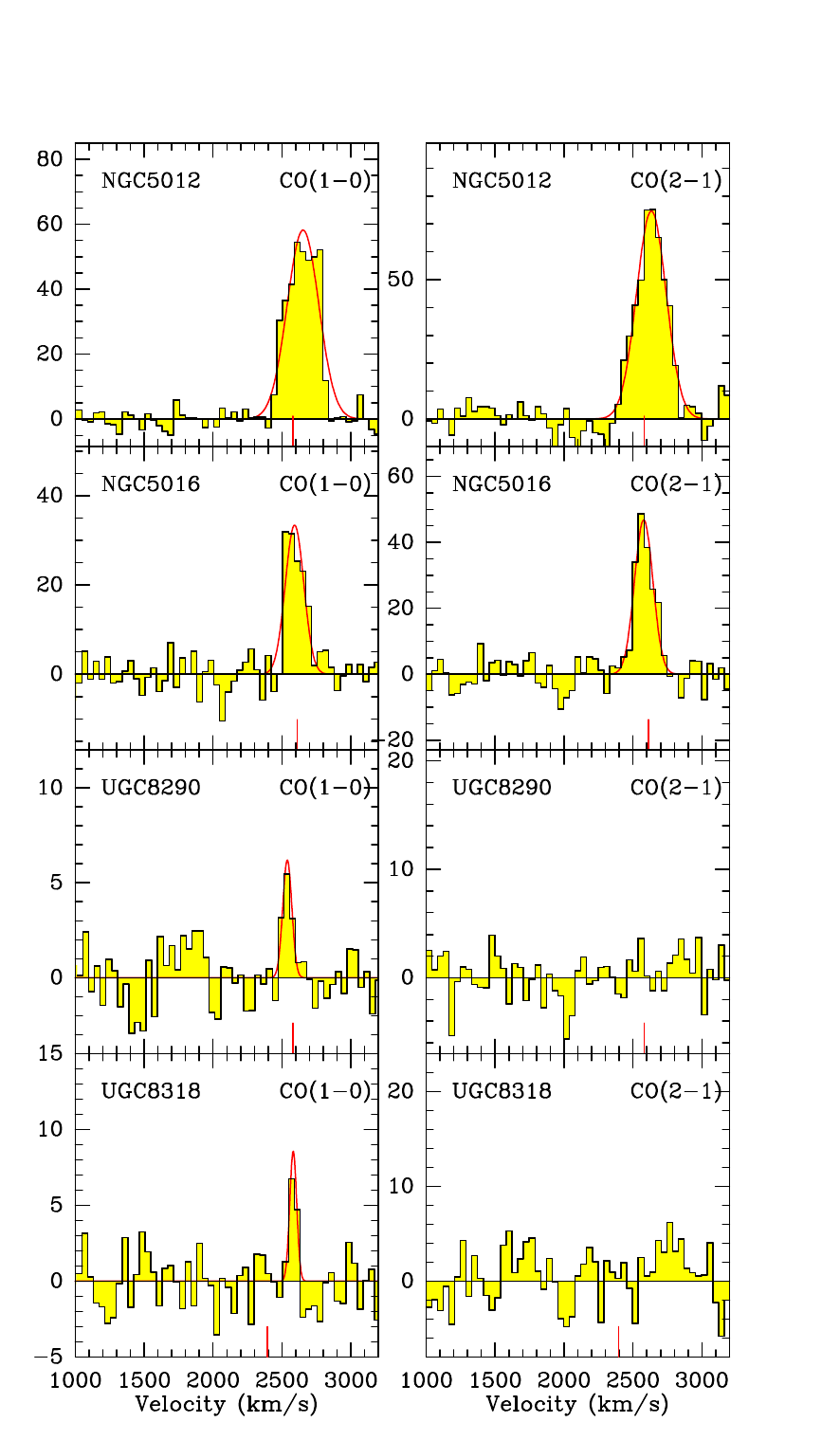}}
\caption{Continued.}
\end{figure*}
\begin{figure*} \centering
\captionsetup[subfigure]{labelformat=empty}
\ContinuedFloat
\subfloat[]{\includegraphics[trim={3.cm 2.cm 3.cm 2cm},clip,page=16,scale=0.9,clip=true]{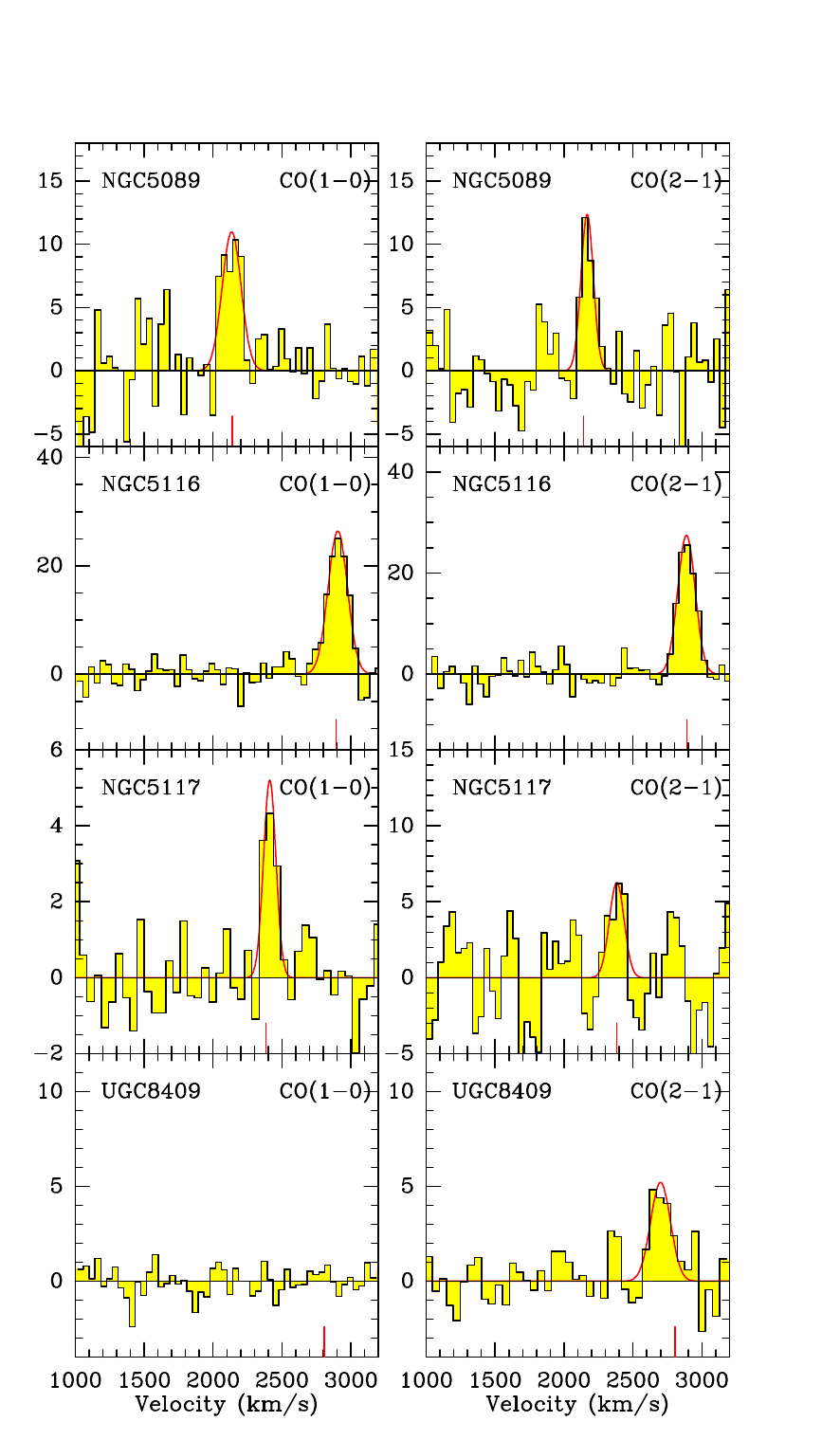}}
\caption{Continued.}
\end{figure*}
\begin{figure*} \centering
\captionsetup[subfigure]{labelformat=empty}
\ContinuedFloat
\subfloat[]{\includegraphics[trim={3.cm 2.cm 3.cm 2cm},clip,page=17,scale=0.9,clip=true]{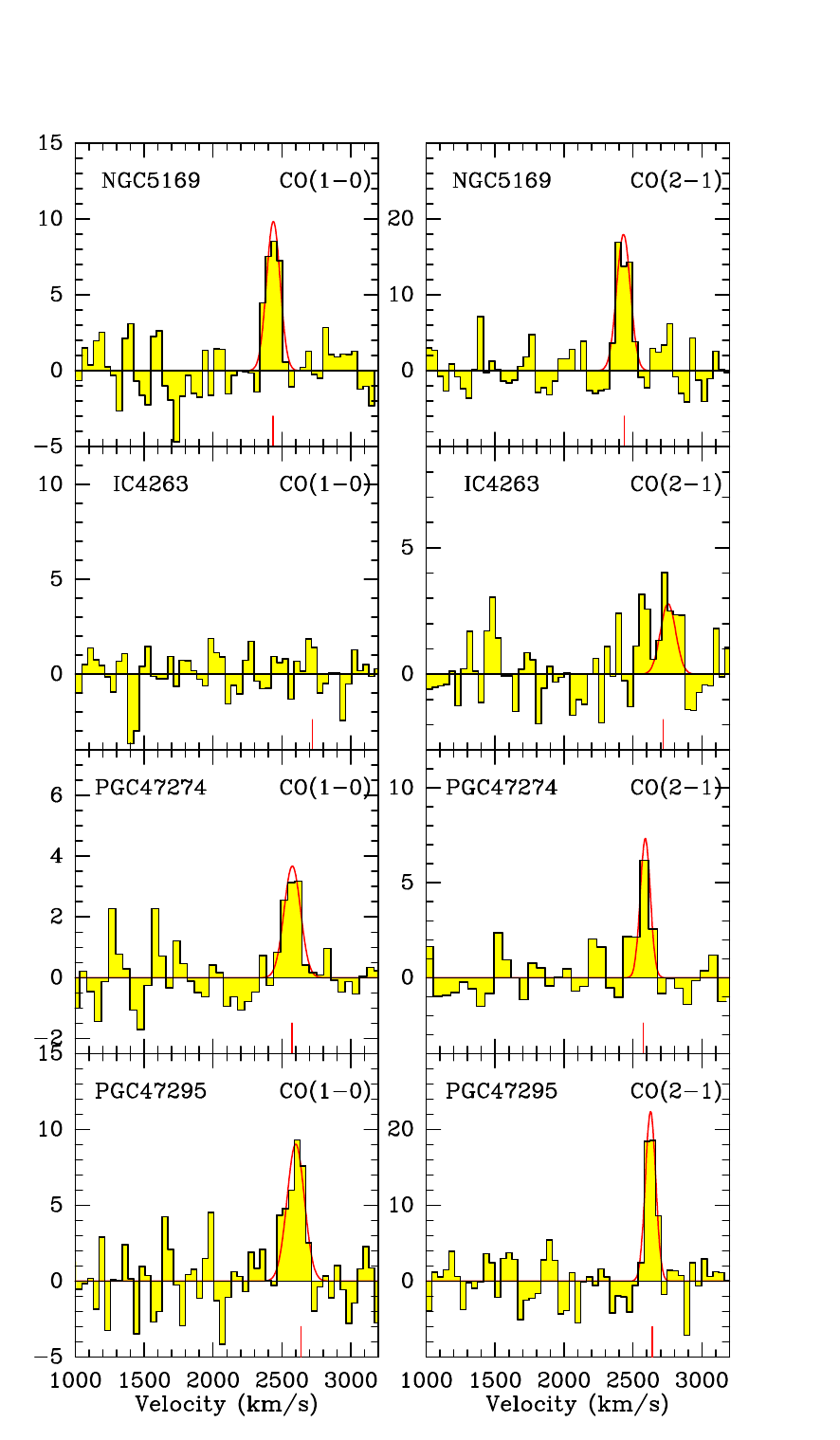}}
\caption{Continued.}
\end{figure*}
\begin{figure*} \centering
\captionsetup[subfigure]{labelformat=empty}
\ContinuedFloat
\subfloat[]{\includegraphics[trim={3.cm 2.cm 3.cm 2cm},clip,page=18,scale=0.9,clip=true]{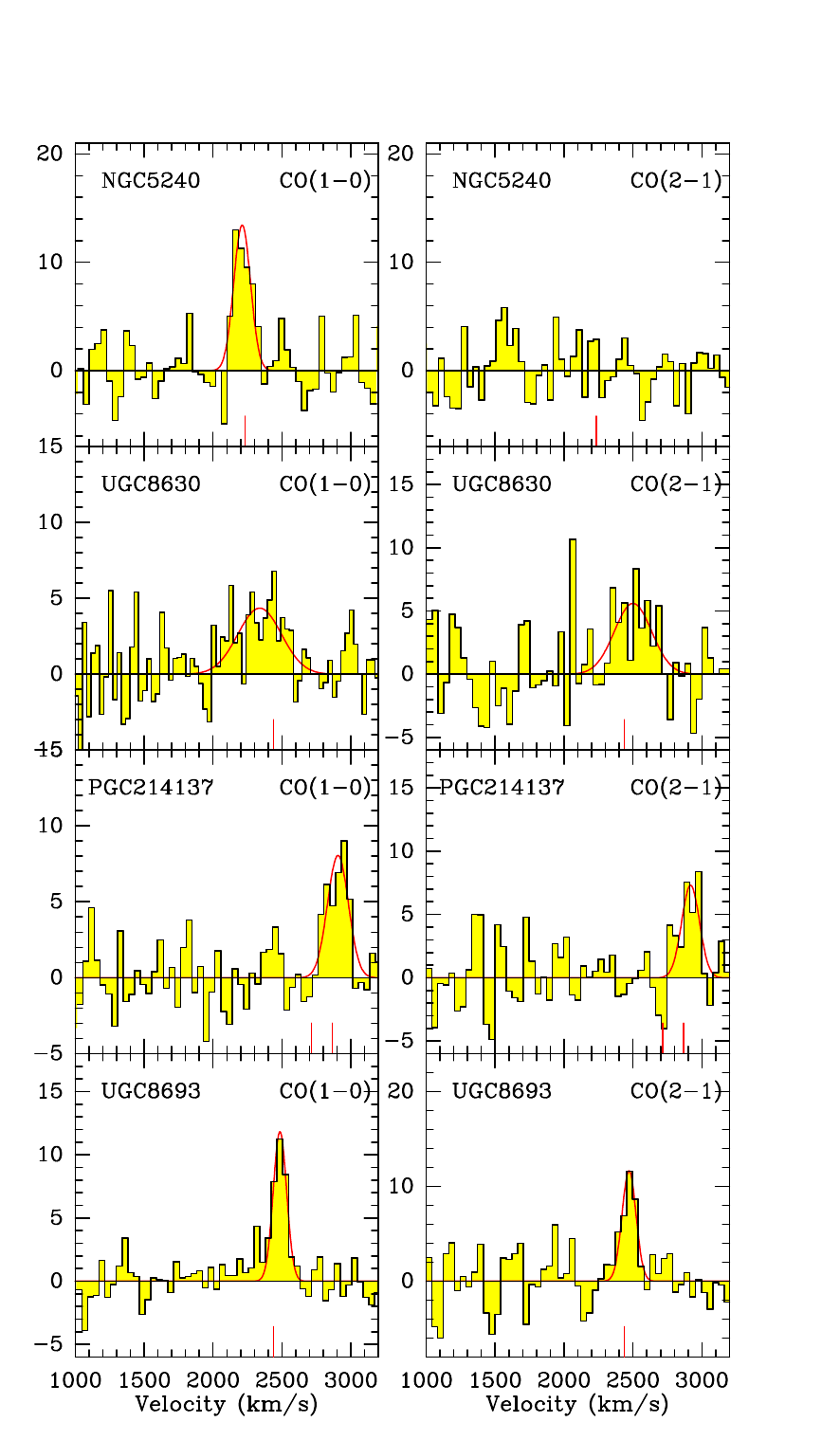}}
\caption{Continued.}
\end{figure*}
\begin{figure*} \centering
\captionsetup[subfigure]{labelformat=empty}
\ContinuedFloat
\subfloat[]{\includegraphics[trim={3.cm 2.cm 3.cm 2cm},clip,page=19,scale=0.9,clip=true]{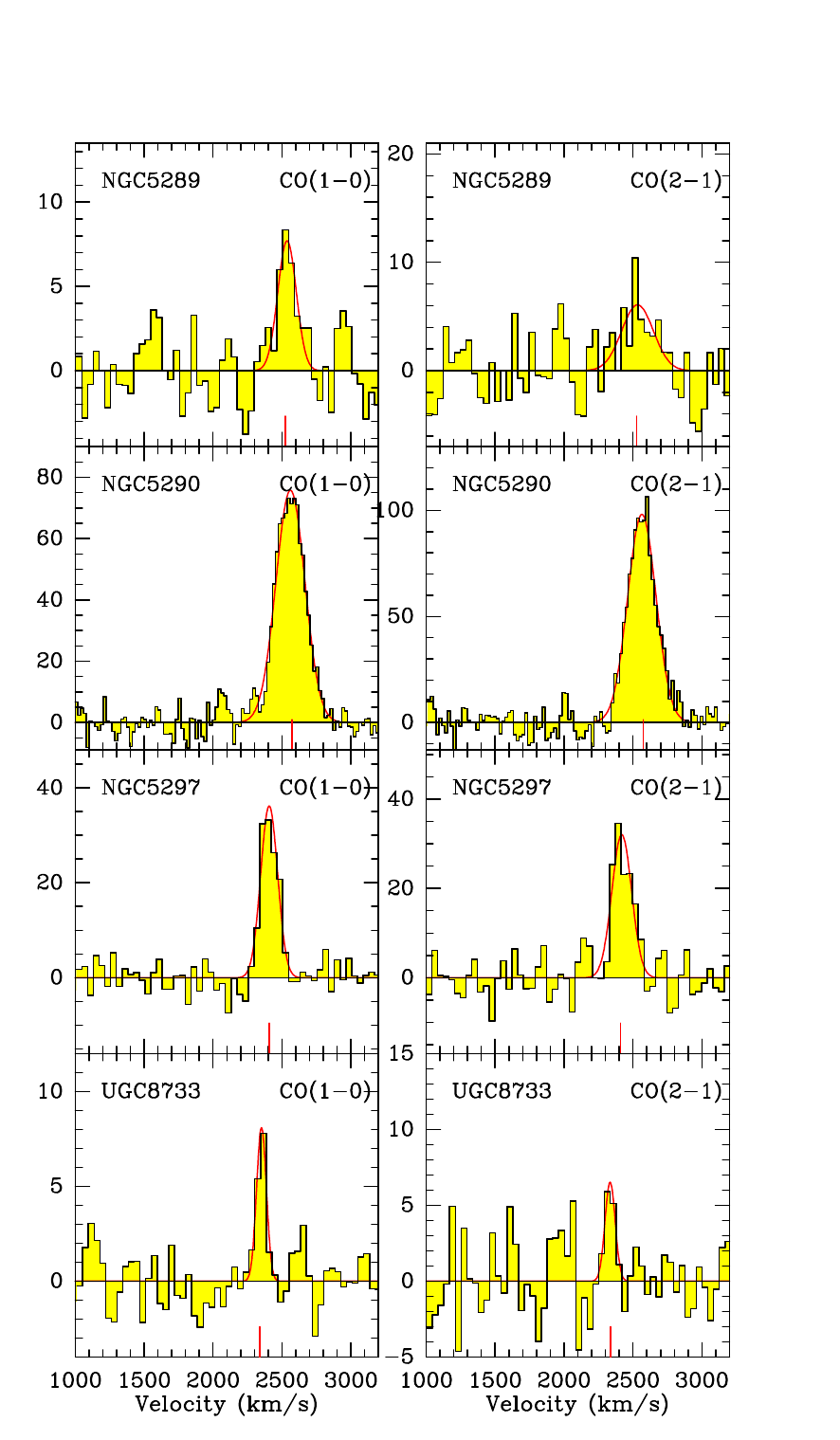}}
\caption{Continued.}
\end{figure*}
\begin{figure*} \centering
\captionsetup[subfigure]{labelformat=empty}
\ContinuedFloat
\subfloat[]{\includegraphics[trim={3.cm 2.cm 3.cm 2cm},clip,page=20,scale=0.9,clip=true]{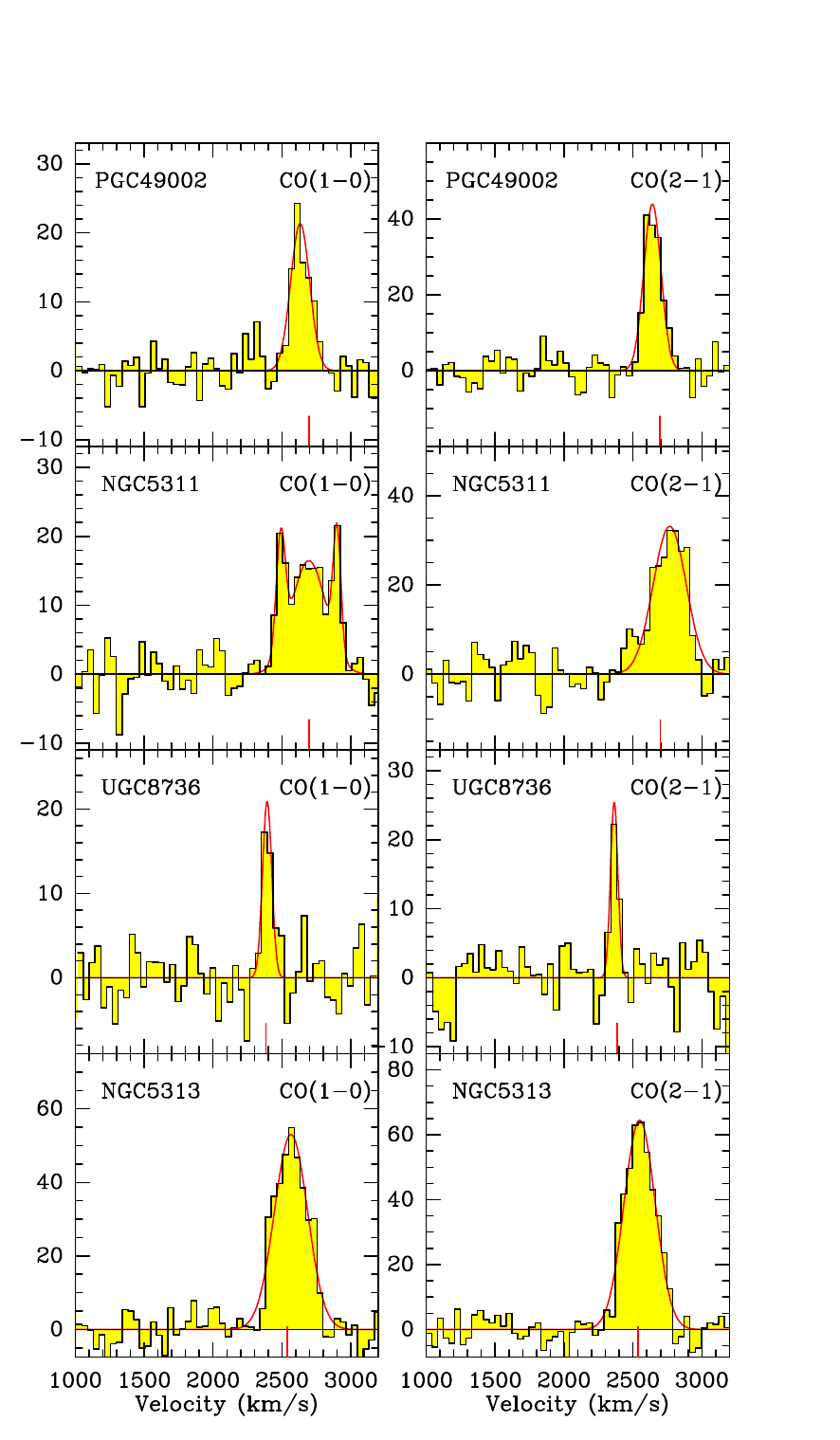}}
\caption{Continued.}
\end{figure*}
\begin{figure*} \centering
\captionsetup[subfigure]{labelformat=empty}
\ContinuedFloat
\subfloat[]{\includegraphics[trim={3.cm 2.cm 3.cm 2cm},clip,page=21,scale=0.9,clip=true]{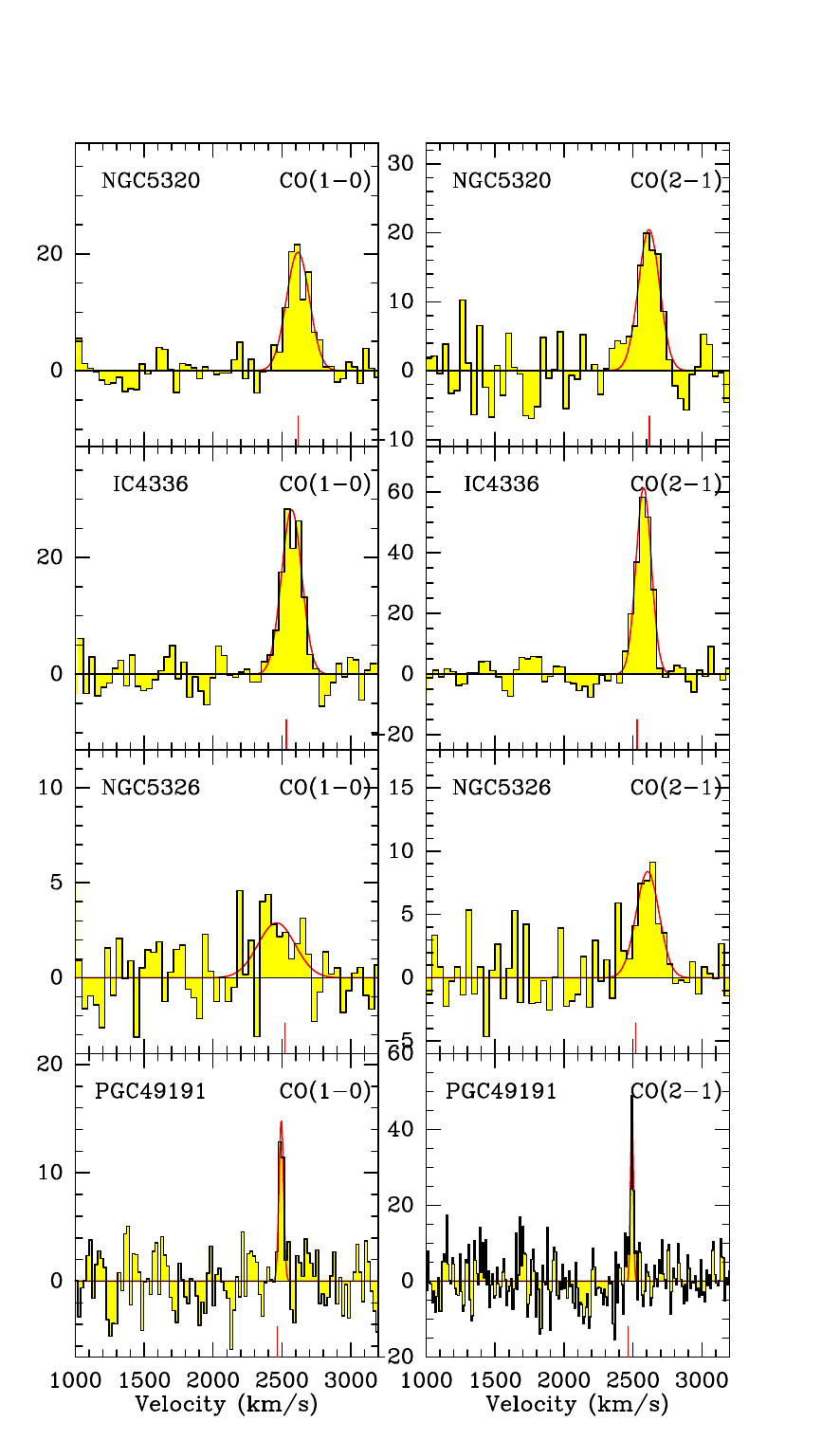}}
\caption{Continued.}
\end{figure*}
\begin{figure*} \centering
\captionsetup[subfigure]{labelformat=empty}
\ContinuedFloat
\subfloat[]{\includegraphics[trim={3.cm 2.cm 3.cm 2cm},clip,page=22,scale=0.9,clip=true]{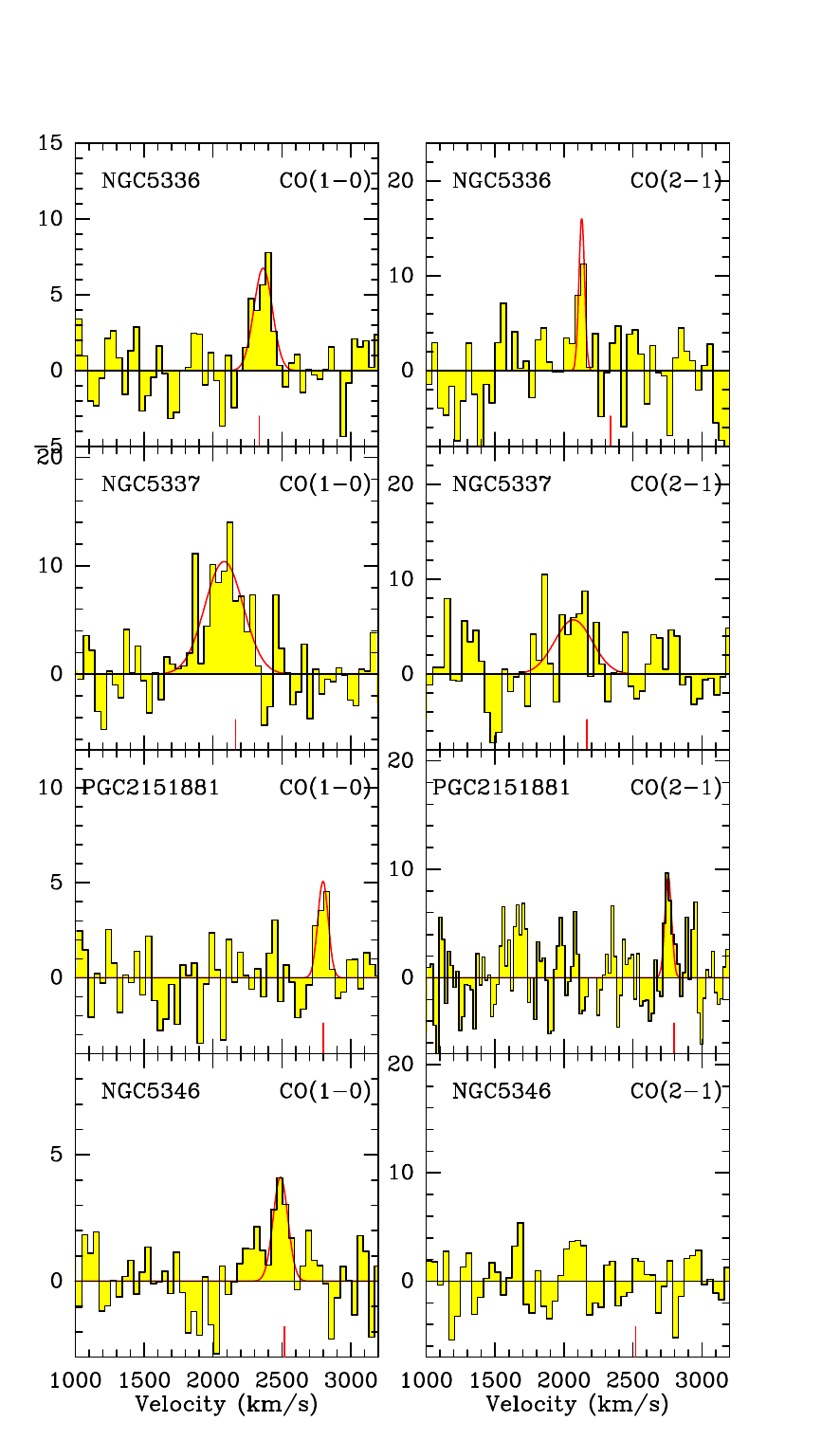}}
\caption{Continued.}
\end{figure*}
\begin{figure*} \centering
\captionsetup[subfigure]{labelformat=empty}
\ContinuedFloat
\subfloat[]{\includegraphics[trim={3.cm 2.cm 3.cm 2cm},clip,page=23,scale=0.9,clip=true]{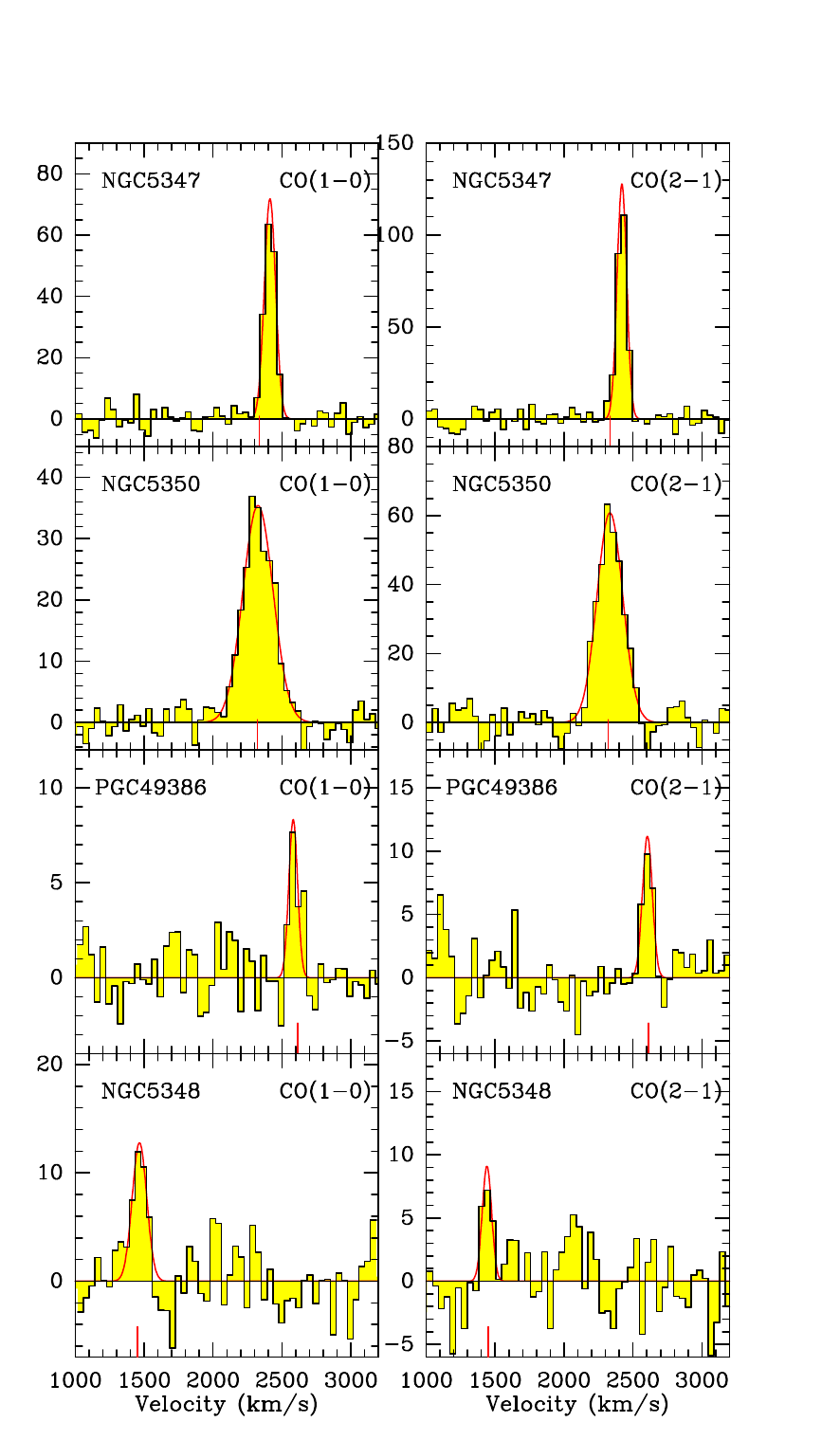}}
\caption{Continued.}
\end{figure*}
\begin{figure*} \centering
\captionsetup[subfigure]{labelformat=empty}
\ContinuedFloat
\subfloat[]{\includegraphics[trim={3.cm 2.cm 3.cm 2cm},clip,page=24,scale=0.9,clip=true]{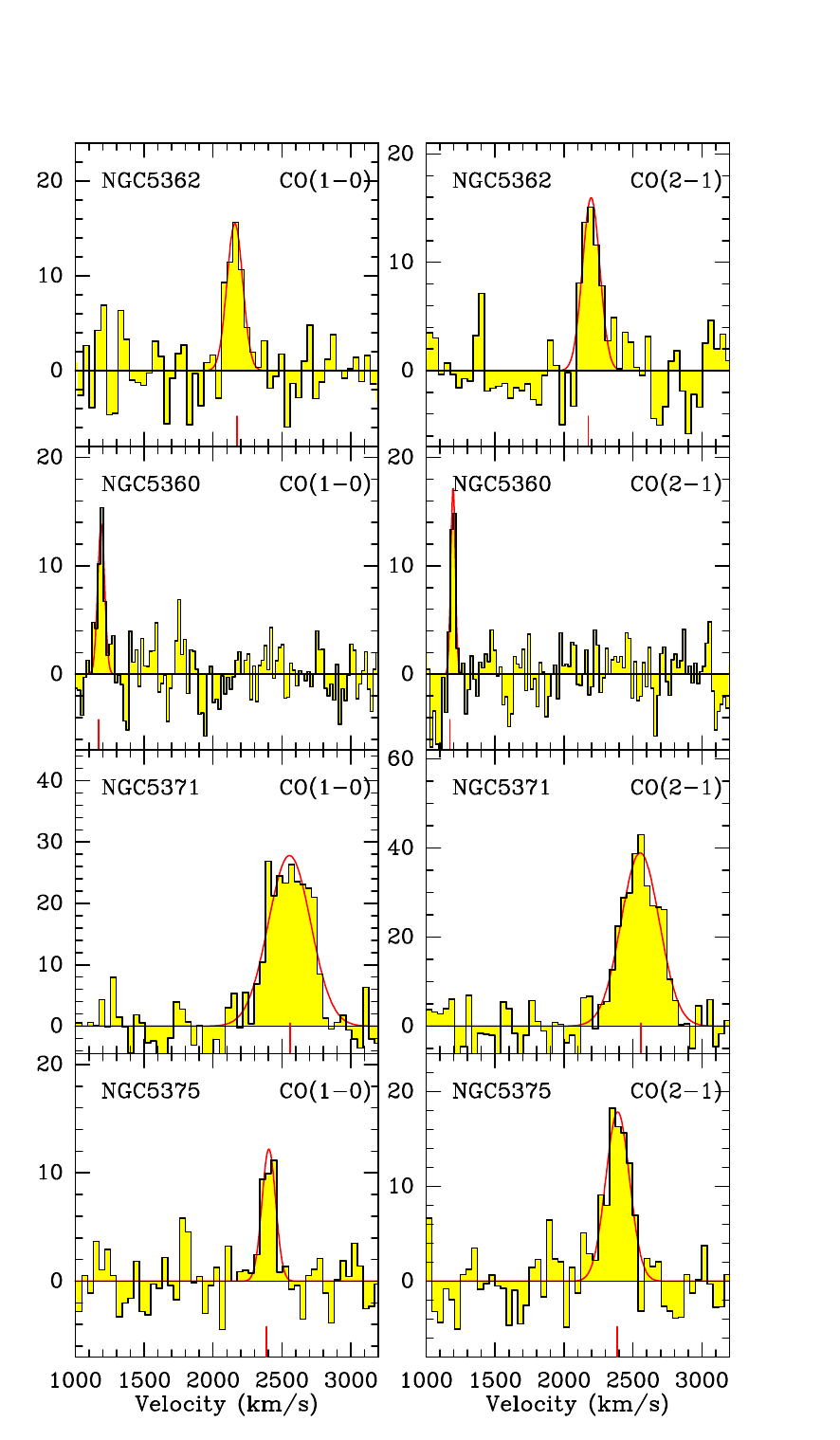}}
\caption{Continued.}
\end{figure*}
\begin{figure*} \centering
\captionsetup[subfigure]{labelformat=empty}
\ContinuedFloat
\subfloat[]{\includegraphics[trim={3.cm 2.cm 3.cm 2cm},clip,page=25,scale=0.9,clip=true]{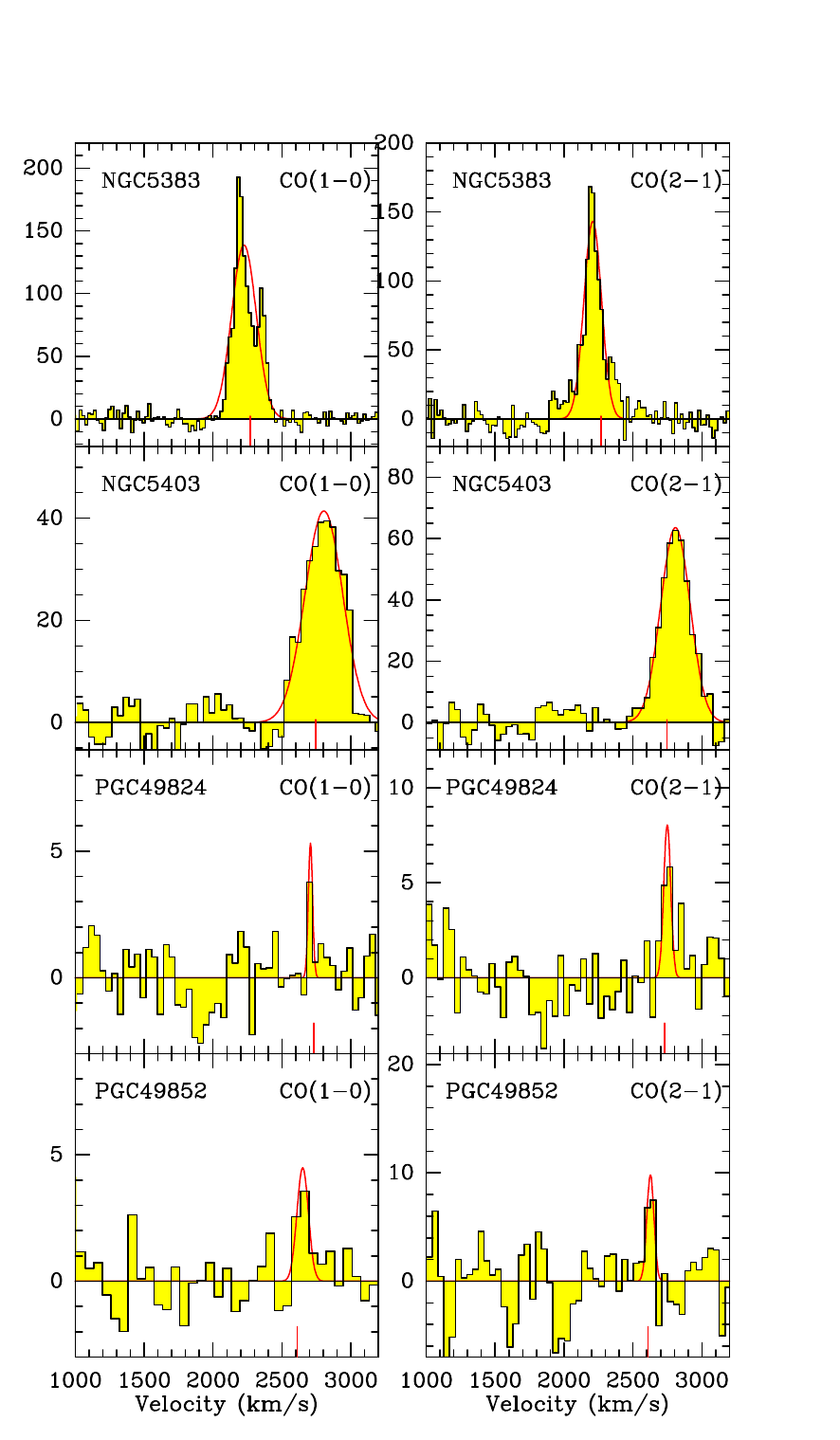}}
\caption{Continued.}
\end{figure*}
\begin{figure*} \centering
\captionsetup[subfigure]{labelformat=empty}
\ContinuedFloat
\subfloat[]{\includegraphics[trim={3.cm 2.cm 3.cm 2cm},clip,page=26,scale=0.9,clip=true]{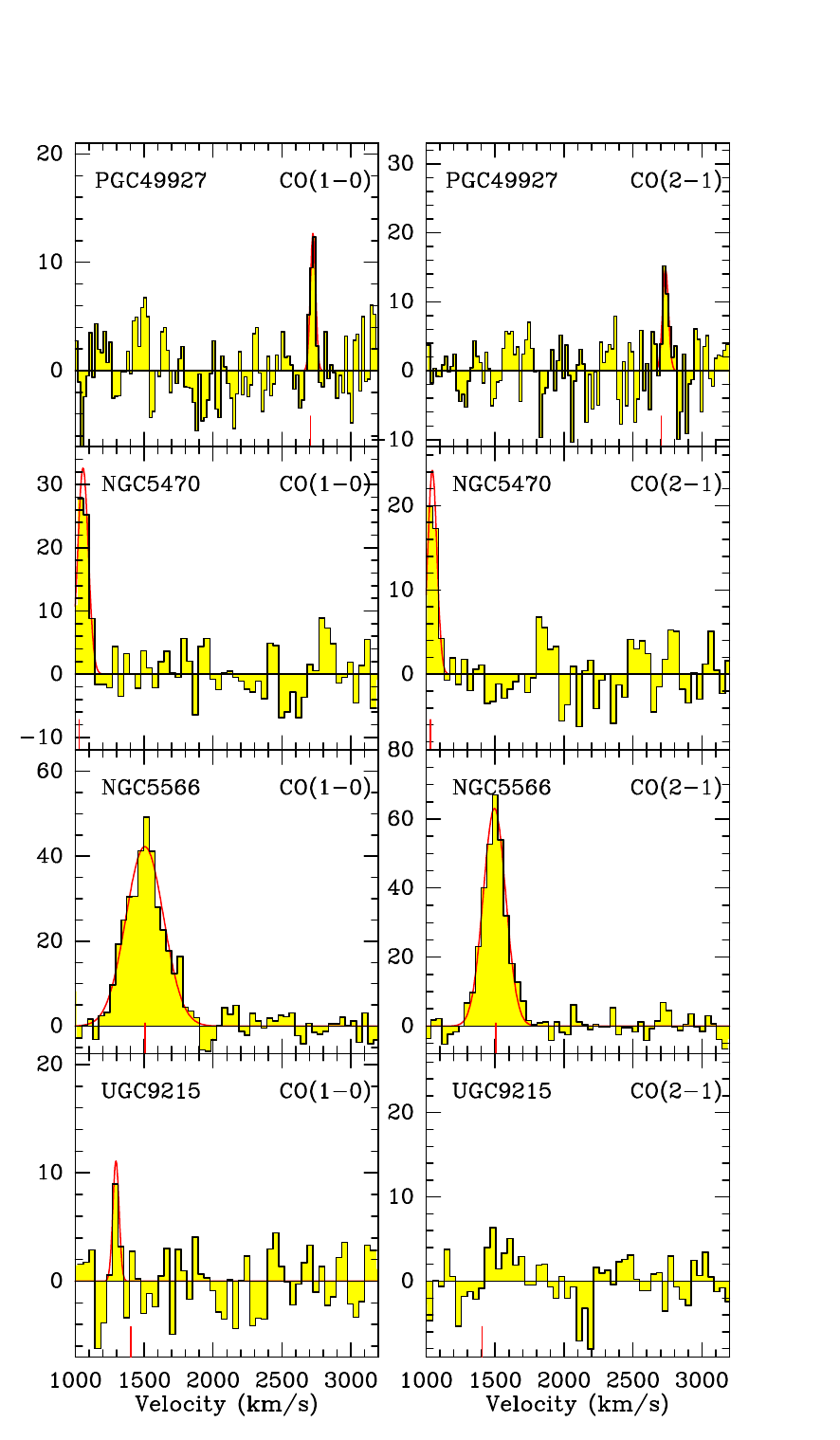}}
\caption{Continued.}
\end{figure*}
\begin{figure*} \centering
\captionsetup[subfigure]{labelformat=empty}
\ContinuedFloat
\subfloat[]{\includegraphics[trim={3.cm 2.cm 3.cm 2cm},clip,page=27,scale=0.9,clip=true]{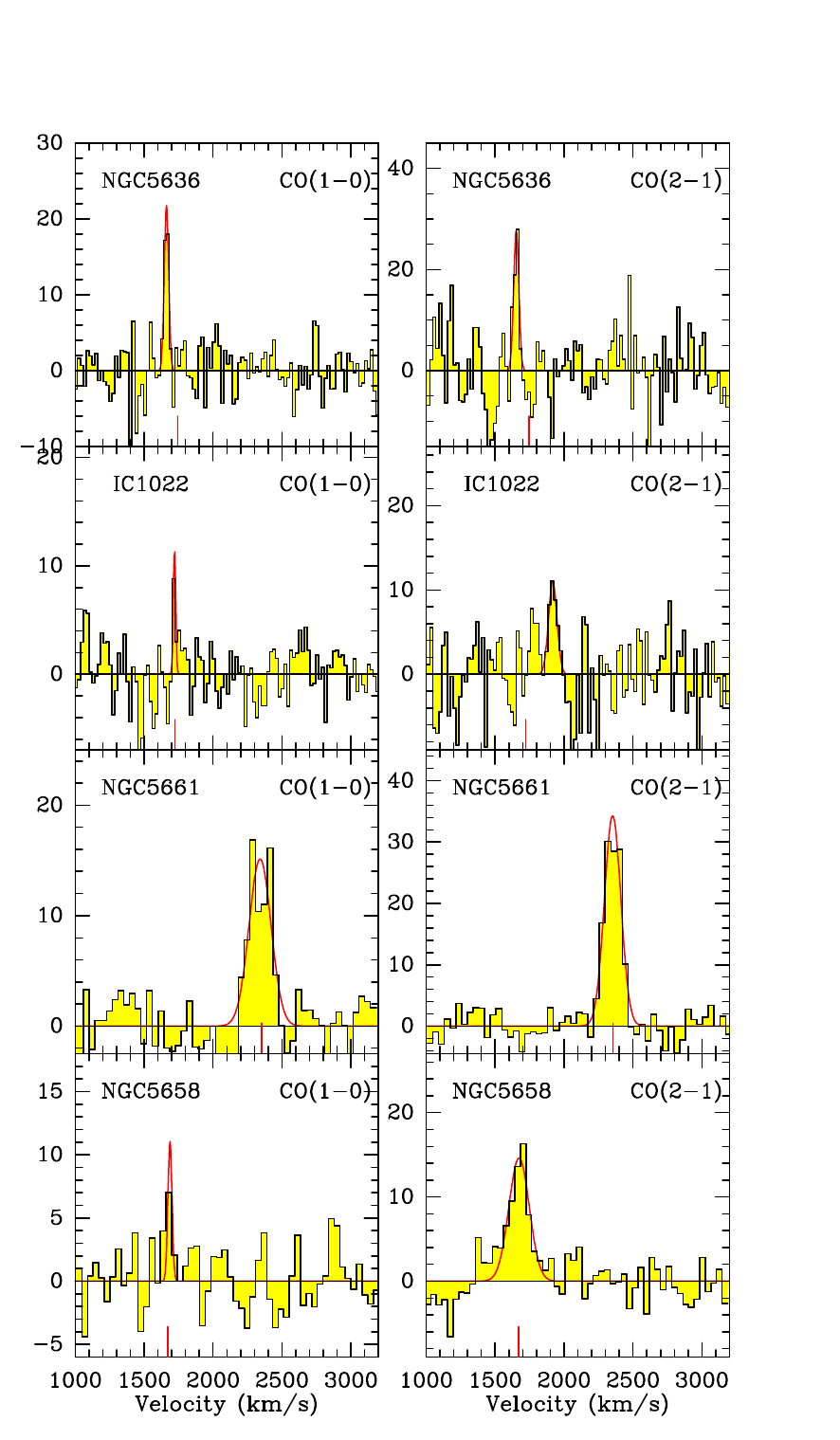}}
\caption{Continued.}
\end{figure*}
\begin{figure*} \centering
\captionsetup[subfigure]{labelformat=empty}
\ContinuedFloat
\subfloat[]{\includegraphics[trim={3.cm 2.cm 3.cm 2cm},clip,page=28,scale=0.9,clip=true]{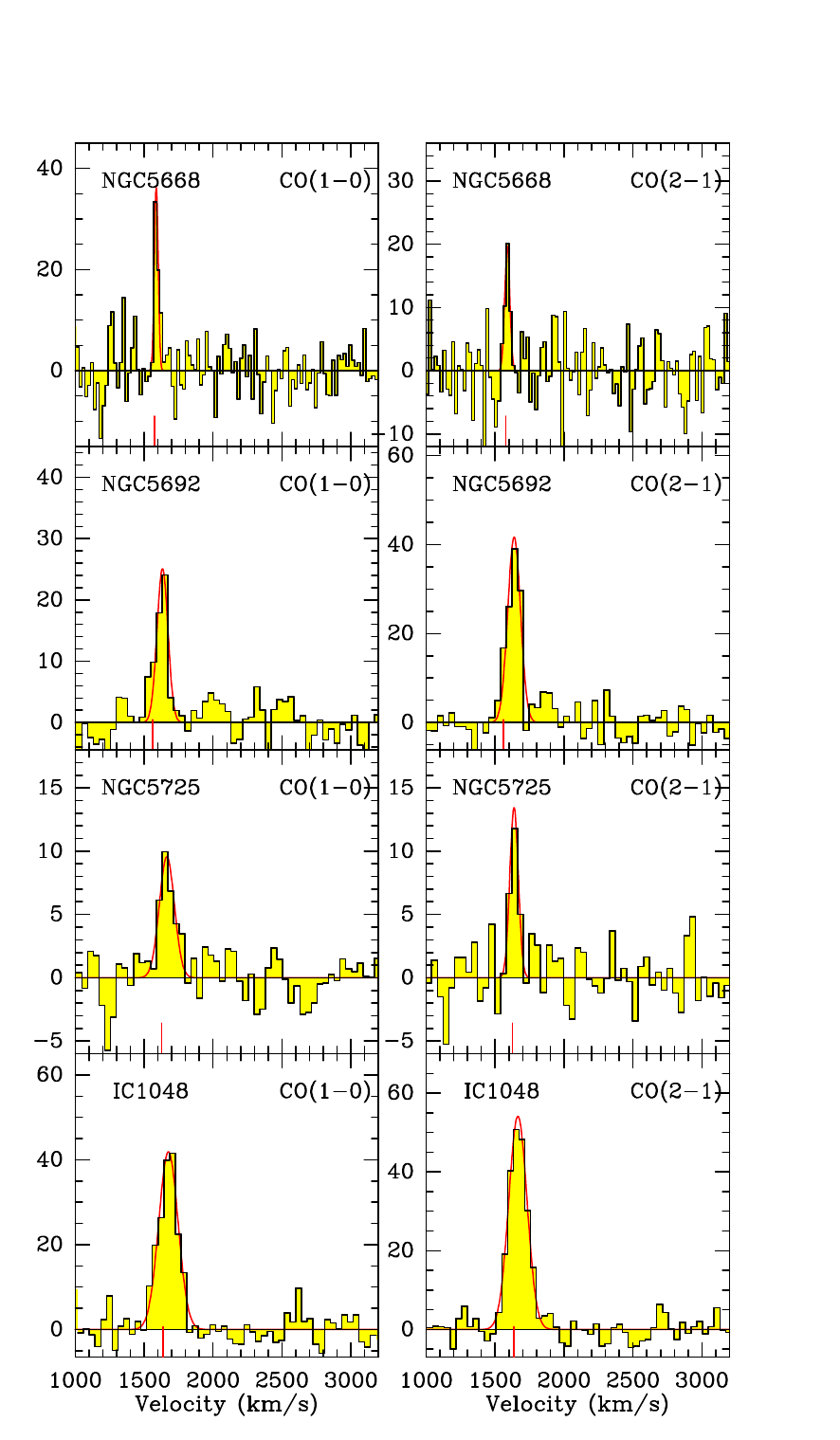}}
\caption{Continued.}
\end{figure*}
\begin{figure*} \centering
\captionsetup[subfigure]{labelformat=empty}
\ContinuedFloat
\subfloat[]{\includegraphics[trim={3.cm 2.cm 3.cm 2cm},clip,page=29,scale=0.9,clip=true]{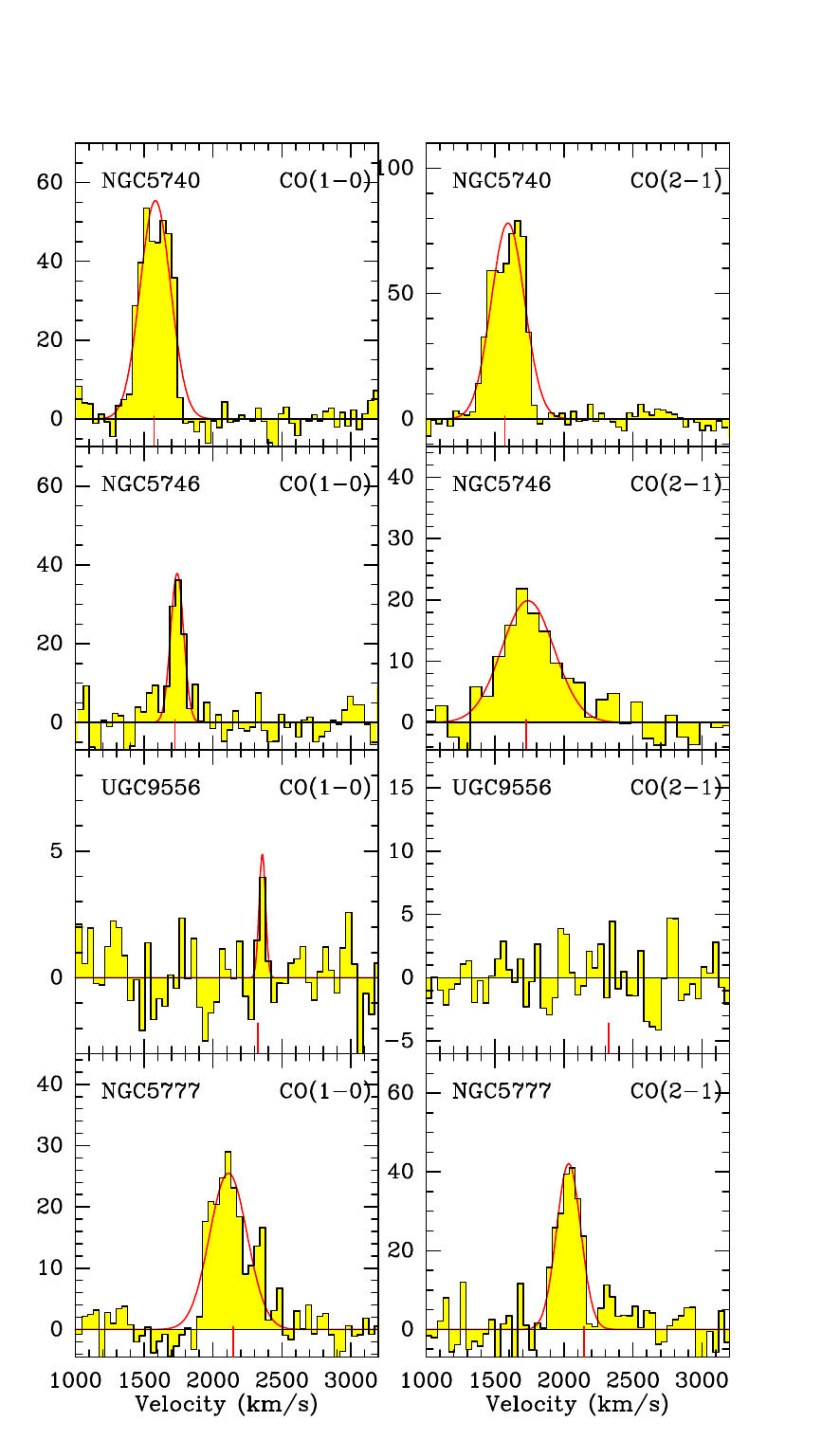}}
\caption{Continued.}
\end{figure*}
\begin{figure*} \centering
\captionsetup[subfigure]{labelformat=empty}
\ContinuedFloat
\subfloat[]{\includegraphics[trim={3.cm 2.cm 3.cm 2cm},clip,page=30,scale=0.9,clip=true]{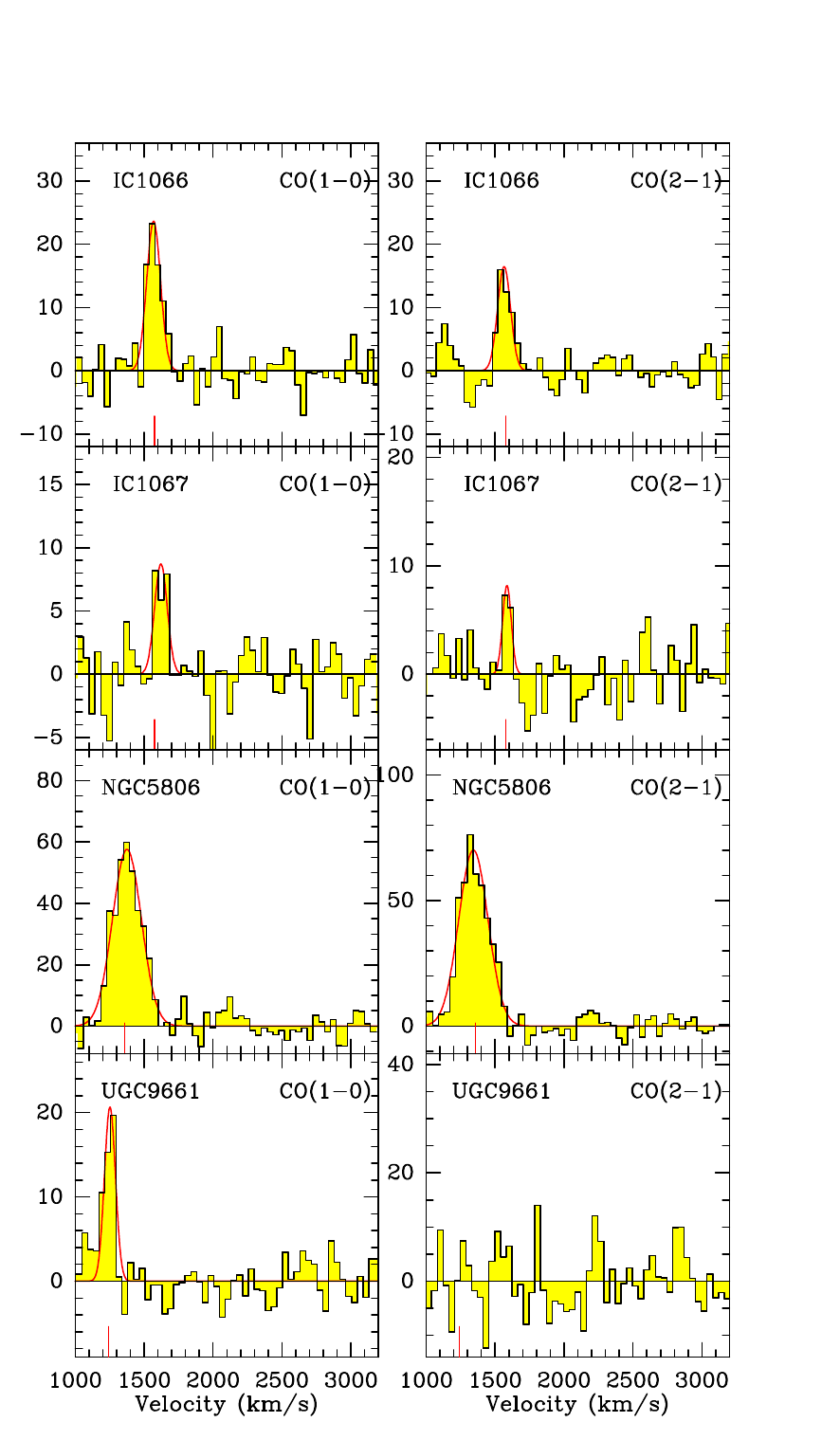}}
\caption{Continued.}
\end{figure*}
\begin{figure*} \centering
\captionsetup[subfigure]{labelformat=empty}
\ContinuedFloat
\subfloat[]{\includegraphics[trim={3.cm 2.cm 3.cm 2cm},clip,page=31,scale=0.9,clip=true]{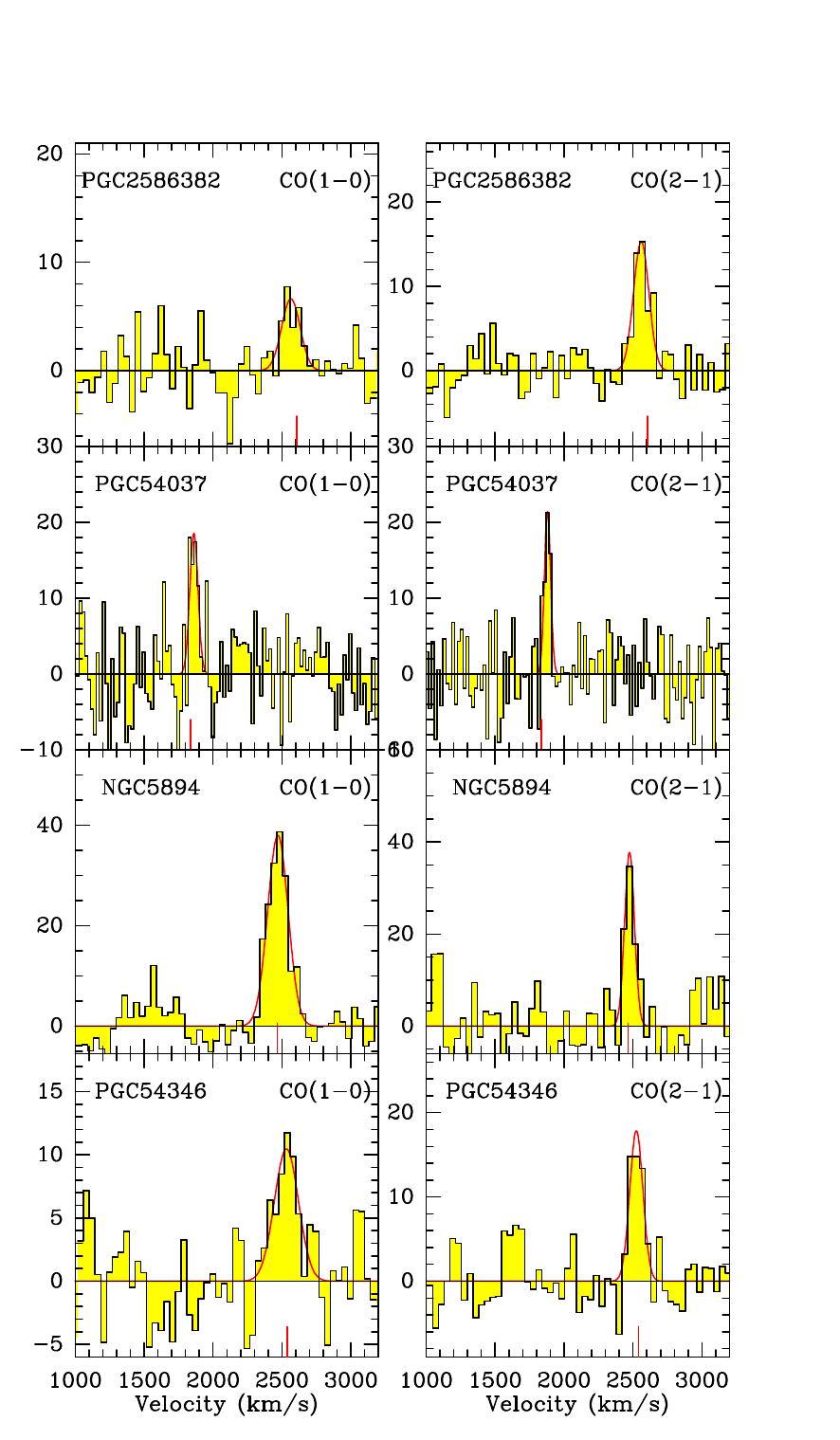}}
\caption{Continued.}
\end{figure*}
\begin{figure*} \centering
\captionsetup[subfigure]{labelformat=empty}
\ContinuedFloat
\subfloat[]{\includegraphics[trim={3.cm 2.cm 3.cm 2cm},clip,page=32,scale=0.9,clip=true]{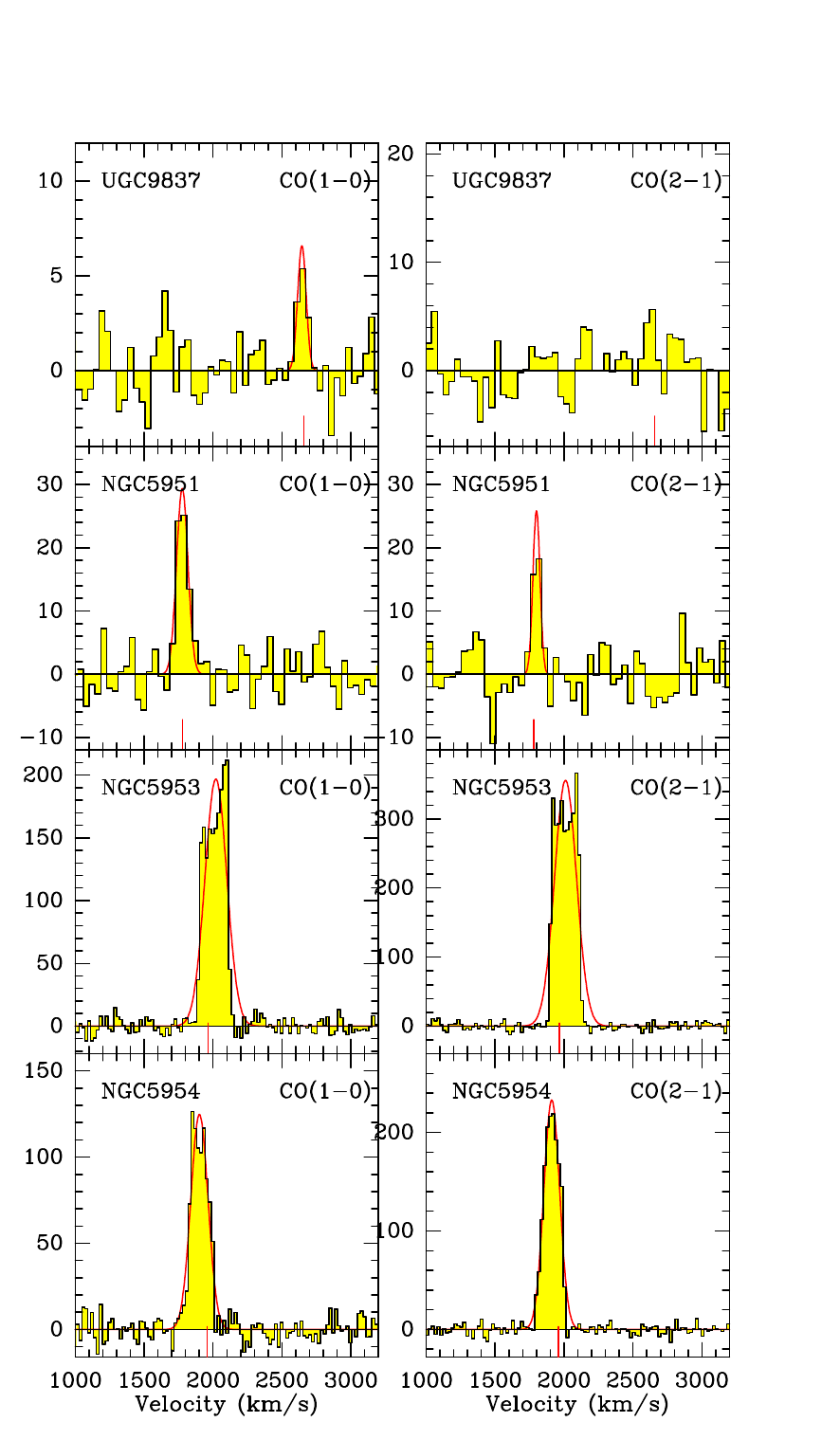}}
\caption{Continued.}
\end{figure*}
\begin{figure*} \centering
\captionsetup[subfigure]{labelformat=empty}
\ContinuedFloat
\subfloat[]{\includegraphics[trim={3.cm 2.cm 3.cm 2cm},clip,page=33,scale=0.9,clip=true]{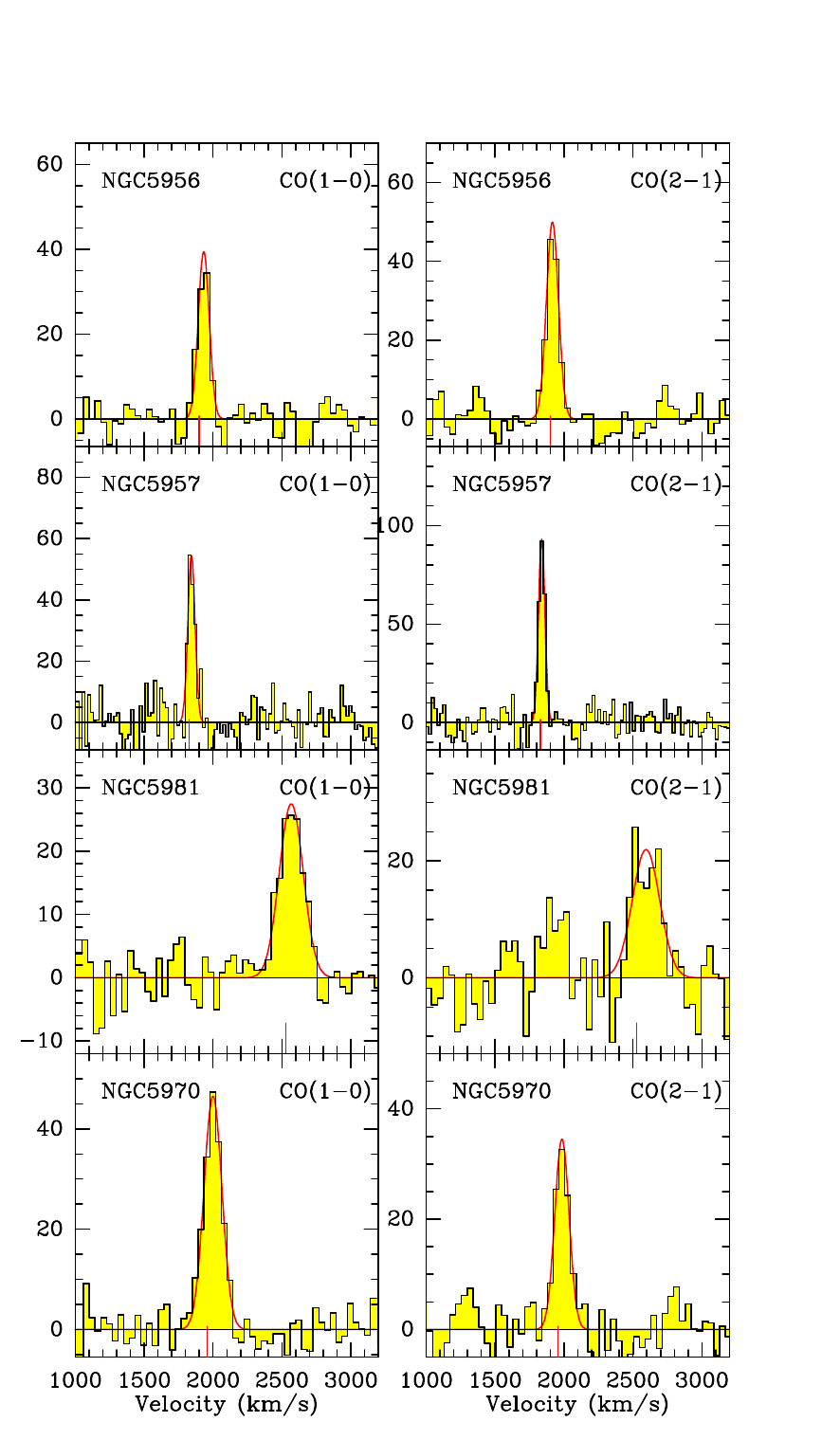}}
\caption{Continued.}
\end{figure*}
\begin{figure*} \centering
\captionsetup[subfigure]{labelformat=empty}
\ContinuedFloat
\subfloat[]{\includegraphics[trim={3.cm 2.cm 3.cm 2cm},clip,page=34,scale=0.9,clip=true]{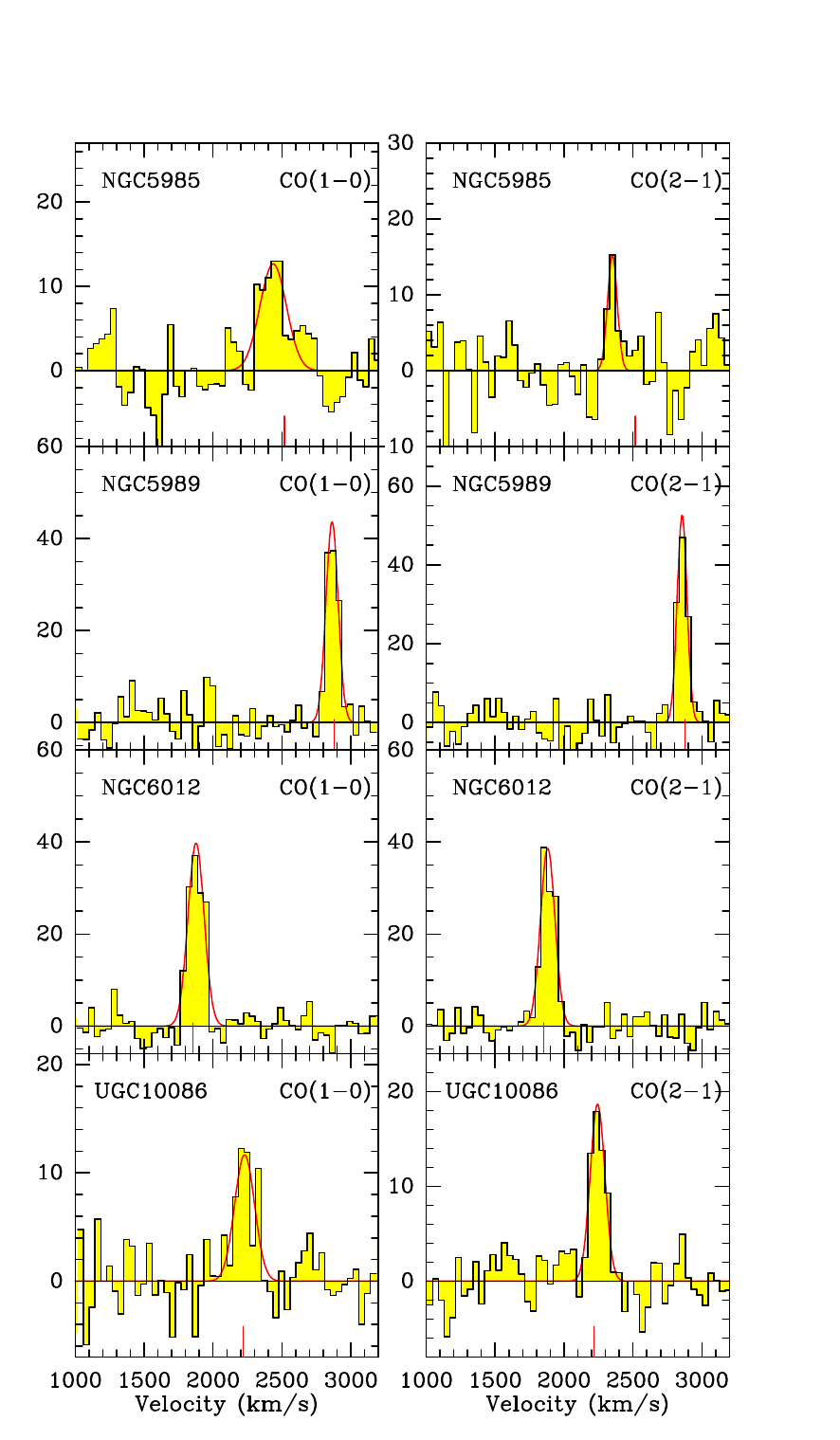}}
\caption{Continued.}
\end{figure*}
\begin{figure*} \centering
\captionsetup[subfigure]{labelformat=empty}
\ContinuedFloat
\subfloat[]{\includegraphics[trim={3.cm 2.cm 3.cm 2cm},clip,page=35,scale=0.9,clip=true]{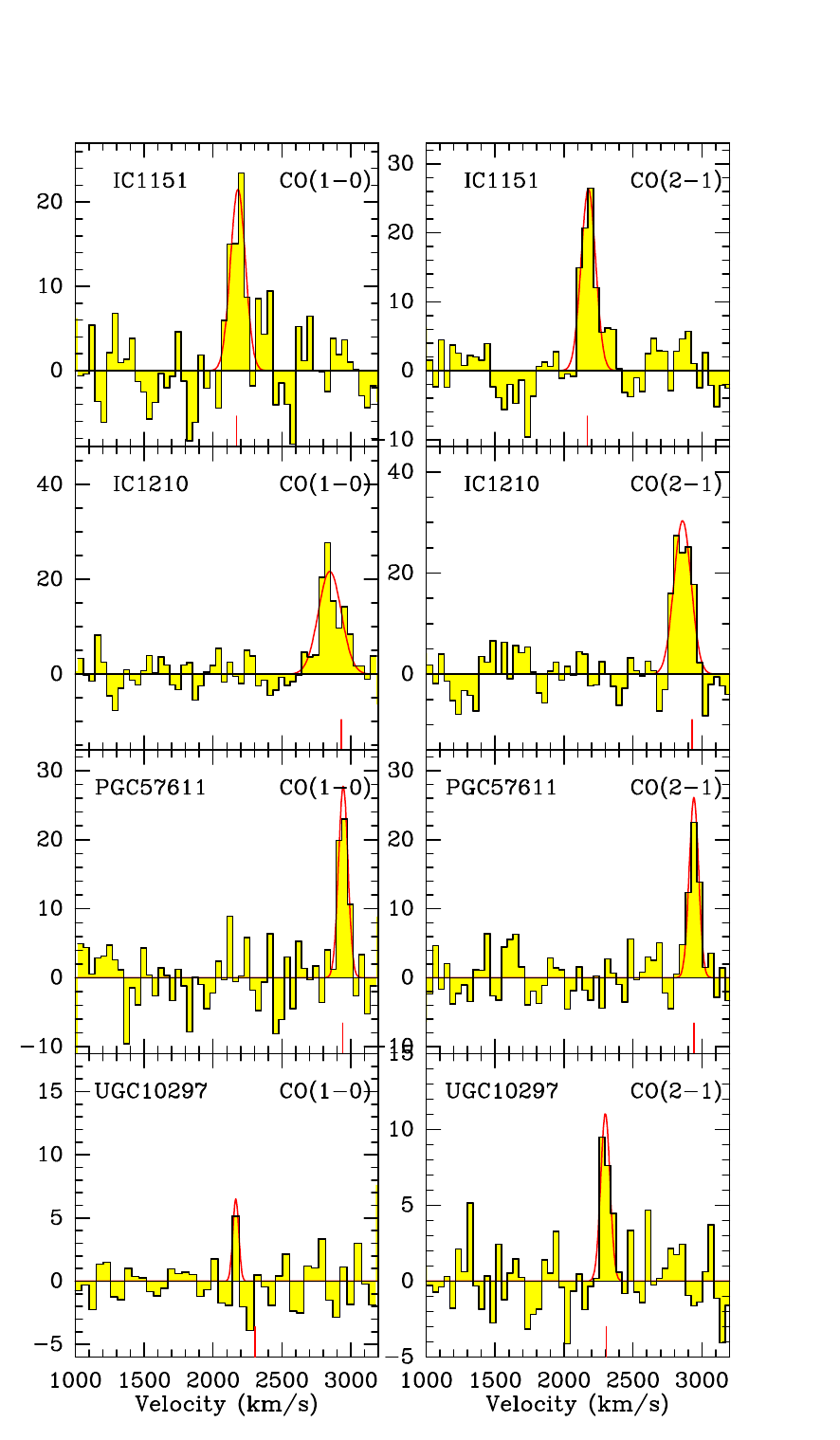}}
\caption{Continued.}
\end{figure*}
\begin{figure*} \centering
\captionsetup[subfigure]{labelformat=empty}
\ContinuedFloat
\subfloat[]{\includegraphics[trim={3.cm 2.cm 3.cm 2cm},clip,page=36,scale=0.9,clip=true]{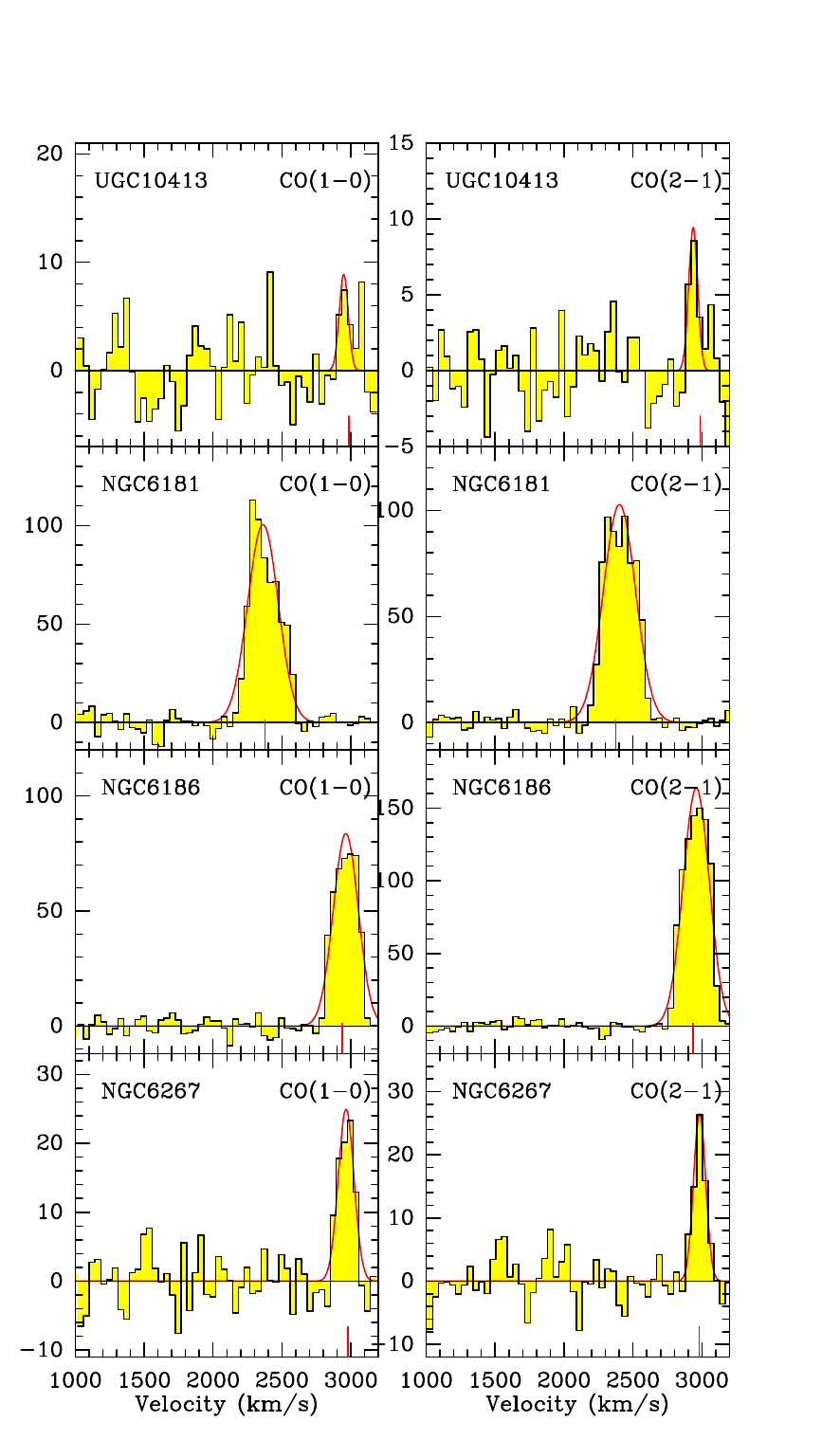}}
\caption{Continued.}
\end{figure*}
\begin{figure*} \centering
\captionsetup[subfigure]{labelformat=empty}
\ContinuedFloat
\subfloat[]{\includegraphics[trim={3.cm 2.cm 3.cm 2cm},clip,page=37,scale=0.9,clip=true]{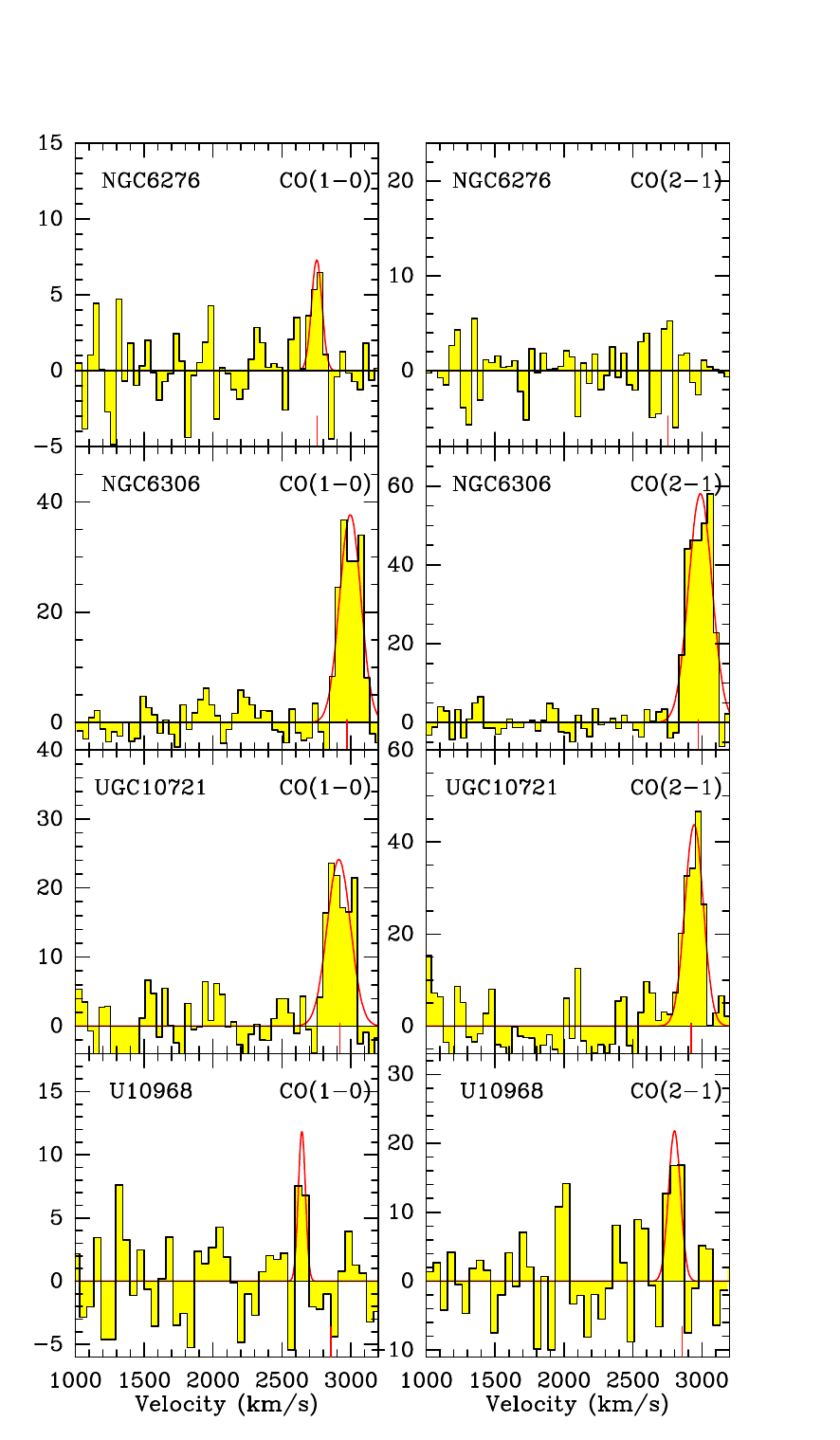}}
\caption{Continued.}
\end{figure*}

\begin{figure*}[!ht]\centering
\captionsetup[subfigure]{labelformat=empty}
\subfloat[]{\includegraphics[trim={3.cm 2.cm 3.cm 3cm},clip,page=1,scale=0.9,clip=true]{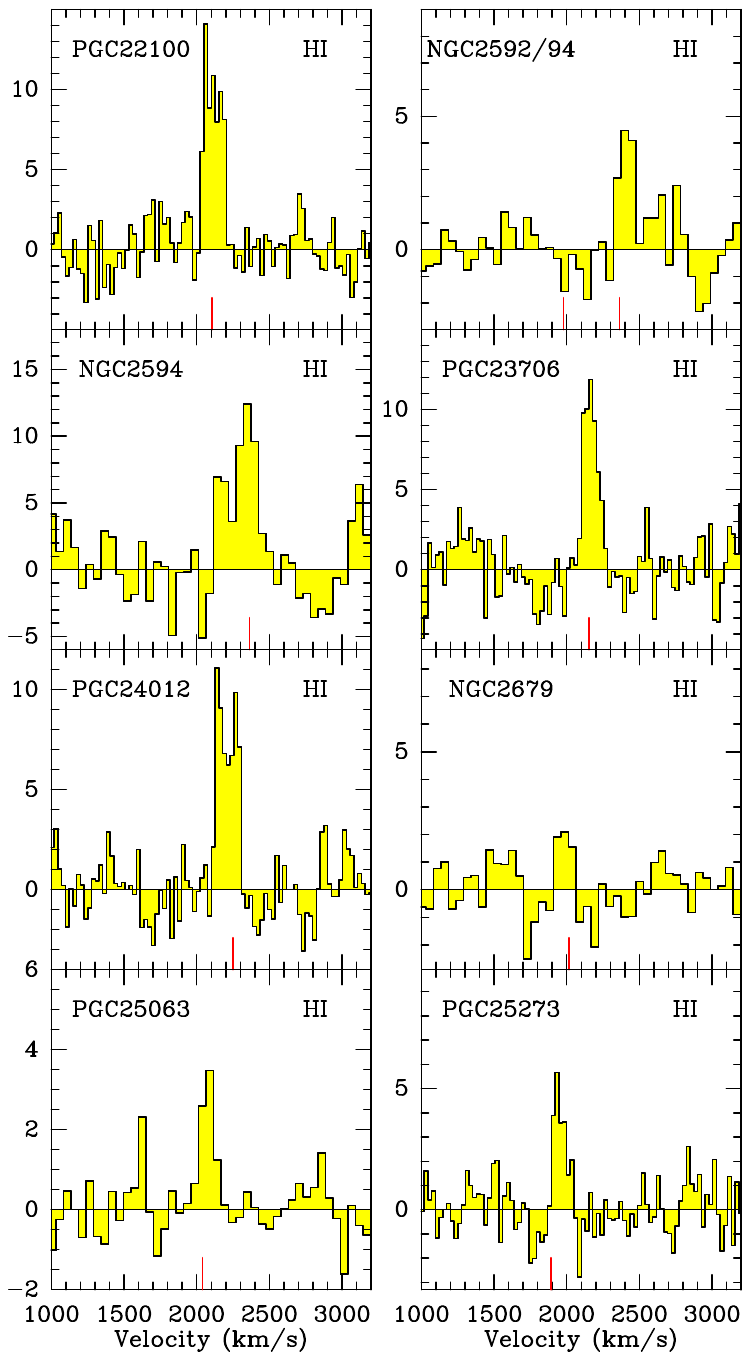}}
\caption{Baseline-subtracted HI spectra from our Nan\c{c}ay campaign for the sources with HI detections. For each spectrum, the x-axis shows the relative velocity, while in the y-axis { the flux is shown in units of mJy}. For each spectrum,  the vertical red segment shows the heliocentric velocity of the galaxy. The HI spectrum of NGC~2592 ($V_{\rm H}=1979$~km/s) is affected by confusion, as indeed the nearby NGC~2594 ($V_{\rm H}=2362$~km/s) is detected within the beam. In the spectrum (denoted as NGC~2592/4) we report the heliocentric velocities of both galaxies as vertical segments.}\label{fig:Nancay_HI_spectra}
\end{figure*}
\begin{figure*} \centering
\captionsetup[subfigure]{labelformat=empty}
\ContinuedFloat
\subfloat[]{\includegraphics[trim={3.cm 2.cm 3.cm 2cm},clip,page=2,scale=0.9,clip=true]{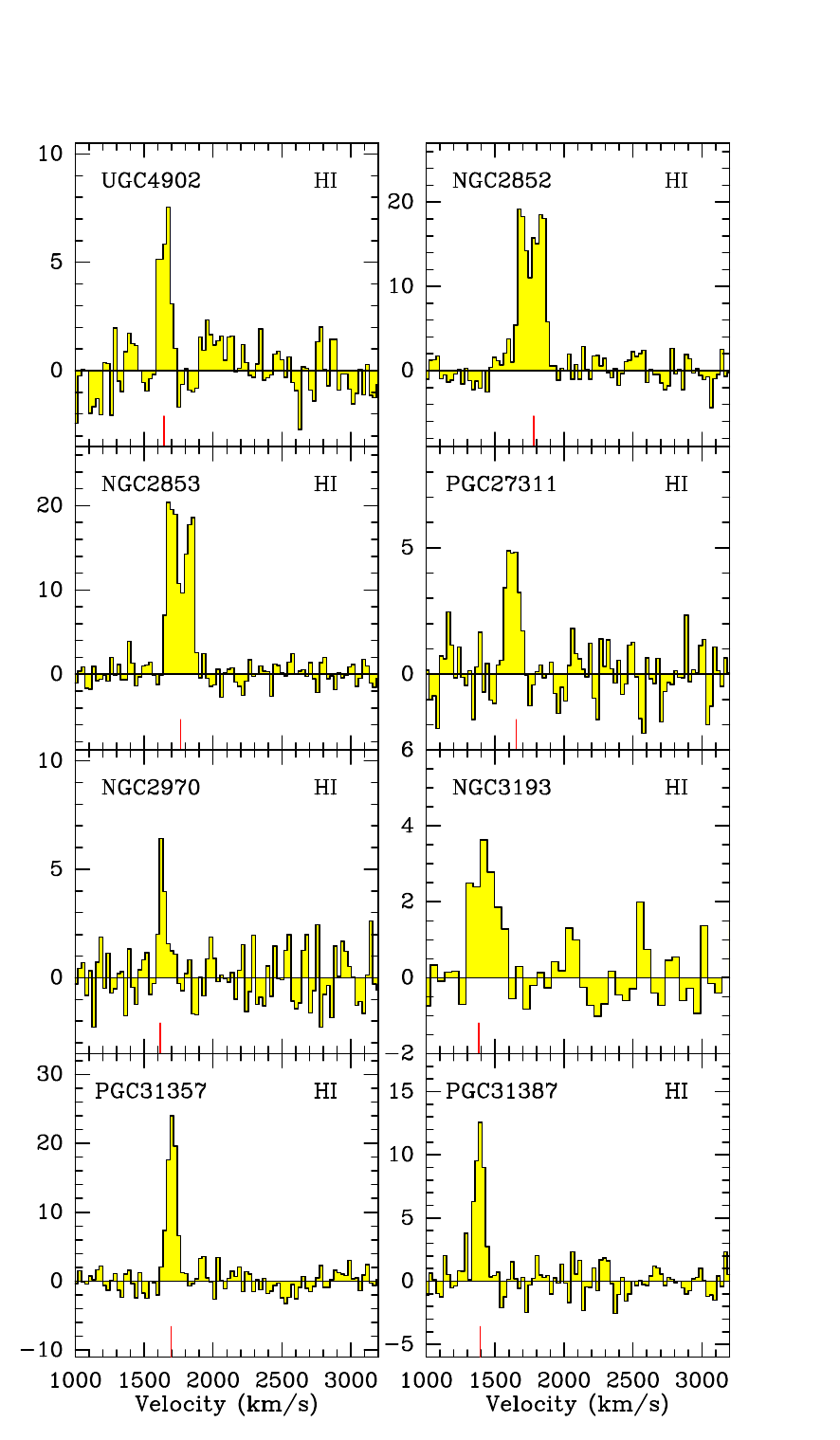}}
\caption{Continued.}
\end{figure*}
\begin{figure*} \centering
\captionsetup[subfigure]{labelformat=empty}
\ContinuedFloat
\subfloat[]{\includegraphics[trim={3.cm 2.cm 3.cm 2cm},clip,page=3,scale=0.9,clip=true]{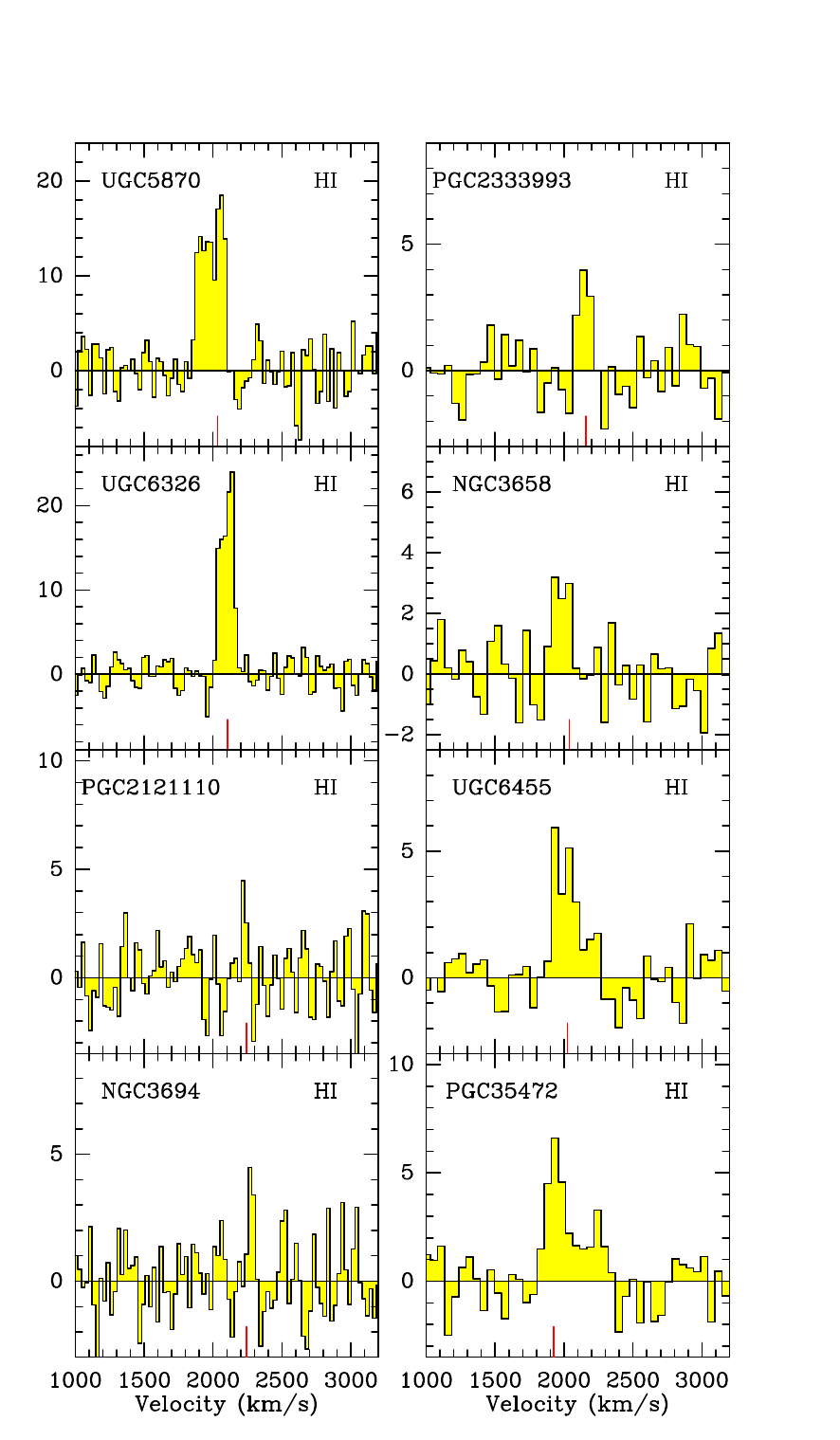}}
\caption{Continued.}
\end{figure*}
\begin{figure*} \centering
\captionsetup[subfigure]{labelformat=empty}
\ContinuedFloat
\subfloat[]{\includegraphics[trim={3.cm 2.cm 3.cm 2cm},clip,page=4,scale=0.9,clip=true]{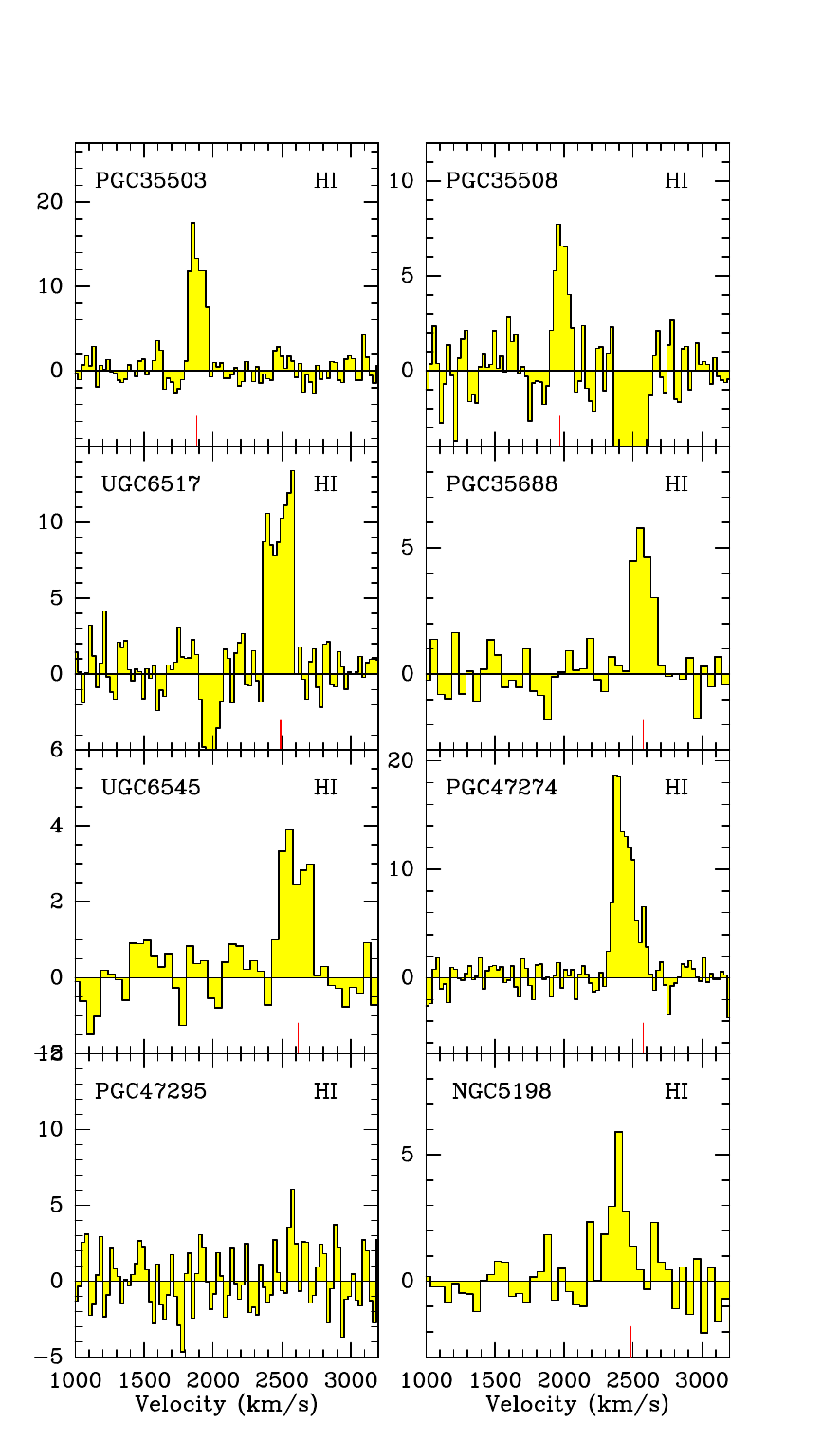}}
\caption{Continued.}
\end{figure*}
\begin{figure*} \centering
\captionsetup[subfigure]{labelformat=empty}
\ContinuedFloat
\subfloat[]{\includegraphics[trim={3.cm 2.cm 3.cm 2cm},clip,page=5,scale=0.9,clip=true]{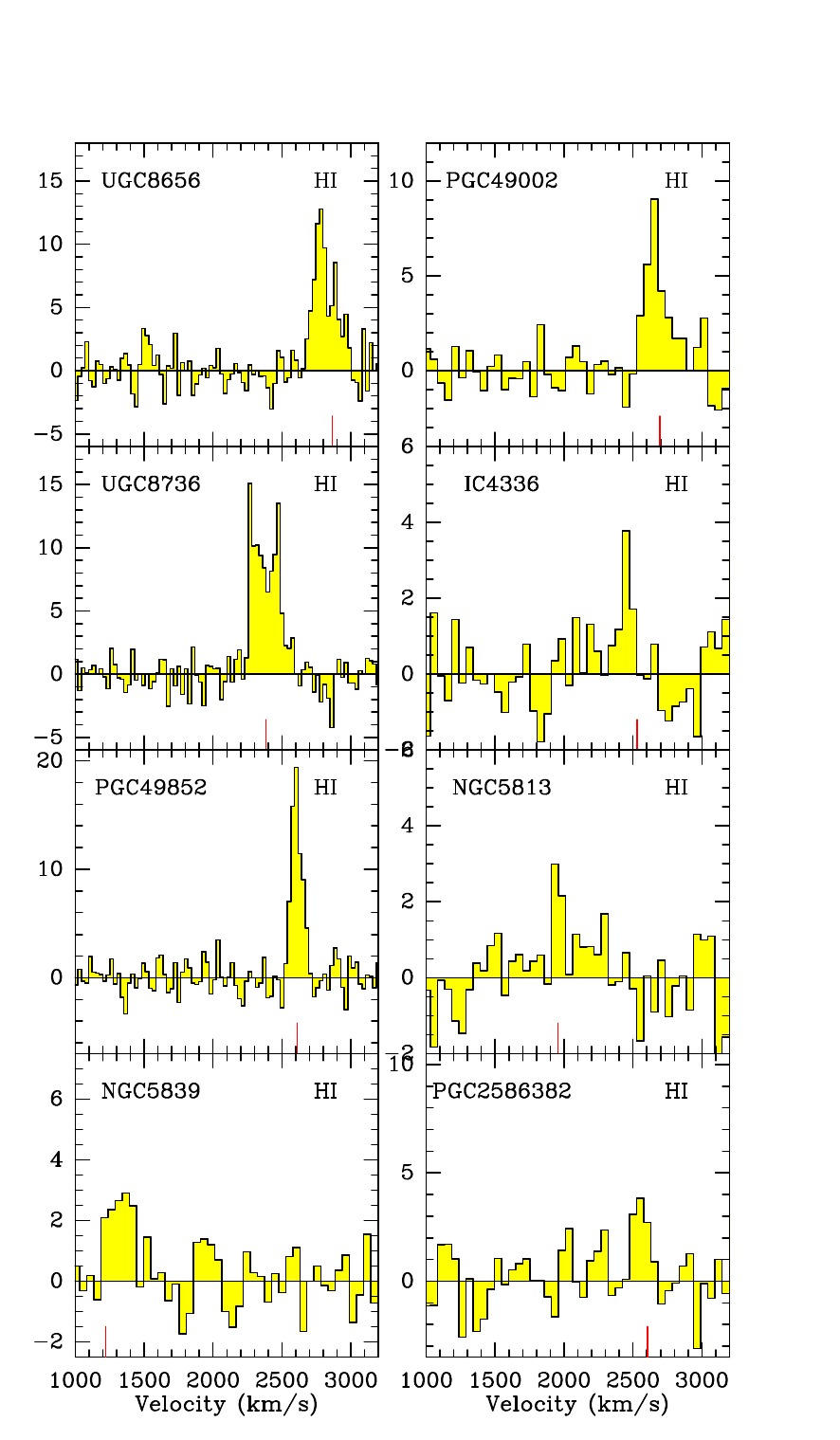}}
\caption{Continued.}
\end{figure*}
\begin{figure*} \centering
\captionsetup[subfigure]{labelformat=empty}
\ContinuedFloat
\subfloat[]{\includegraphics[trim={3.cm 2.cm 3.cm 2cm},clip,page=6,scale=0.9,clip=true]{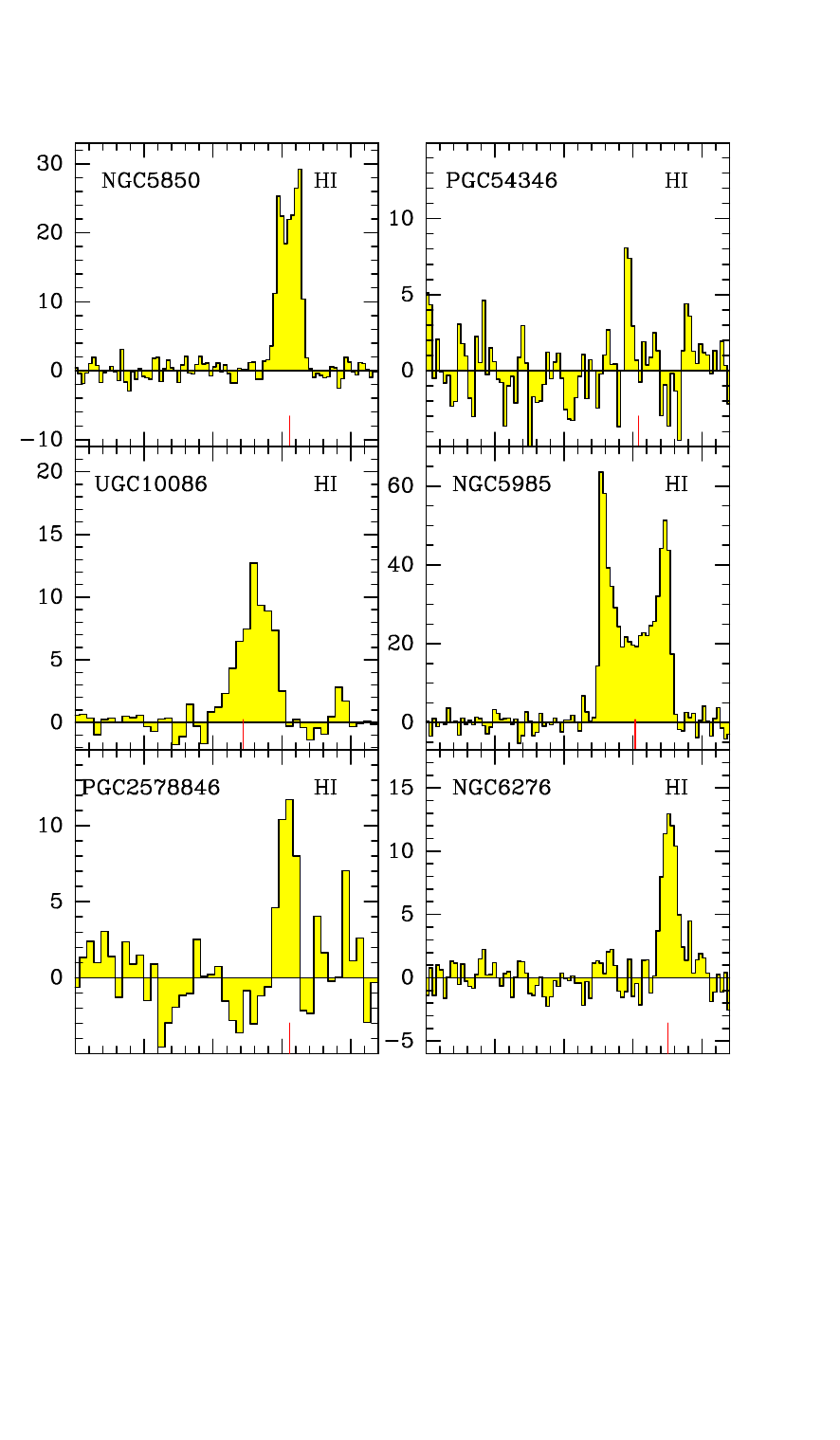}}
\caption{Continued.}
\end{figure*}

\FloatBarrier

\section{Star formation, stellar mass, and gas content diagnostic}\label{sec:SFR_gas_diagnostic}

 \begin{figure}[]\centering
\captionsetup[subfigure]{labelformat=empty}
\subfloat[]{\hspace{0.cm}\includegraphics[trim={0cm 0cm 0cm 
0cm},clip,width=0.5\textwidth,clip=true]{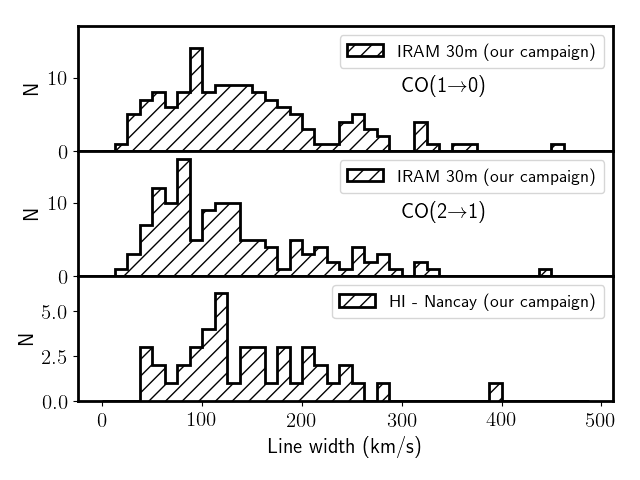}}
\caption{{ Line width distributions} for CO(1$\rightarrow$0), CO(2$\rightarrow$1), and HI from our IRAM-30m and Nan\c{c}ay campaigns.}
\label{fig:histo_FWHM_CO_HI}
\end{figure}

\begin{figure*}[]\centering
\captionsetup[subfigure]{labelformat=empty}
\subfloat[]{\hspace{0.cm}\includegraphics[trim={0.2cm 0cm 2.8cm 
0cm},clip,width=0.33\textwidth,clip=true]{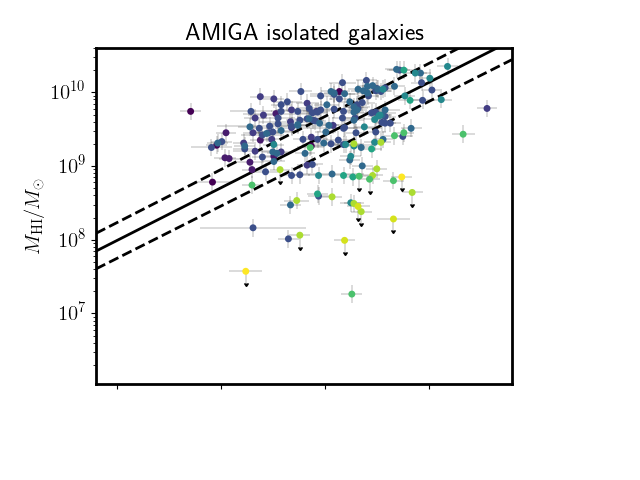}}
\subfloat[]{\hspace{-0.1cm}\includegraphics[trim={2.cm 0cm 1cm 
0cm},clip,width=0.33\textwidth,clip=true]{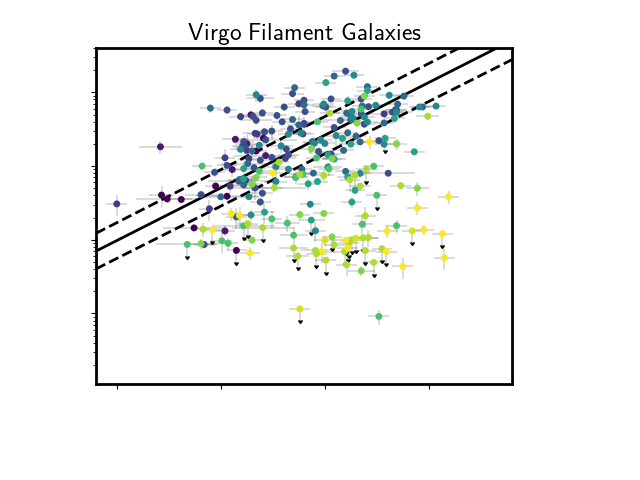}}
\subfloat[]{\hspace{-0.9cm}\includegraphics[trim={2.cm 0cm 1cm 
0cm},clip,width=0.33\textwidth,clip=true]{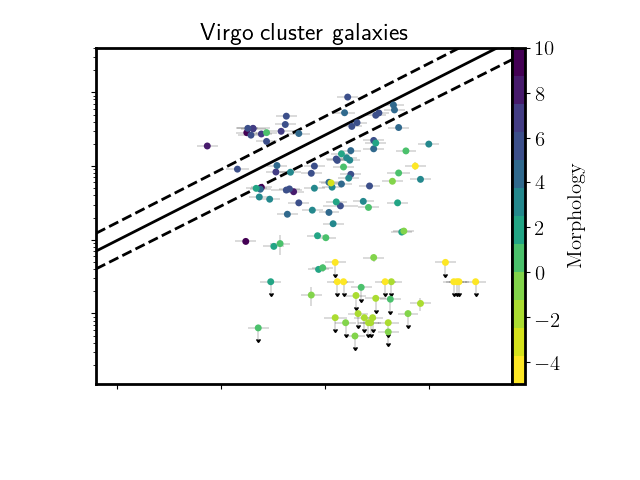}} \\
\vspace{-1.9cm}
\subfloat[]{\hspace{0.cm}\includegraphics[trim={0.2cm 0cm 2.8cm 1cm},clip,width=0.33\textwidth,clip=true]{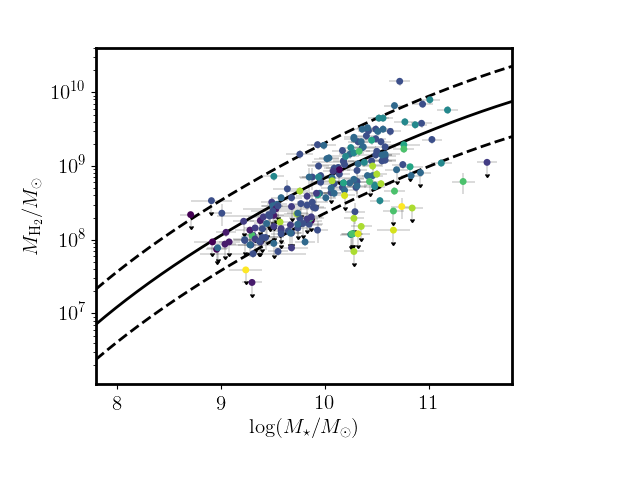}}
\subfloat[]{\hspace{-0.1cm}\includegraphics[trim={2.cm 0cm 1cm 1cm},clip,width=0.33\textwidth,clip=true]{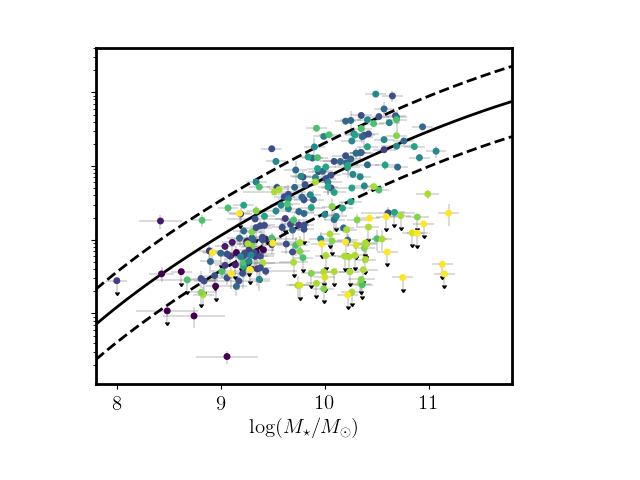}}
\subfloat[]{\hspace{-0.9cm}\includegraphics[trim={2.cm 0cm 1cm 
1cm},clip,width=0.33\textwidth,clip=true]{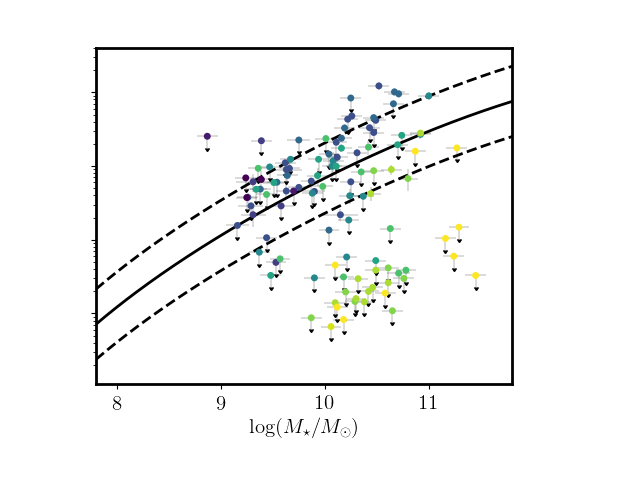}}
\caption{HI mass (top row) and H$_2$ mass (bottom row) plotted against $M_\star$ for the Virgo filament sources in our sample (center), AMIGA isolated galaxies (left), and Virgo cluster galaxies (right). Sources are color-coded according to their morphological classification. Solid lines correspond to local prescriptions at the MS, while 
dashed lines denote the uncertainties. For HI  we used the prescription by \citet{DeLooze2020}, while for H$_2$ that  by \citet{Tacconi2018}, calibrated using a Galactic conversion factor $\alpha_{\rm CO}=4.3~M_\odot$~(K~km~s$^{-1}$~pc$^{2}$)$^{-1}$.}\label{fig:MHI_MH2_vs_Mstar}
\end{figure*}

In this section we report diagnostic plots between gas content, stellar mass, and star formation. The aim of the analysis is to show the consistency of all essential galaxy properties used in this work (i.e., $M_\star$, SFR, $M_{\rm HI}$, and $M_{\rm H_2}$). 

\subsection{Line widths, gas masses, and stellar masses}

Figure~\ref{fig:histo_FWHM_CO_HI} shows the { line width} distribution of the sources observed and detected in CO(1$\rightarrow$0), CO(2$\rightarrow$1), or HI with our campaigns. The associated median values are { FWHM=}$129^{+103}_{-62}$~km~s$^{-1}$ for CO(1$\rightarrow$0), { FWHM=}$107^{+97}_{-55}$~km~s$^{-1}$ for CO(2$\rightarrow$1), ${{\rm W_{50}}=139^{+75}_{-56}}$~km~s$^{-1}$ for HI. The reported values are fairly consistent between each other and safely below the value of 300~km~s$^{-1}$ that we adopted to estimate conservative upper limits, as discussed in  Sect.~\ref{sec:observations}. { The CO line widths were derived from single Gaussian fits (see Sect.~\ref{sec:CO_IRAM30m_observations}). However, the galaxy NGC~5311 is an exception, as the CO(1$\rightarrow$0) emission was fit with three Gaussian components. For this source, the FWHM=$(452\pm34)$~km~s$^{-1}$ reported in Table~\ref{tab:CO_properties_IRAM30m} corresponds to line width for the central component in the spectrum (Fig.~\ref{fig:30m_spectrum}), while the other two are much narrower in velocity.}

In Fig.~\ref{fig:MHI_MH2_vs_Mstar} we report HI and H$_2$ gas masses as a function of the stellar mass for our sample of Virgo filament galaxies as well as the AMIGA field isolated galaxies  and Virgo cluster galaxies for a comparison.
In the $M_{\rm HI}$ versus $M_\star$ plots we overlay the relation by \citet{DeLooze2020} for local field galaxies evaluated at the MS using the SFR versus $M_\star$ prescription by \citet{Leroy2019}.
In the $M_{\rm H_2}$ versus $M_\star$ plots
we overlay instead the local $M_{\rm H_2}$ versus $M_\star$ relation by \citet{Tacconi2018} for MS galaxies, calibrated using the Galactic conversion factor $\alpha_{\rm CO}=4.3~M_\odot$~(K~km~s$^{-1}$~pc$^{2}$)$^{-1}$ used in this work. We note that  \citet{Tacconi2018} adopted a metallicity-dependent conversion.

Within the range of stellar masses $\log(M_\star/M_\odot)\sim9-11$ considered, LTGs overall follow the field scaling relations for both HI and H$_2$, although with a large associated dispersion.
\footnote{For LTGs the median logarithmic difference 
between $M_{\rm HI}$ and the MS prediction is  $(0.18^{+0.31}_{-0.53})$~dex (AMIGA), $(-0.01^{+0.46}_{- 0.51})$~dex (filaments), and $(-0.48^{+0.75}_{- 0.74})$~dex (Virgo cluster). For $M_{\rm H_2}$ the median offsets are: $(-0.14^{+0.38}_{-0.30})$~dex (AMIGA), $(-0.18^{+0.52}_{-0.37})$~dex (filaments), and $(0.20^{+0.35}_{-0.77})$~dex (Virgo cluster).}
On the other hand, the H$_2$ and HI content of ETGs is significantly lower than what is expected for MS galaxies at a given stellar mass.\footnote{For ETGs the median logarithmic difference  between $M_{\rm HI}$ and the MS prediction is  
$(-1.11^{+0.55}_{-0.24})$~dex (AMIGA), $(-1.37^{+1.00}_{-0.47})$~dex (filaments), and $(-2.61^{+0.67}_{-0.33})$~dex (Virgo cluster). For $M_{\rm H_2}$ the median offsets are: $(-0.55^{+0.36}_{-0.41})$~dex (AMIGA), $(-1.06^{+0.63}_{-0.34})$~dex (filaments), and $(-1.78^{+1.32}_{-0.16})$~dex (Virgo cluster).}
These results imply that LTGs in the three different environments (field, filaments, cluster) overall follow the local MS relations for field galaxies. However, this does not apply to ETGs. Indeed, as discussed further in the text (Sect.~\ref{sec:Mstar_SFR_environment}), while the majority of LTGs are within the MS,  ETGs are preferentially below the MS and in the quenching phase.

For the three different environments (field, filaments, cluster) there are thus strong similarities in what concerns the distribution and split of ETGs and LTGs, separately, when comparing  the gas mass { versus} $M_\star$ with the SFR versus $M_\star$ plots. This motivated us to further investigate how the gas content traces the star formation specifically in filament galaxies, as described below.

\subsection{SFR versus gas mass}
Figures~\ref{fig:scatter_plot_SFR_vs_MHI} and \ref{fig:scatter_plot_SFR_vs_MH2} display the SFR against the HI and H$_2$ gas masses, respectively, for filament galaxies as well as for AMIGA and Virgo cluster galaxies for comparison. Similarly to our previous plots, we overlay here the scaling relations for MS field galaxies by \citet{DeLooze2020} and \citet{Tacconi2018}, respectively. 

Filament galaxies overall follow the field MS scaling relations between the SFR and gas mass. The agreement with the scaling relations is found for both HI and H$_2$, even if the latter traces better the ongoing star formation. This is true for the field \citep{Bigiel2008,Schruba2011,Leroy2013}, and we show with this work that this is also valid for filament galaxies. The rms of the SFR around the scaling relation is found to be $\sim0.53$~dex when considering the  SFR versus $M_{\rm H_2}$ scatter plot. It is thus lower than that of $\sim0.67$~dex around the SFR versus $M_{\rm HI}$ relation. 

Furthermore, for all three considered environments, while in the gas mass versus $M_\star$ and in the SFR versus $M_\star$ plots LTGs and ETGs are fairly split, in the SFR versus gas mass (HI, H$_2$) plane they both nicely follow the MS relations for field galaxies, with only some exceptions discussed below. This suggests a self-consistency of the adopted $\alpha_{\rm CO}$ conversion factor, as well as an overall universality in the way H$_2$ gas is consumed to form new stars. This is translated into a limited scatter for the star formation efficiency or for its inverse, the depletion time-scale, as further discussed in Sect.~\ref{sec:depletion_time}.
The above-mentioned exceptions are those galaxies in filaments, as well as in AMIGA and in the cluster, with only upper limits to the gas content, at the low-gas-mass end $\lesssim10^8~M_\odot$. As seen in Figs.~\ref{fig:scatter_plot_SFR_vs_MHI} and \ref{fig:scatter_plot_SFR_vs_MH2}, this region of low gas masses (HI, H$_2$) is mostly populated by ETGs, in filaments and also in the cluster, while for the AMIGA sample it is overall underpopulated, because isolated ETGs are rare. 

Interestingly, a fraction of these ETGs with low $M_{\rm H_2}\lesssim10^8~M_\odot$ also have  low SFR, both in filaments and in the cluster, while others have instead high SFR with respect to the MS, in particular in the Virgo cluster. The former low SFR sources correspond to ETGs in the phase of quenching, where the H$_2$ gas reservoir has been consumed, or is in the process of exhaustion. The latter are instead a population of more star forming ETGs, with low H$_2$ gas content and thus relatively low depletion time (upper limits) of $\tau_{\rm dep}\lesssim10^{8-9}$~yr. These correspond to ETGs that are still forming stars, but are  in a rapid phase of quenching, similarly to their higher-$z$ star forming analogs found  in cluster cores \citep{Castignani2020}; they are experiencing a rapid phase of quenching and will possibly turn, in less than 1~Gyr, into {\it red and dead} galaxies, which are commonly seen in cluster cores.
On the other hand, the observed population of filament and Virgo cluster galaxies with low $M_{\rm HI}\lesssim10^8~M_\odot$ can be explained as galaxies having already experienced the removal of the HI envelope, likely via ram-pressure stripping in dense environments.
Both in filaments and in Virgo, this  population with low $M_{\rm HI}$ is mostly comprised of ETGs, for which the HI content is impacted more than for LTGs.  We refer to the text for further discussion.

\begin{figure*}[h!]\centering
\captionsetup[subfigure]{labelformat=empty}
\subfloat[]{\hspace{0.cm}\includegraphics[trim={0.5cm 0cm 2.8cm 
0cm},clip,width=0.33\textwidth,clip=true]{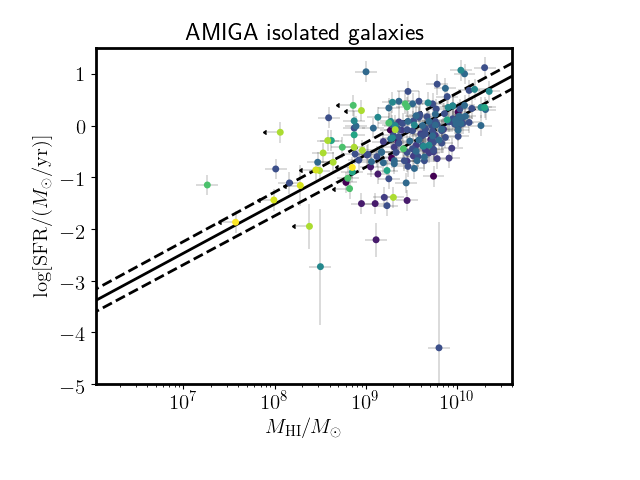}}
\subfloat[]{\hspace{-0.1cm}\includegraphics[trim={2.3cm 0cm 1cm 
0cm},clip,width=0.33\textwidth,clip=true]{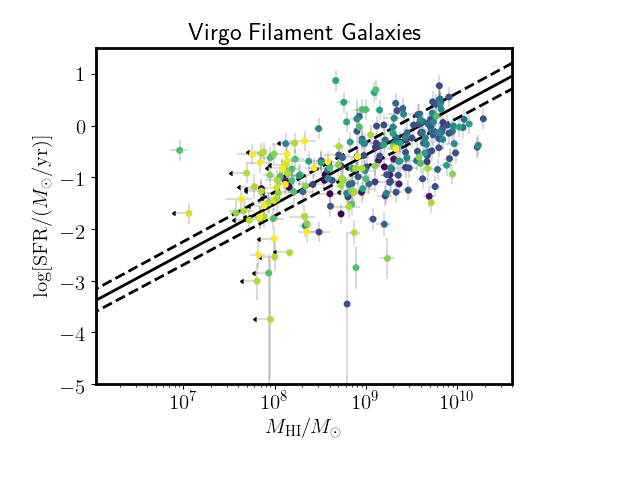}}
\subfloat[]{\hspace{-0.9cm}\includegraphics[trim={2.3cm 0cm 1cm 
0cm},clip,width=0.33\textwidth,clip=true]{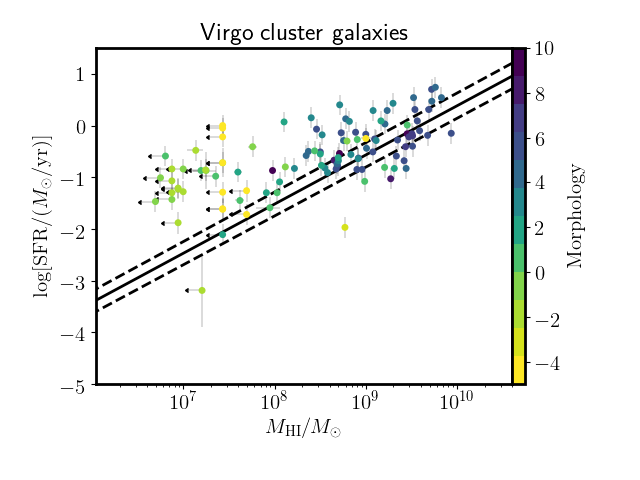}}
\caption{SFR vs. HI mass scatter plot for the Virgo filament sources in our sample (center), AMIGA isolated galaxies (left), and Virgo cluster galaxies (right). Sources are color-coded according to their morphological classification. Solid and dashed lines correspond to the local prescription and model uncertainties by \citet{DeLooze2020} for MS galaxies.}\label{fig:scatter_plot_SFR_vs_MHI}
\end{figure*}

\begin{figure*}[h!]\centering
\captionsetup[subfigure]{labelformat=empty}
\subfloat[]{\hspace{0.cm}\includegraphics[trim={0.5cm 0cm 2.8cm 
0cm},clip,width=0.33\textwidth,clip=true]{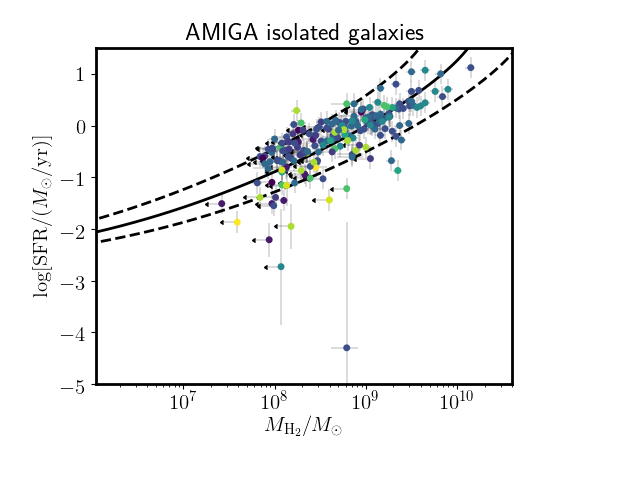}}
\subfloat[]{\hspace{-0.1cm}\includegraphics[trim={2.3cm 0cm 1cm 
0cm},clip,width=0.33\textwidth,clip=true]{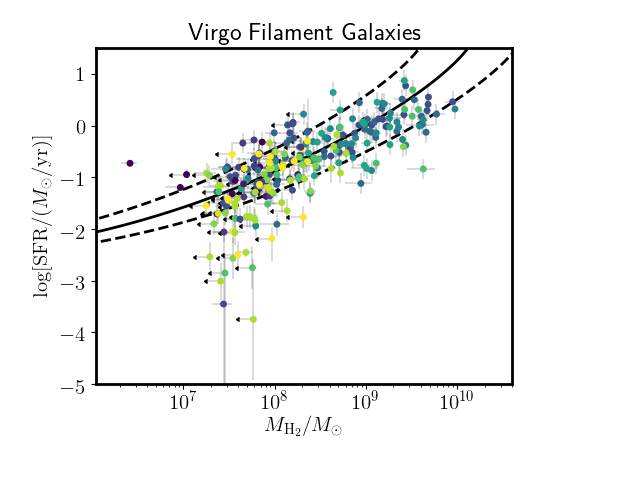}}
\subfloat[]{\hspace{-0.9cm}\includegraphics[trim={2.3cm 0cm 1cm 
0cm},clip,width=0.33\textwidth,clip=true]{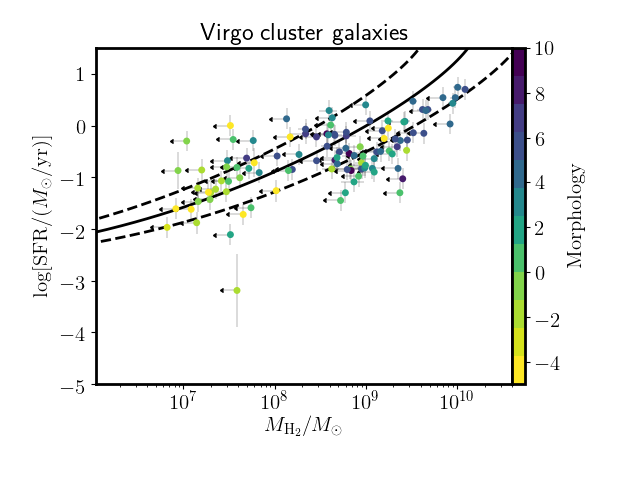}}
\caption{SFR vs. H$_2$ mass scatter plot for the Virgo filament sources in our sample (center), AMIGA isolated galaxies (left), and Virgo cluster galaxies (right). Sources are color-coded according to their morphological classification. Solid and dashed lines correspond to the local prescription and model uncertainties by \citet{Tacconi2018} for MS galaxies, calibrated to a Galactic $\alpha_{\rm CO}=4.3~M_\odot$~(K~km~s$^{-1}$~pc$^{2}$)$^{-1}$.}\label{fig:scatter_plot_SFR_vs_MH2}
\end{figure*}

\end{appendix}

\end{document}